%% file: ms.tex
\documentclass[12pt, letterpaper, twoside, openany]{report}   
\pdfoutput=1   
\usepackage{import}                                 
\usepackage{styleGuide}                          

\begin{document}

\import{./texFiles/frontMatter/}{title.tex}
\import{./texFiles/frontMatter/}{abstract.tex}

\newpage
\import{./texFiles/frontMatter/}{acknowledgements.tex}

\begin{singlespace}
  \tableofcontents
  \listoffigures
  \listoftables
\end{singlespace}

\chapter{Introduction}
\import{./texFiles/mainMatter/introduction/}{backgroundAndMotivation.tex}    
\import{./texFiles/mainMatter/introduction/}{outline.tex}                    

\chapter{Construction of a Detailed LES Model for Flow Over a Permeable Bed}
\import{./texFiles/mainMatter/chapter1/}{introduction.tex}                   
\import{./texFiles/mainMatter/chapter1/}{geometryAndNumericalModel.tex}      
\import{./texFiles/mainMatter/chapter1/}{doubleAveraging.tex}                
\import{./texFiles/mainMatter/chapter1/}{initialStudiesAndResults.tex}       

\chapter{Numerical Aspects of the LES Model: Investigation and Results}
\import{./texFiles/mainMatter/chapter2/}{introduction.tex}                   
\import{./texFiles/mainMatter/chapter2/}{redoGeometryMeshModel.tex}          
\import{./texFiles/mainMatter/chapter2/}{domainSize.tex}                     
\import{./texFiles/mainMatter/chapter2/}{turbulenceModel.tex}                
\import{./texFiles/mainMatter/chapter2/}{conclusions.tex}                

\chapter{Implementation of Particle Tracking within Nalu}
\import{./texFiles/mainMatter/chapter3/}{introduction.tex}                   
\import{./texFiles/mainMatter/chapter3/}{naluStructure.tex}                  
\import{./texFiles/mainMatter/chapter3/}{overview.tex}                       
\import{./texFiles/mainMatter/chapter3/}{particleSearch.tex}                 
\import{./texFiles/mainMatter/chapter3/}{evolutionAlg.tex}                   
\import{./texFiles/mainMatter/chapter3/}{verification.tex}                   

\chapter{Conclusions and Future Work}
\import{./texFiles/mainMatter/conclusion/}{conclusions.tex}                  

\bibliographystyle{unsrt}
\bibliography{./Thesis_Literature.bib}
\end{document}

%% file: texFiles/frontMatter/title.tex
\newenvironment{mycenter}[1][\topsep]
  {\setlength{\topsep}{#1}\par\kern\topsep\center}
  {\par\kern\topsep}

\begin{mycenter}[5pt]
  Northwestern University
\end{mycenter}
\begin{mycenter}[5pt]
\end{mycenter}
\begin{mycenter}[5pt]
  Large Eddy Simulation of Flow Interactions Between a Turbulent Free-Stream and a Permeable Bed
\end{mycenter}
\begin{mycenter}[5pt]
\end{mycenter}
\begin{mycenter}[5pt]
  A DISSERTATION
\end{mycenter}
\begin{mycenter}[5pt]
\end{mycenter}
\begin{mycenter}[5pt]
  SUBMITTED TO THE GRADUATE SCHOOL\\ IN PARTIAL FULFILLMENT OF THE REQUIREMENTS
\end{mycenter}
\begin{mycenter}[5pt]
\end{mycenter}
\begin{mycenter}[5pt]
  for the degree
\end{mycenter}
\begin{mycenter}[5pt]
\end{mycenter}
\begin{mycenter}[5pt]
  MASTER OF SCIENCE
\end{mycenter}
\begin{mycenter}[5pt]
\end{mycenter}
\begin{mycenter}[5pt]
  Field of Mechanical Engineering
\end{mycenter}
\begin{mycenter}[5pt]
\end{mycenter}
\begin{mycenter}[5pt]
  By
\end{mycenter}
\begin{mycenter}[5pt]
  Benjamin H. Sonin
\end{mycenter}
\begin{mycenter}[5pt]
\end{mycenter}
\begin{mycenter}[5pt]
  EVANSTON, ILLINOIS
\end{mycenter}
\begin{mycenter}[5pt]
\end{mycenter}
\begin{mycenter}[5pt]
  July 2017
\end{mycenter}
\thispagestyle{empty}
\newpage

%% file: texFiles/frontMatter/abstract.tex
\section*{Abstract}
Systems comprising a turbulent channel flow overlaying a permeable bed can be found in a variety of 
industrial and natural applications (e.g. urban planning, fracking, submerged vegetation). One
important realization of this system is at the bottom of rivers, where
surface waters of the river permeate the deposited sediment and exchange their contents (e.g. nutrients)
with subsurface waters. Obtaining a complete picture of this process, known as hyporheic exchange, is 
important for the understanding and maintenance of water quality and river ecology \cite{Stonedahl2010}.

Given the wide range of length scales (mm to km) and corresponding time scales of interest, reduced-order 
numerical models (e.g. stochastic particle tracking, advection-dispersion equations) are often used to study 
the transport of mass and momentum in this system \cite{Berkowitz2006,Chakraborty2009}. 
However, the many scales associated with surface/subsurface mixing make the development of a 
high-fidelity model for turbulent scalar transport a formidable challenge \cite{Boano2014}.

In an effort to capture transport behavior at the pore scale, this work is focused on development of a detailed,
three-dimensional flow model for the Large Eddy Simulation (LES) of turbulent flow over a permeable bed.
A rigorous assessment of the computational mesh, domain size and turbulence model is performed. Concurrently,
the double-averaging methodology is used to study the large-scale, persistent flow behavior in the presence
of spatio-temporal heterogeneity. It is found that accurate resolution of the largest turbulent scales
is necessary to develop a complete picture of interfacial momentum transport, and that these large
structures are felt deeper within the bed.

Additionally, the development of a particle tracking software module is detailed. 
This module provides a foundation for the extraction of Lagrangian dispersion information from the LES,
which will eventually be used to physically inform reduced-order models.

%% file: texFiles/frontMatter/acknowledgements.tex
\section*{Acknowledgements}
First and foremost, I would like to thank my adviser, Professor Gregory J. Wagner, for his focused mentorship 
and understanding during my time at Northwestern University. He has been influential in my development as a graduate student,
and without his help, there would be no dissertation to write. I also owe a sincere thank you to those serving 
on my dissertation committee, Professor Aaron I. Packman and Professor Wing Kam Liu. The perspective and guidance
I have gained from you both has helped me to become well rounded, and you have instilled in me a deep respect
for motivations which fuel the work at hand. Finally, I must express gratitude towards my family and friends. 
Your constant support through phone calls and coffee breaks means more to me than I can put into words.
This work has been supported by the Army Research Office (ARO) grant W911NF-15-1-0569, 
Physically-based tempered fractional-order operators for efficient multiscale simulations.

%% file: texFiles/mainMatter/introduction/backgroundAndMotivation.tex
\section{Background and Motivation}\label{sec:thesisIntro}
The work in this thesis is focused on the numerical study of systems in which a rigid permeable 
bed is bounded on one side by an unobstructed surface flow. Here, a \textit{permeable bed} is defined 
as some structure containing adjacent, connected pores throughout which fluid may flow, and the 
term \textit{rigid} is used to specify that the solid constituents of this permeable bed do not move in time. 
Such a system is the subject of study in myriad environmental and engineering applications, 
including submerged vegetation \cite{WHITE2007}, urban planning \cite{Giometto2016} and 
hydraulic fracturing \cite{Shiozawa2016}.

\begin{figure}
  \centerline{\includegraphics[width=0.5\textwidth]{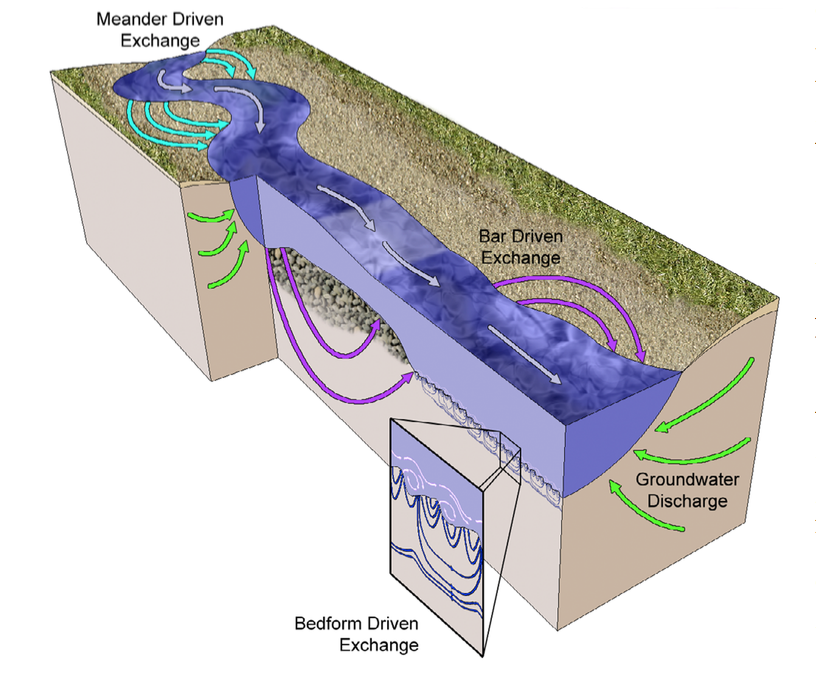}}
  \caption{Hyporheic exchange zones within a river. Taken from \cite{Stonedahl2010}.}
\label{fig:hyporheicZone}
\end{figure}

Another application, which has proved to be a fundamental influence of the work presented in this thesis, 
appears in the study of fluvial systems. Throughout rivers, mass and momentum are exchanged between the 
surface waters of the river channel and the groundwater, or subsurface flow, when fluid permeates the 
bedforms and deposed sediment at the bottom of the channel. Quantifying the complex interactions between 
the surface and subsurface flows and their effects on scalar transport within this region, referred to as 
the \textit{hyporheic zone} and shown in Figure \ref{fig:hyporheicZone}, is critical to the understanding
and study of river ecology and water quality \cite{Stonedahl2010}. Boano et al. recently authored  a 
thorough review of physical mechanisms, numerical models and environmental implications associated 
with hyporheic exchange \cite{Boano2014}. A dominant theme of this work is that the many temporal and 
spatial scales associated with phenomena of interest (e.g. nutrient delivery to sediments) make 
the development of a detailed numerical model which accounts for the broad range of physical activity
seen in this region challenging.

Profiles of concentration break through curves, which generally exhibit heavy tailing, 
have been modeled with some success using
stochastic particle tracking models and fractional advection-dispersion equations (fADE)
\cite{Berkowitz2006,Schumer2009,Chakraborty2009}. Additionally, there has been 
work using the
results of particle tracking simulations to parameterize lower-order models, known as 
\textit{upscaling} \cite{Sund2015}. However, such numerical models often must be parameterized
\textit{ad hoc} and lack physical information at the individual pore scale, where geometry-induced 
flow structure (e.g. vortices) are thought to be a primary cause of the heavy tailing seen
in experiments \cite{Cardenas2008a}.
%

\begin{figure}
  \centerline{\includegraphics[width=0.5\textwidth]{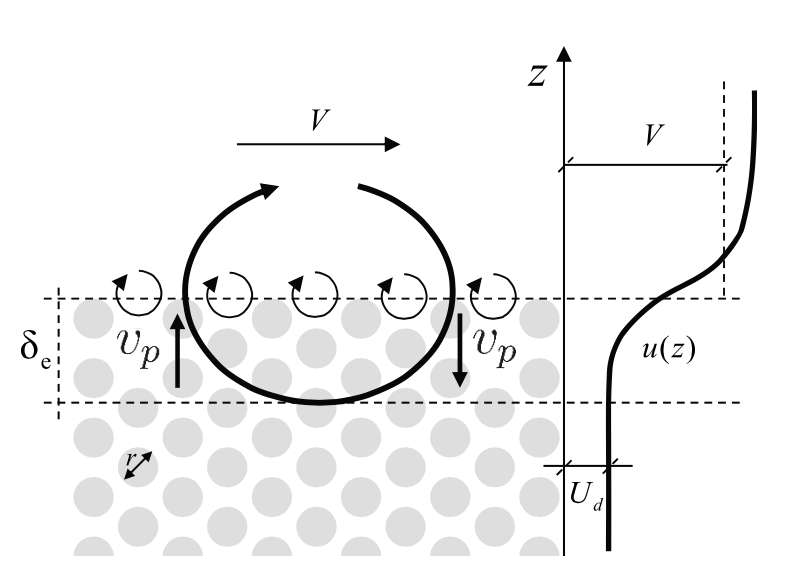}}
  \caption{A conceptual schematic of momentum transport in a permeable bed bounded by a free surface flow.
           Eddies (curled arrows) of different scales are seen moving fluid in and out of the 
           the transition region (a layer of length $\delta_{e}$ experiencing high turbulent activity)
           with a penetration velocity $v_{p}$. The average
           streamwise velocity changes from a constant surface magnitude $V$ to a constant Darcy velocity
           $U_{d}$ within the bed. Taken from \cite{Manes2012}.}
\label{fig:conceptualTransport}
\end{figure}

Significant effort has been allocated towards both understanding and quantifying the interactions
between surface and subsurface waters, and more broadly the interfacial momentum transport, at
the pore scale. An illustration of the driving processes behind momentum transport at the 
surface/subsurface interface is shown in Figure \ref{fig:conceptualTransport}. Interest is often
directed at the length over which turbulent structures penetrate into the bed, $\delta_{e}$, and the 
depth-wise variation in an averaged streamwise velocity profile, $u(z)$. This average
is frequently taken to be a time-space average, rigorously defined in \cite{Nikora2007a}, which eases the 
study of large-scale, persistent structure in such a heterogeneous environment. While in a laminar 
flow regime $\delta_{e}$ has been found to be on the order of a single grain diameter and insensitive 
to the bulk Reynolds number, derived from the mean surface velocity and surface flow height
\cite{Goharzadeh2005}, the introduction of free-stream turbulence
adds complexity to the structure and development of these features.
%

Experiments have shown that the 
time-space averaged velocity is found to decrease dramatically from the surface into the transition region, 
achieving a minimum in the first layer of pores, then increasing towards a constant velocity deep within 
the bed \cite{Pokrajac2007,Manes2009}. Additionally, the friction factor, or ratio of 
frictional to inertial forces, in flows over permeable
beds has been found to increase with $Re$ far beyond the plateau seen in flows over rough walls
\cite{Manes2011a}, and is suspected to be a result of increased turbulent penetration into the bed as 
turbulent structures grow in size \cite{Manes2012}. This penetration significantly alters
the near-wall, or more truly the near-interface, turbulent structure from what is seen in rough wall 
flows, prompting the construction of several modified laws to better describe velocity scaling in 
the logarithmic layer of the flow. It also raises concern that the viscous wall unit may not be 
appropriate for scaling inner variables in such a system \cite{Manes2011,BenMeftah2016,Breugem2006}. 

In addition to these experiments, simulations have proved to be an invaluable tool for 
investigating the driving processes behind surface/subsurface flow coupling. 
Direct numerical simulation (DNS) of turbulent flow over a channel of regularly packed cubes
confirmed that large vortical turbulent structures near the permeable wall are generated
from Kelvin-Helmholtz (KH) type instabilities, which significantly contribute to the skin friction
felt by the surface flow \cite{Breugem2006}. The same study, in concert with
previous work \cite{Breugem2005}, also promotes the use of the 
permeability Reynolds number, $Re_{K}$, and the roughness Reynolds number, $Re_{D}$,
as means to quantify the influence of bed permeability and roughness elements on the flow
field, respectively. DNS has been used to compare flow over permeable and impermeable beds of 
staggered cubes, also confirming the prevalence of interfacial KH instabilities. 
Although, a significant contribution from sweep events (i.e. downward fluid motions with high 
streamwise momentum) to the Reynolds shear stress occurs just 
below the interface, the Reynolds shear stress decays almost entirely over the first grain diameter 
\cite{Kuwata2015,Kuwata2016}. The influence of these hydrodynamic characteristics on heat transfer 
has recently been studied in a similar packed cube geometry, finding enhanced turbulent heat transfer
above the porous interface, due to the presence of large eddies, and large temperature fluctuations
deep within the bed, caused by pressure waves propagating from the transition region \cite{Chandesris2016}. 

While tremendous progress has been been made in understanding these systems, challenges
moving forward are evident. Experimental investigation has large, physical constraints on its ability
to capture spatial and temporal information. Although simulations may compliment experimental work, 
having a much higher spatio-temporal resolution, the computational resources necessary to perform
DNS limit the system of interest to low Reynolds numbers and relatively simple geometries.
One alternative to DNS, which offers a high degree of detail in the resolved flow field without 
imposing the same intractability problems of DNS, is Large Eddy Simulation (LES) \cite{SMAGORINSKY1963,
Deardorff1970}.

The driving principle behind LES is to apply a low-pass filter to the Naiver-Stokes equations and 
model the effect of the unresolved motions with some closure scheme. 
Given that the fine scale motions are not directly computed, the case for LES 
as a useful tool grows stronger when momentum and mass transport is driven by the larger, resolved 
flow structures \cite{Pope2004}. This is certainly suggested by the previously mentioned 
literature for the case of flows over permeable beds. Thus, LES seems to be an reasonable candidate tool 
for investigating a system so rich in structure and host to various phenomena 
(e.g. flow separation and reattachment, surface/subsurface mixing).

There has been limited use of LES to examine flows
over fixed, coarse-gravel beds. Notably, Stoesser et al. simulated turbulent
flow over a three-layer deep bed of spheres, providing evidence that LES is able to predict several 
experimental and DNS results for this type of system (e.g. streamwise velocity profiles and penetration of 
pressure waves deep into the bed) \cite{Stoesser2007}. However, this work does not address the impact of 
many numerical modeling decisions (e.g. turbulence model, domain size) on the resultant flow statistics. 
Although LES is generally tractable, yet expensive to run, the cost associated with 
rigorously testing the numerical model may be orders of magnitude more than a single simulation with respect 
to compute time \cite{Pope2004}. Regardless of this expense, several studies have cemented the need to 
examine variations in the flow field due to changes in turbulence model, domain size, mesh size, etc.,
particularly when using periodic boundary condition in both the streamwise and spanwise directions
\cite{Temmerman2003,Mendez2008,Frohlich2005}.

The goal of the work presented in this thesis is to advance the understanding 
of the driving mechanisms behind interfacial momentum exchange in turbulent flows over permeable beds, 
and how these mechanisms may impact scalar transport. Primarily, progress towards this goal is 
achieved through the development of a suitable numerical model for a highly detailed LES of the 
surface/subsurface system and analysis of its predicted dynamics. Secondary to this project is the 
creation of a Lagrangian particle tracking model and its implementation within an open source fluid dynamics
code base, with an eye on extracting and upscaling dispersion information from the detailed
LES. 

%% file: texFiles/mainMatter/introduction/outline.tex
\section{Outline}

This thesis is the culmination of two currently distinct projects that are united under the
purpose of interrogating the hydrodynamics governing hyporheic exchange. Thus, the first two
chapters following this introduction may be read independently from the third, although future work will integrate these
components.

Chapter 2 is concerned with developing and performing a Large Eddy Simulation (LES) of turbulent 
flow over a permeable bed. The characteristic geometry used in all computational experiments will be 
presented, followed by a discussion on the domain discretization and numerical treatment of the governing 
fluid equations. Finally, the double-averaging methodology for quantities of interest is introduced and 
initial simulation results are discussed.

Chapter 3 follows Chapter 2 chronologically, focusing on several challenges encountered 
while performing the LES. Particular focus is paid to the mesh refinement, issues associated with 
obtaining a representative volume element (RVE) and selection of turbulence closure. Concurrent with
these discussions, results from several different runs of the LES are examined to both gain insights into 
the processes governing momentum transport and highlight the lingering hurdles in modeling 
flow over permeable beds.

Disparate from the previous two chapters, Chapter 4 introduces the addition of a particle tracing module 
to the Nalu code base. A brief introduction to Nalu is given, followed by a thorough presentation of the 
particle module's organization, key algorithms and current performance.

Finally, Chapter 5 summarizes primary conclusions drawn from the numerical experiments and provides 
suggestions for future work on both the LES and particle module.

%% file: texFiles/mainMatter/chapter1/introduction.tex
\section{Introduction}
This chapter discusses the initial development of a detailed Large Eddy Simulation (LES) model
of turbulent flow over a permeable bed of rigid, simple-cubic packed spheres. First, considerations
with respect to the bed geometry, computational mesh and numerical treatment of the governing
fluid equations are discussed. The double-averaging methodology, as discussed in 
\cite{Nikora2007a} and \cite{Nikora2007b}, is presented and manipulated according to the features
of the LES model. Finally, preliminary results are shown for two simulations using different
turbulence models which highlight the need for continued efforts towards creating a suitable
numerical model of the specified system.

%% file: texFiles/mainMatter/chapter1/geometryAndNumericalModel.tex
\section{Geometry and Numerical Model}\label{sec:numericalModel}

\subsection*{Geometry}\label{sec:geomAndModel}
Simulations of turbulent, open-channel flow over a permeable bed are the focus of this chapter,
and as such, modeling efforts begin with the development of an appropriate domain geometry. 
In an effort to numerically replicate the experimental configurations seen in \cite{Blois2012}, 
a geometry fit to treat the surface and subsurface flows within a single domain has 
been constructed. The upper half of the domain models open channel flow above
a permeable wall, while the lower half contains a rigid porous medium, formed from spheres 
of diameter D = 3.8 cm. Each sphere has a 2 mm gap, $d_{g}$, between itself and its nearest neighbor,
which leads to a simple cubic packed structure for the porous medium. This gap is useful 
to avoid challenges associated with singularities in the geometry at contact points while remaining small enough to 
allow for comparing obtained simulation results against a vast body of experimental literature 
(e.g. \cite{Manes2009,Blois2012,Pokrajac2009}).
To simplify discussions concerning bed geometry moving forward, a unit cell is defined to be a cube of length 
L = 4.0 cm which is concentric with a given sphere when in the bed.
\begin{figure}[!tbp]
    \subfloat[2x2x4]{\includegraphics[width=0.45\textwidth]{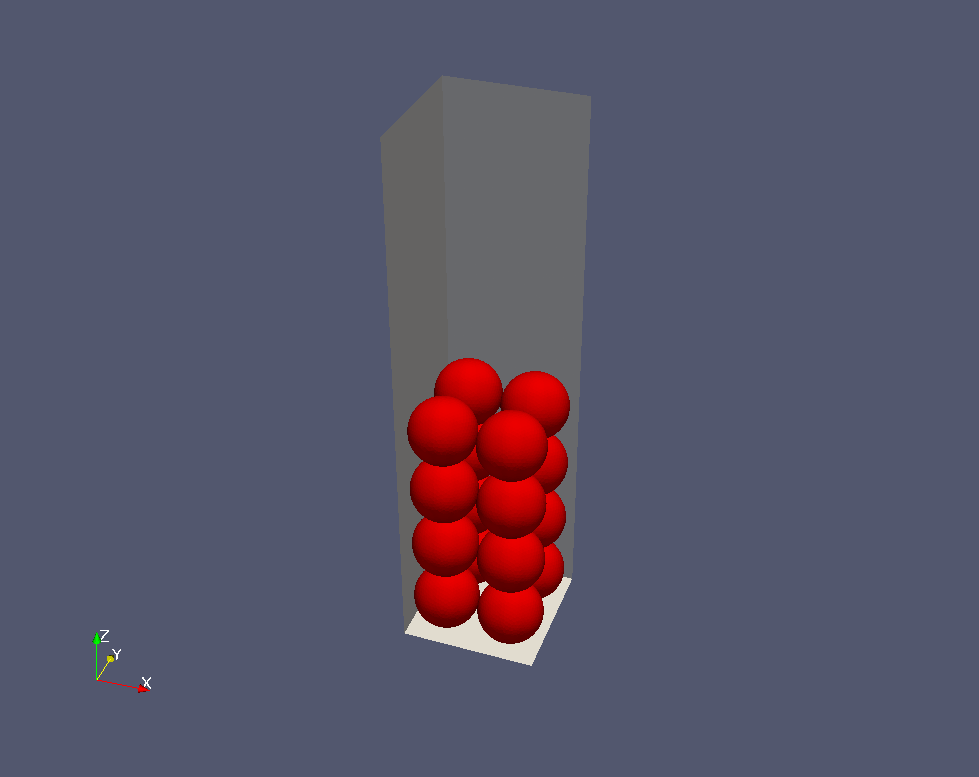}}
    \hfill
    \subfloat[5x3x4]{\includegraphics[width=0.45\textwidth]{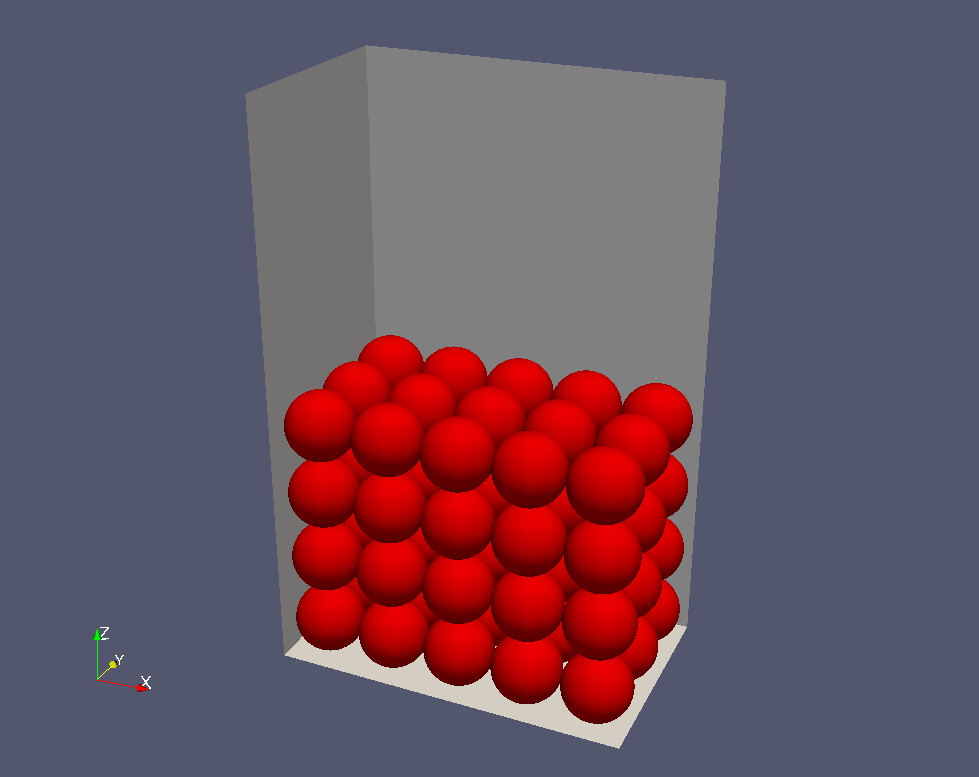}}
    \hfill
    \subfloat[10x5x4]{\includegraphics[width=0.45\textwidth]{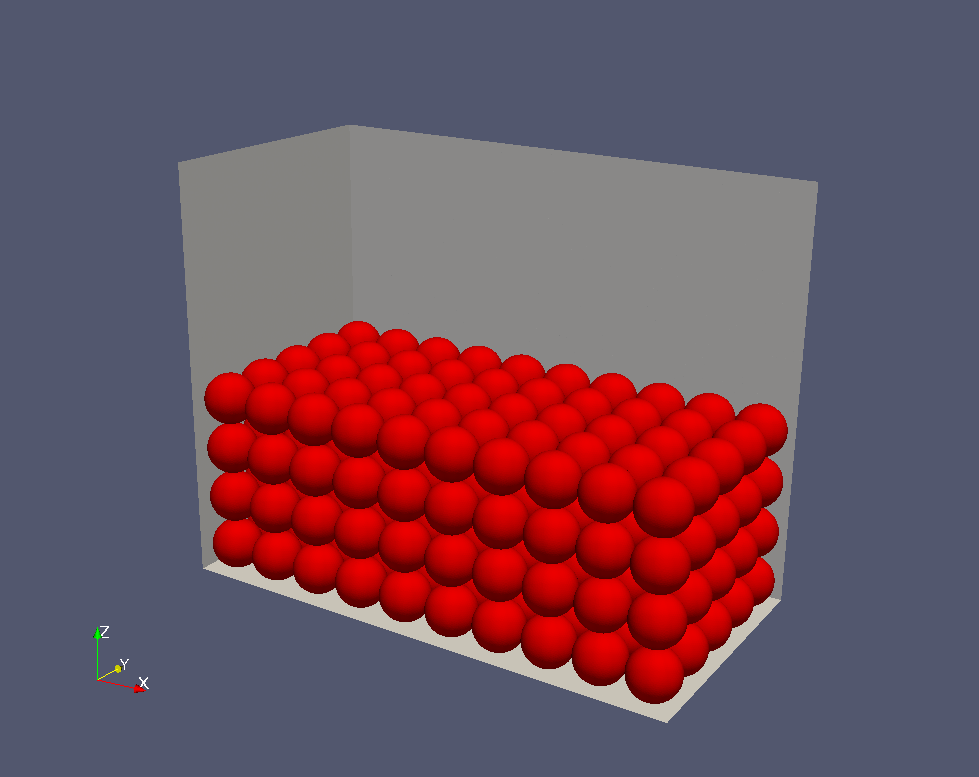}}
    \hfill
    \subfloat[15x7x4]{\includegraphics[width=0.45\textwidth]{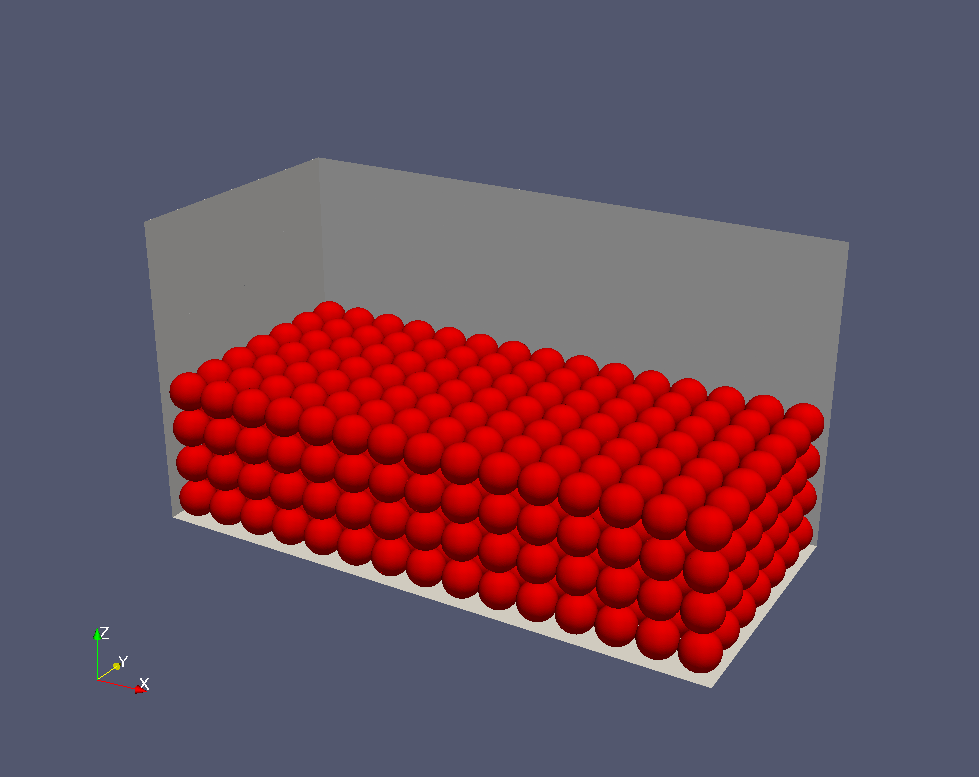}}
    \caption{Several geometries used throughout the thesis are shown. All domains have a height of twice the 
             bed depth.}
    \label{fig:geometries}
\end{figure}

The task of determining a suitable domain size to capture all of the relevant physics at play is not a trivial one.
A primary consideration in selecting appropriate domain dimensions is the need to resolve the largest turbulent scales 
while maintaining a feasible computational work load. Although initial geometries presented in this chapter reflect a
dominant focus on mimicking the experimental configuration mentioned above, proper resolution of turbulence will be 
discussed thoroughly in Section \ref{sec:domainSize}. Figure \ref{fig:geometries} shows several example 
configurations which will be examined throughout the following chapter in an effort to address this concern.

As shown in Figure \ref{fig:geometries}, the right-handed coordinate system is adopted here,
where the $x-$ axis is oriented along the streamwise direction, the $y-$
axis is oriented along the spanwise direction and the $z-$ axis (hereby deemed the wall-normal direction)
is perpendicular to the bottom surface with a positive unit vector pointing outward from the domain's top 
surface. Tensor notation is used with the Einstein summation convention, where $x_{i}$ ($i$ may assume a value of
1, 2 or 3) corresponds to the directions $x$, $y$, and $z$, respectively. Similarly, the velocity components 
$u$, $v$ and $w$ are defined by $u_{1}$, $u_{2}$ and $u_{3}$, respectively. Finally, planes referred to
in terms of \textit{maximum} or \textit{minimum} porosity denote cross-sections of the entire flow domain
taken in the $x-z$ plane which bisect the gap between spheres or the spheres, respectively.

In all cases, boundary conditions have been selected to create a model representative of an infinite bed.
Thus, periodic boundary conditions are applied to the streamwise and spanwise dimensions.
A slip boundary condition is applied to the top surface. By applying 
a slip boundary condition to the bottom surface as well, interfacial dynamics similar to those of 
an infinitely deep bed may be realized with a limited number of spheres in the wall-normal dimension, as 
there is no influence of a boundary layer from the bottom of the computational domain. Several studies 
(e.g. \cite{Pokrajac2009,Stoesser2007}) find that the influence of turbulence on the subsurface flow
is minimal beyond a depth of two to three layers of spheres, so four layers are cautiously taken to be sufficient 
for resolving all dominant behavior while limiting computational cost.

\subsection*{Details Regarding the Computational Mesh}
\begin{figure}[!tbp]
    \centering  
    \subfloat[]
        {\includegraphics[width=0.5\textwidth]{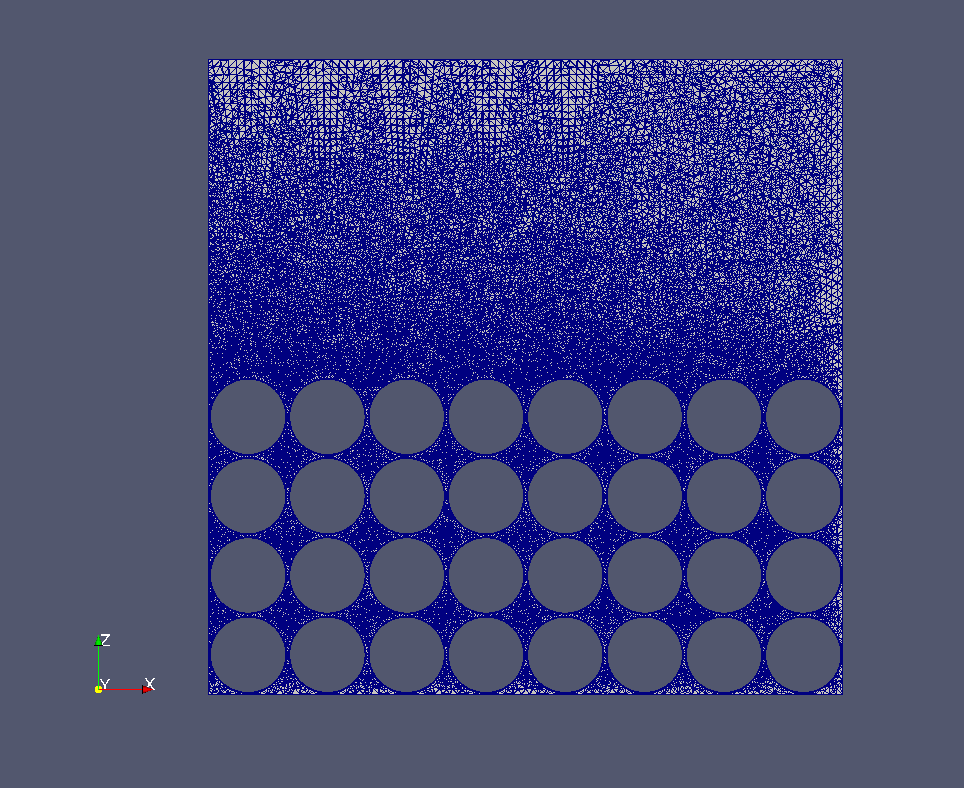}
          \label{fig:initialMeshA}}
    \subfloat[]
        {\includegraphics[width=0.205\textwidth]{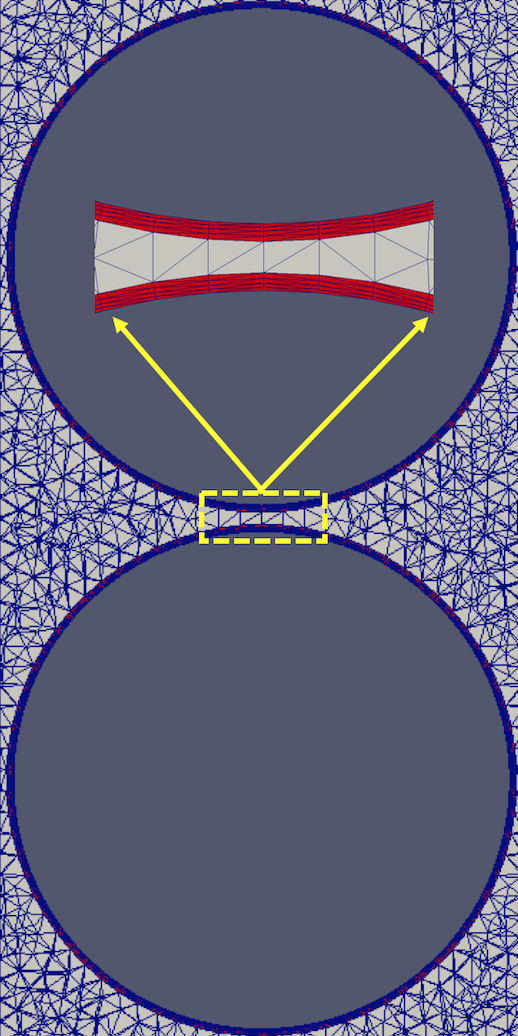}
          \label{fig:initialMeshB}}
    \caption{Partial cross-sections of the unstructured mesh showing: 
      \protect\subref{fig:initialMeshA} the relative mesh density throughout the domain and  
      \protect\subref{fig:initialMeshB} a close-up view of the boundary layer elements.}
    \label{fig:initialMesh}
\end{figure}

With an aim to elucidate the hydrodynamic processes driving the interfacial transport, an unstructured,
conforming mesh consisting of both wedge and tetrahedral element types was applied to the
computational domain. Figure \ref{fig:initialMeshA} shows a partial cross-section
of this mesh through the plane of minimum porosity, while Figure \ref{fig:initialMeshB} provides a clarifying image 
of the meshed region between two spheres.
%

To mitigate the challenges associated with incorporating grains, or spherical voids, into the model (e.g. enforcing the fluid-solid
interface and resolving a thin viscous sub-layer), a conforming mesh has been applied to the spherical voids, 
which allows one to minimize the complexity of the numerical model by directly applying no-slip boundary conditions 
to the spheres' surfaces, rather than applying these conditions through an immersed-boundary or interface 
capturing method. Additionally, several layers of wedge elements envelop the spheres in an effort to obtain 
high-resolution LES near the solid boundaries. 

In order to determine how many layers of wedge elements are necessary to resolve the boundary layer 
activity while maintaining a feasible numerical work load, the viscous length scale, $l_{*}$,  which 
governs the viscous sub-layer may be computed by Equation \ref{eq:wallUnit}, where $\tau_{w}$ and $u_{*}$ are
the wall shear-stress and shear velocity, respectively.
\begin{equation}
  u_{*} = \sqrt{\frac{\tau_{w}}{\rho}}, \quad\quad l_{*} = \frac{\nu}{u_{*}}
  \label{eq:wallUnit}
\end{equation}
Literature recommends that elements nearest to the wall have a 
height no greater than $z^{+} = \mathcal{O}(1)$, where $z^{+} = z/l_{*}$ is the wall-unit
\cite{Temmerman2003,Frohlich2005,Stoesser2007}.
Calculation of a meaningful wall-unit relies on the use of a $\tau_{w}$ which accurately represents the 
wall shear-stress experienced by fluid particles. Using a control-volume argument for a 
body-force driven surface flow (Figure \ref{fig:geometries}), an approximation for $\tau_{w}$ at the top of the 
bed, in a plane-averaged sense, can be obtained by Equation \ref{eq:wallShear}, where $b$ and $H$ are the 
body-force and surface-flow height, respectively. 
\begin{equation}
  \tau_{w} = bH  
  \label{eq:wallShear}
\end{equation}
While this formulation provides an initial guess at $\tau_{w}$,
it is important to note that the presence of wall permeability and only tangential contact with the spheres significantly
complicates the calculation of a wall-unit and obscures the result's meaning, discussed further in \cite{Manes2011}.

In the following simulations, a per volume body force of $1.1$ $kg/m^{2}s^{2}$ is used to drive the flow. 
All models have a surface-flow height of $H = 0.161$m, a molecular viscosity of $\mu = 1.002\cdot{10^{-3}}$ 
$kg/(m\cdot{s})$ and 
a density of $\rho = 1\cdot{10^{3}}$ $kg/m^{3}$. Thus, a viscous wall-unit of $l_{*} = 7.58\times{10^{-5}}$ is
obtained. Applying a four-element thick boundary layer to each sphere, where each layer grows in height by a factor
of 1.2 and the initial element height is set to $d_{g}/20$, the wall-nearest element height and boundary layer mesh 
height are $z^{+} = 1.32$ and $z^{+} = 7.09$, respectively. An alternative method for computing this quantity comes
from taking the two wall-nearest nodes, computing a shear velocity gradient between them and obtaining a local wall 
shear stress using the molecular viscosity. Doing this several times across multiple datasets also recovers 
$z^{+} = \mathcal{O}(1)$. As both methods respect the suggestions of previously mentioned literature, this boundary-layer
mesh is taken to be satisfactory for resolving near-wall detail.

\subsection*{A Brief Discussion on Reynolds Number}
When modeling flow over a permeable wall, several length scales must be taken into consideration.
These scales include the surface flow height, H, the grain diameter, D, and the bed permeability,
K. In order to compare hydrodynamic quantities for similar systems with varied geometric or fluid 
properties, a Reynolds number which captures the length and velocity scale corresponding to the physics
of interest must be developed. One can imagine that near the surface/subsurface interface, where all of
the mentioned length scales are present, the determination of a single Reynolds number which adequately
characterizes the flow is challenging.

Frequently, a bulk Reynolds number, defined by the mean flow, $U_{b}$, surface height, H, and kinematic viscosity,
$\nu$, is used(e.g. \cite{Blois2012,Stoesser2007,Pokrajac2009}). 
This measure, defined in Equation \ref{eq:bulkRe}, has the 
benefit of being relatively easy to compute within experiments and carries with it a traditional meaning.
\begin{equation}
  Re = \frac{U_{b}H}{\nu}
  \label{eq:bulkRe}
\end{equation}
One disadvantage of the bulk Reynolds number, however, is that it does not account for the effects of
wall permeability or roughness. By defining the bed permeability according to Equation \ref{eq:permeability}, 
permeability and roughness Reynolds numbers may be computed according to Equations \ref{eq:permRe} and 
\ref{eq:roughRe}, respectively \cite{Breugem2006}. 
\begin{equation}
  K = \frac{D^{2}\epsilon^{3}}{180(1-\epsilon)^{2}}
  \label{eq:permeability}
\end{equation}
\begin{equation}
  Re_{K} = \frac{u_{*}\sqrt{K}}{\nu}
  \label{eq:permRe}
\end{equation}
\begin{equation}
  Re_{D} = \frac{u_{*}D}{\nu}
  \label{eq:roughRe}
\end{equation}
where $K$ is the permeability and $\epsilon$ is the constant bed porosity. The permeability
Reynolds number can be considered a comparison of the relative importance of an effective
pore diameter and the length scale associated with the viscous sublayer along individual grains.
The roughness Reynolds number holds a similar meaning, substituting the height of the roughness
elements for the pore diameter \cite{Breugem2006}.

Given the previously defined geometric parameters, including the approximate calculation for $u_{*}$, 
values of $Re_{K} \approx 34 $ and $Re_{D} \approx 504$ are obtained. One can expect to isolate the 
effects of permeability if $Re_{K} > 1$ and $Re_{D} << 70$ \cite{Manes2011}. As both Reynolds
numbers are an order of magnitude larger than these limits, the effects of permeability and roughness 
should both be present and hard to distinguish from one another. Simulations with different geometric
parameters will be necessary to further study the independent influences of roughness and permeability
on momentum transport, though this work is beyond the scope of this thesis.

\subsection*{Discretization and Governing Equations}
Throughout the entire domain, the fluid is assumed Newtonian and incompressible, governed by 
the LES model consisting of the continuity and momentum equations, respectively defined as:
%
\begin{equation}
  \frac{\partial u_{i}}{\partial x_{i}}=0
  \label{eq:MassConservation}
\end{equation}
%
\begin{equation}
 \rho \frac{\partial u_{i}}{\partial t}+\rho \frac{\partial( u_{i}u_{j})}{\partial x_{j}}=
   -\frac{\partial p}{\partial x_{i}}+\frac{\partial\tau_{ij}}{\partial x_{j}}
   +\frac{\partial\tau_{ij}^{SGS}}{\partial x_{j}}+b_{i}
\label{eq:MomentumEquation}
\end{equation}
where $u_{i}$ stands for the space-filtered velocity,  $\rho$ is the fluid density and $b_{i}$ is the body 
force used to drive the flow. The  deviatoric components
of the resolved stress and unresolved subgrid stress are defined by $\tau_{ij}$ and $\tau_{ij}^{SGS}$, 
respectively. While these subgrid stresses are computed directly, the subgrid turbulent kinetic energy 
is taken to be an additional normal stress in the pressure term, as the majority of turbulent kinetic 
energy exists within the resolved scales.

The resolved shear stress is defined as $\tau_{ij}=2\mu S_{ij}$, where $\mu$ denotes the molecular viscosity
of the fluid and $S_{ij}$ is the resolved rate of deformation, defined as
%
\begin{equation}
  S_{ij} = \frac{1}{2} \bigg( \frac{\partial{u_{i}}}{\partial{x_{j}}} +  
                              \frac{\partial{u_{j}}}{\partial{x_{i}}} \bigg)
\end{equation}
Subgrid or modeled contributions to the fluid shear stress are defined as $\tau_{ij}^{SGS} = 2\mu_{t}S_{ij}$,
where $\mu_{t}$ is the eddy viscosity, computed by a chosen LES closure model.

\begin{figure}
  \centerline{\includegraphics{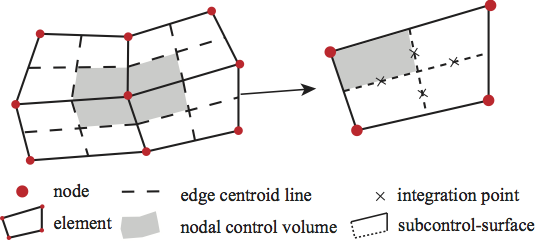}}
  \caption{A nodal control volume formed from the assembly of four subcontrol volumes.}
\label{fig:quadElement}
\end{figure}

To perform the LES, we use the control volume finite element method (CVFEM) \cite{Schneider1987}, 
a numerical scheme which draws upon the strengths of both finite element and control volume methods.
The form of the CVFEM mesh is demonstrated in Figure \ref{fig:quadElement}.  A geometry is first discretized with 
finite elements. Lines are drawn through the centroids of each shared face, following the faces' normal
vectors. These lines then create a new grid, called the mesh dual, where each finite element is divided into 
sub-control volumes (SCV) and a set of sub-control surfaces (SCS) amongst neighboring elements defines a 
control volume about the node. Governing equations (e.g. \ref{eq:MassConservation} and \ref{eq:MomentumEquation}) 
are then integrated over the nodal control volume, which results in algebraic discrete equations for a set 
of unknown variables, collocated at the node. Using Gauss' theorem, volume integration is transformed into area 
integration over the bounding subcontrol-surfaces, approximated by Gauss quadrature. For flux terms, 
including diffusion and convection, reconstruction at the integration points of subcontrol-surfaces is performed 
using the local element's nodal shape functions. Additionally, an upwinding scheme and variational multiscale method 
are applied for the advection and pressure stabilizations, respectively.

Within this thesis, simulations of turbulent flow are performed with Nalu \cite{Domino2015},
a generalized unstructured and massively parallel code base which implements the LES-CVFEM 
formulation discussed above. 

\subsection*{Turbulence Models}
In this work, both the standard Smagorinsky turbulence model \cite{SMAGORINSKY1963} and the 
Wall-Adapting Local Eddy-Viscosity (WALE) model \cite{Ducros1998} are applied to the flow model in an effort to clarify 
each closure's effect on the observed hydrodynamics. 
The constant-coefficient Smagorinsky model computes the turbulent viscosity as:
%
\begin{equation}
  \mu_t = \rho (C_{s}\Delta)^2 |S|
  \label{eq:smagorinsky}
\end{equation}
where $C_{s}$ is set to 0.17 and $|S| = \sqrt{2S_{ij}S_{ij}}$. While relatively simple to implement,
this sub-grid stress model is known to produce turbulent viscosity even in the viscous sub-layer
of wall elements where flow is nearly laminar \cite{Temmerman2003}.

One alternative to the Smagorinsky model is the WALE model, specifically formulated to capture
the proper scaling of turbulent viscosity in the near-wall limit, $\nu_{t} \sim z^{+^{3}}$ 
\cite{Ducros1998}. The WALE model computes the turbulent viscosity as:
%
\begin{equation}
  \mu_t = \rho (C_{w}\Delta)^2 \frac{(S_{ij}^{d}S_{ij}^{d})^{\frac{3}{2}}}
                               { (S_{ij}S_{ij})^{\frac{5}{2}} + (S_{ij}S_{ij})^{\frac{5}{4}} }
  \label{eq:wale}
\end{equation}
%
\begin{equation}
  S_{ij}^{d} = \frac{1}{2} \bigg( \frac{\partial{u_{i}}}{\partial{x_{k}}}  
                                  \frac{\partial{u_{k}}}{\partial{x_{j}}} +
                                  \frac{\partial{u_{j}}}{\partial{x_{k}}} 
                                  \frac{\partial{u_{k}}}{\partial{x_{i}}} \bigg)
\end{equation}
where the constant $C_{w}$ is set to 0.325. In both cases, the grid filter is computed as $\Delta = V^{1/3}$ 
where $V$ denotes the volume of the local control volume.

%% file: texFiles/mainMatter/chapter1/doubleAveraging.tex
\section{The Double-Averaged Navier Stokes (DANS) Equations}\label{sec:dansFormulation}
Two decomposition methodologies will be applied to instantaneous variables obtained from the LES,
both of which will aid in developing a meaningful analysis and interpretation of flow data. First, 
Reynolds decomposition is used to separate fields into mean and fluctuating components, 
which results in the Reynolds-Averaged Naiver-Stokes (RANS) equations. Although the RANS equations provide 
a method for navigating temporal variations in the flow, the flow both near and within 
the bed exhibits a high degree of spatial heterogeneity, which makes the RANS analysis an imperfect tool. 
Therefore, the double decomposition methodology is also utilized. Using this technique, the time-averaged
variables from the Reynolds decomposition are further decomposed into mean and fluctuating fields in space.
This analysis leads to Double-Averaged Naiver-Stokes (DANS) equations \cite{Nikora2007a}. 
One particular benefit of the DANS analysis is that it allows for calculation of  both form and
viscous drags, which is critical to analysis of the subsurface flow. Each method of decomposition 
is introduced formally in the following subsections.

\subsection*{Reynolds Decomposition Methodology}
For a general flow variable, $\theta(\boldsymbol{x},t)$ the Reynolds decomposition is defined as
$ \theta=\bar{\theta}+\theta^{\prime} $, where an overbar denotes time averaging, and 
the prime represents the deviation from the mean field. This temporal decomposition 
satisfies several well known rules (e.g. \cite{schlichting2000}), given in Equation
\ref{eq:ransRules}.
\begin{eqnarray}
  \overline{\overline{\theta}}                    = \overline{\theta}                                 \nonumber \\ 
  \overline{\theta + \phi}                        = \overline{\theta} + \overline{\phi}               \nonumber \\
  \overline{\frac{\partial{\theta}}{\partial{s}}} = \frac{\partial{\overline{\theta}}}{\partial{s}}   \nonumber \\ 
  \overline{\overline{\theta}\cdot\phi}           = \overline{\theta}\cdot\overline{\phi}
  \label{eq:ransRules}
\end{eqnarray}
where $\phi = \phi(\boldsymbol{x},t)$ and $s$ is an independent variable (i.e. $x_{i}$ or $t$).
One should note that the final rule, $\overline{\overline{\theta}\cdot\phi} = \overline{\theta}\cdot\overline{\phi}$
is only satisfied when the flow has sufficient scale separation \cite{Galmarini1999}. 
Effectively, this means that the time scale over which the mean flow changes must be significantly larger
than the time scale at which turbulent fluctuations persist. Given the tendency in the field to take the 
assumption that this criterion is met, along with the length of time over which time-averaged statistics
will be developed in the following simulations, the assumption of sufficient temporal scale separation
is taken here.

Applying the Reynolds decomposition to the instantaneous variables in time domain
governed by Equations \ref{eq:MassConservation} and \ref{eq:MomentumEquation} leads to
LES-based Reynolds-averaged Naiver-Stokes equations, which are referred to as RA-LES equations,
as follows.
%
\begin{equation}
  \frac{\partial \bar{u}_{i}}{\partial x_{i}}=0
  \label{eq:RA-MassConservation}
\end{equation}
and
%
\begin{equation}
  \rho \frac{\partial \bar{u}_{i}}{\partial t}                  +
  \rho \frac{\partial( \bar{u}_{i}\bar{u}_{j})}{\partial x_{j}} =
  -\frac{\partial \bar{p}}{\partial x_{i}}                      +
  \frac{\partial\overline{\tau_{ij}}}{\partial x_{j}}           +
  \frac{\partial\overline{\tau_{ij}^{SGS}}}{\partial x_{j}}     +
  \frac{\partial \tau_{ij}^{R}}{\partial x_{j}}+\bar{b}_{i}
  \label{eq:RA-MomentumEquation}
\end{equation}
where
%
\begin{equation}
  \tau_{ij}^{R}=-\rho \overline{u_{i}^{\prime}u_{j}^{\prime}}
\end{equation}
is the Reynolds stress and $\bar{b}_{i}=b_{i}$ when $b_{i}$ holds a constant value.
Cross-terms produced from the decomposition (e.g. the Leonard stress) have been disregarded 
in accordance with the assumption that a satisfactory separation of scales exists.

Additionally, the divergence of resolved and modeled shear stresses, respectively, are computed as:
%
\begin{equation}
  \frac{\partial \overline{\tau_{ij}}}{\partial x_{j}} = 
  \nu \frac{\partial^{2}\bar u_{i}}{\partial x_{j}^{2}}
\end{equation}
%
\begin{equation}
  \frac{\partial \overline{\tau_{ij}^{SGS}}}{\partial x_{j}} = 
  2\frac{\partial \overline{\nu_{t}S_{ij}}}{\partial x_{j}}
\end{equation}
%
\subsubsection*{Computing a Running Time Average}
Formulations for a running temporal mean must be developed for each term 
of interest. The LES computes only the instantaneous fields, and maintaining the full time history
for each term of interest over the course of the simulation would be prohibitive due to
memory requirements. Here, an example of this derivation is given both for the 
mean velocity and Reynolds stress. While higher-order terms may be more 
complicated to compute, the underlying method for accumulating the time-averaged values 
is similar. In the following derivations, the superscript $N$, $T^{N}$ and $\Delta{t}^{N}$ denote the 
time level, time at time level $N$ and the time step size at time level $N$, respectively. A unit
density is taken here for simplicity.

The mean fluid velocity may be expressed discretely as:
\begin{equation}
  \overline{u}_{i}^{N} = \frac{1}{T^{N}}\sum_{k=1}^{N} u_{i}^{k}\Delta{t}^{k}
\end{equation}
Extracting the current velocity from the summation and performing some manipulation allows one to 
obtain the current mean velocity with compact temporal support, needing information from only the current
and previous time step.
\begin{eqnarray}
  \overline{u}_{i}^{N} = \frac{1}{T^{N}}(u^{N}\Delta{t}^{N} + \sum_{k=1}^{N-1} u_{i}^{k}\Delta{t}^{k}) \nonumber \\
  \overline{u}_{i}^{N} = \frac{1}{T^{N}}(u^{N}\Delta{t}^{N} + \frac{T^{N-1}}{T^{N-1}} 
                                                              \sum_{k=1}^{N-1} u_{i}^{k}\Delta{t}^{k}) \nonumber \\
  \overline{u}_{i}^{N} = \frac{1}{T^{N}}(u^{N}\Delta{t}^{N} + T^{N-1}\overline{u}_{i}^{N-1})
\end{eqnarray}

Calculation of the Reynolds stress requires a bit more work. Again, the term of interest is first presented 
in discrete form, assuming the time-averaged velocity has already been computed at the current time level.
For the purpose of this derivation only, let $\tau_{ij} = \overline{u_{i}^{'}u_{j}^{'}}$.
\begin{eqnarray}
  \tau_{ij}^{N} = \overline{u_{i}^{\prime{N}}u_{j}^{\prime{N}}}                                   \nonumber \\
  \tau_{ij}^{N} = \overline{u_{i}^{N}u_{j}^{N}} - \overline{u}_{i}^{N}\overline{u}_{j}^{N}      \nonumber \\
  \tau_{ij}^{N} = \frac{1}{T^{N}} \sum_{k=1}^{N}u_{i}^{k}u_{j}^{k}\Delta{t}^{k} - 
      \overline{u}_{i}^{N}\overline{u}_{j}^{N} 
\end{eqnarray}
Now, the current velocities may be extracted from the summation, as done previously, to obtain the 
running mean.
\begin{eqnarray}
  \tau_{ij}^{N} = \frac{1}{T^{N}} (u_{i}^{N}u_{j}^{N}\Delta{t}^{N} + \frac{T^{N-1}}{T^{N-1}}
        \sum_{k=1}^{N-1}u_{i}^{k}u_{j}^{k}\Delta{t}^{k}) - 
        \overline{u}_{i}^{N}\overline{u}_{j}^{N}                              \nonumber \\
  \tau_{ij}^{N} = \frac{1}{T^{N}} (u_{i}^{N}u_{j}^{N}\Delta{t}^{N} + T^{N-1}\tau_{ij}^{N-1} + 
        T^{N-1}\overline{u}_{i}^{N-1}\overline{u}_{j}^{N-1}) - 
        \overline{u}_{i}^{N}\overline{u}_{j}^{N}          
\end{eqnarray}
%
\subsection*{Double Decomposition Methodology}\label{sec:DDM}
When performing double decomposition, averaging fields in a different order (i.e. time-space or space-time)
does not necessarily produce equivalent derivations of the DANS equations. As noted in \cite{Nikora2007a}, 
the averaging methods commute in the specific case of a fixed bed, although time-space averaging is more 
suitable in general for describing rough-bed hydrodynamics, due to its consistency with 
traditions in turbulence research. Therefore, focus will be placed on the DANS equations 
derived with time-space averaging order in this work.
 
For our purposes, the spatial average is computed as the volume average over a thin slab, although formally 
the volume of interest may more general. For an arbitrary flow variable, $\theta$, the spatial 
average is taken as:
%
\begin{equation}
  \langle \theta \rangle(\boldsymbol{x},t) = \frac{1}{V_f}\int_{V_{f}} \theta \mathrm{d}V ,\quad
  \langle \theta \rangle_{s} (\boldsymbol{x},t) = \frac{1}{V_0}\int_{V_{f}} \theta \mathrm{d}V
  \label{eq:SpatceAverageOperator1}
\end{equation}
where the presence and lack of the subscript $s$ denotes the superficial and intrinsic averages, respectively. 
These averages differ in that the superficial volume average normalizes the integration over the fluid volume 
by the entire averaging volume,
while the intrinsic average normalizes by the fluid volume. All formulations produced here will focus on use of the 
intrinsic volume average, to be consistent with existing literature (e.g. \cite{Nikora2007a,Breugem2006}).
By defining a geometry indicator function as the fluid volume fraction in the averaging domain, these 
averages may be related as:
%
\begin{equation}
  \phi_{s} = \frac{V_{f}}{V_{o}}, \quad
  \langle \theta \rangle _{s} = \phi_{s}\langle \theta \rangle
  \label{eq:geometryIndicator}
\end{equation}

Similar to the traditional Reynolds decomposition, the spatial decomposition separates a variable into its 
spatial mean and deviation from this mean:
\begin{equation}
  \theta = \langle{\theta}\rangle + \hat{\theta}
  \label{eq:spatialDecomp}
\end{equation}
where the $\langle\theta\rangle$ and $\hat\theta$ denote the mean and fluctuating components, respectively.
With the spatial decomposition defined, it may be applied to a time-averaged flow variable to obtain the
double decomposition:
%
\begin{equation}
  \bar{\theta}=\langle \bar{\theta} \rangle +\tilde{\theta}
  \label{eq:doubleDecomposition}
\end{equation}
where $\tilde{\theta}$ indicates the spatial perturbation from the time-space averaged variable 
represented by $\langle \bar{\theta} \rangle$ \cite{Nikora2007a}. One should note that the spatial average, just like 
the time average, is a linear operation, and thus satisfies the rule 
$\langle\theta+\phi\rangle = \langle\theta\rangle + \langle\phi\rangle$.

Extending this framework to include differential operators, both in time and space, 
Whitaker's transport and spatial averaging theorems \cite{Whitaker1999}, respectively,
enable one to study transport from a volume-averaged perspective. The transport theorem allows for the expansion 
of the temporal derivative as:
%
\begin{equation}
\langle \frac{\partial \bar\theta}{\partial t} \rangle_{s} =
   \frac{\partial \langle \bar\theta \rangle_{s}}{\partial t}
    +\frac{1}{V_{o}}\int\int_{S_{int}} \bar\theta\boldsymbol{v}\cdot\boldsymbol{n}\mathrm{d}S
\label{eq:TransportTheorem}
\end{equation}
where $S_{int}$ stands for the contact area between the fluid and the solid phase (spheres) 
inside the averaging volume $V_{o}$, and $\boldsymbol{n}$ is the unit normal at $S_{int}$ 
that points from the solid phase into the fluid. On account of the no-slip condition applied at the 
surface of the spheres, $\boldsymbol{v}$, the velocity at the surface of solid phase, is identically zero. 
Accordingly, the surface integral disappears and the transport equation may be simplified as:
%
\begin{equation}
\langle \frac{\partial \bar\theta}{\partial t} \rangle_{s} =
   \frac{\partial \langle \bar\theta \rangle_{s}}{\partial t}
  \label{eq:DA-TimeDif}
\end{equation}
which highlights the commutability of spatial-averaging and temporal differentiation for a
stationary bedform.
The spatial-averaging theorem provides a similar expansion for the spatial derivative:
\begin{equation}
\langle \frac{\partial \bar\theta}{\partial x_{i}} \rangle_{s} =
   \frac{\partial \langle \bar\theta \rangle_{s}}{\partial x_{i}}
   - \frac{1}{V_{o}}\int\int_{S_{int}}\bar\theta n_{i}\mathrm{d}S
  \label{eq:DA-SpaceDif}
\end{equation}

Using Equation \ref{eq:geometryIndicator}, relationships for the differentiation of double-averaged 
quantities which have been defined with respect to the intrinsic volume average are readily obtained 
from the above equations as:
%
\begin{equation}
\langle \frac{\partial \bar\theta}{\partial t} \rangle =
   \frac{1}{\phi_{s}}\frac{\partial \phi_{s}\langle \bar\theta \rangle}{\partial t}
\label{eq:DA-TimeDif-Intrinsic}
\end{equation}
and
%
\begin{equation}
\langle \frac{\partial \bar\theta}{\partial x_{i}} \rangle =
   \frac{1}{\phi_{s}}\frac{\partial \phi_{s}\langle \bar\theta \rangle}{\partial x_{i}}
   - \frac{1}{V_{f}}\int\int_{S_{int}}\bar\theta n_{i}\mathrm{d}S
\label{eq:spatialAveTheorem}
\end{equation}
%
\subsection*{Dimension of the Averaging Domain}

Within this work, the primary goal in taking a volume average is to capture the large-scale 
hydrodynamic features which arise from the presence of the bed topography, 
while smoothing out much of the small-scale flow structure coming from the spatial heterogeneity
within the bed. 

Considering the configurations shown in Figure \ref{fig:geometries} and the conceptual model provided in Figure 
\ref{fig:conceptualTransport}, the variation in the flow field as a function of depth is of critical interest. 
An averaging volume enabling one to observe such behavior may be obtained by first taking a plane which covers 
the streamwise and spanwise dimensions in full. This surface may then be uniformly extruded a length $L^{\star}$ 
in the wall-normal direction to create the desired volume. By setting $L^{\star} = L$ and taking volume 
averages throughout depth of the entire domain, three physical regions are recovered, distinct in their 
porosity. 

\begin{figure}
  \centerline{\includegraphics[width=0.5\textwidth]{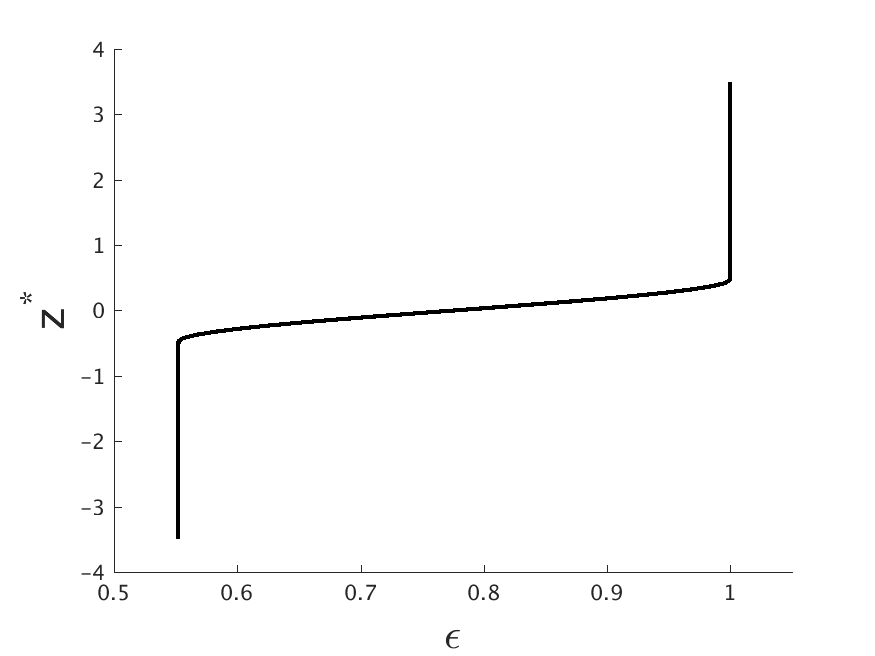}}
  \caption{Average fluid volume fraction computed with an averaging volume of dimension $L_{x} \times L_{y} \times L$.}
\label{fig:porosity}
\end{figure}

The subscripts S, B and T are used to distinguish these surface, bed and 
transition regions, respectively. As shown in Figure \ref{fig:porosity}, where the volume-averaged porosity 
is plotted against the depth of the averaging volume's centroid, the surface flow primarily holds a 
volume-averaged porosity of $\epsilon_{S} = 1$ and may be described as a homogeneous fluid region.
Deep within the bed, the subsurface domain is characterized by a porosity $\epsilon_{B} = 0.5511$, and 
can be referred to as the homogeneous porous region. When the centroid of an averaging volume reaches 
a depth $z^{*} \in (-0.5, 0.5)$, the volume enters a region of variable averaged porosity, 
termed the transition region. This region is described by averaged porosity $\epsilon_{T} = \epsilon(z)$, 
and will be of particular interest when analyzing simulation results. One should note that by taking an 
averaging volume of these dimensions, the three regions discussed in Figure \ref{fig:conceptualTransport} have been 
made concrete, at least to a first approximation, by the local value of the averaged porosity. 

Useful properties are associated with data that has been averaged in this way. First and foremost, any 
flow variable subject to double-averaging becomes a function of only depth (i.e. 
$\langle\overline{\theta(x,y,z,t)}\rangle = \langle\overline{\theta}\rangle(z)$. 
Additionally, it should be clear that as the geometry indicator 
takes a unit value in the homogeneous fluid region, the intrinsic and superficial average become equivalent
(i.e. $\langle\theta\rangle = \langle\theta\rangle_{s}$).

Moving forward, double-averaged quantities will commonly be plotted against the depth of the centroid of 
the averaging volume. To simplify the description of this location, a depth $z^{*}$ may be defined 
in the following way:
\begin{equation}
  z^{*}(z) = \frac{z-z_{0}}{L}
\end{equation}
where $z_{0}$ is the zero position of the transformed coordinates ($z_{0}$ = 0.16 m in these simulations) 
and L = 0.04 m is the height of the 
averaging volume. This transformation sets the zero-height of the domain at half of the original
domain height, and further distinguishes the surface and subsurface flows roughly according to 
positive and negative depths, respectively. Note that this transformed coordinate will be used for both
double-averaged and instantaneous flow data.

\subsection*{Double-Averaged LES Equations}
Application of the volume-averaging operator and the theorems defined in Section \ref{sec:DDM}
to each term in RA-LES equations (\ref{eq:RA-MassConservation} and \ref{eq:RA-MomentumEquation}),
while enforcing the boundary conditions specified in Section \ref{sec:geomAndModel},
results in the  double-averaged continuity and momentum equations, obtained, respectively, as:
%
\begin{equation}
\frac{\partial \phi_{s} \langle \bar u_{i} \rangle}{\partial x_{i}}=0
\label{eq:DA-ContinutyEq}
\end{equation}
and
%
\begin{equation}
  \frac{\rho}{\phi_{s}}\frac{\partial\phi_{s}\langle\overline{u}_{i}\rangle}{\partial{t}}                 +
  \frac{\rho}{\phi_{s}} \frac{\partial \phi_{s} \langle \bar u_{i} \bar u_{j} \rangle}{\partial x_{j}}    =
  -\frac{1}{\phi_{s}}\frac{\partial\phi_{s}\langle \overline{p} \rangle}{\partial x_{i}}                  +
  \frac{1}{\phi}\frac{\partial\phi_{s} \langle \overline {\tau_{ij}} \rangle}{\partial x_{j}}             +
  \langle \overline {\tau_{ij,j}^{SGS}}\rangle                                                            +
  \frac{1}{\phi_{s}}\frac{\partial \phi_{s} \langle \overline{\tau_{ij}^{R}} \rangle}{\partial x_{j}}     -
  f_{i}^{p}                                                                                               +
  f_{i}^{v}                                                                                               +
  \langle \bar{b_{i}} \rangle
  \label{eq:DA-MomentumEquation}
\end{equation}
where $f_{i}^{p}$ represents the form drag per unit fluid volume:
%
\begin{equation}
  f_{i}^{p} = -\frac{1}{V_{f}}\int\int_{S_{int}} \bar p n_{i} \mathrm{d}S
\end{equation}
and $f_{i}^{v}$ denotes the viscous drag per unit fluid volume:
%
\begin{equation}
  f_{i}^{v} = -\frac{1}{V_{f}}\int\int_{S_{int}} \overline{2\mu S_{ij}} n_{j} \mathrm{d}S
\end{equation}

Using \ref{eq:spatialAveTheorem}, these drag terms may also be calculated in the form:
%
\begin{equation}
  f_{i}^{p} = \langle \frac{\partial \bar p}{\partial x_{i}} \rangle -
  \frac{1}{\phi_{s}} \frac{\partial \phi_{s} \langle \bar p \rangle}{\partial x_{i}}
  \label{eq:FormDragFormulation}
\end{equation}
and
%
\begin{equation}
  f_{i}^{v} = \langle \frac{\partial \overline{2\mu S_{ij}}}{\partial x_{j}} \rangle -
  \frac{1}{\phi_{s}} \frac{\partial \phi_{s} \langle \overline{2\mu S_{ij}} \rangle}{\partial x_{j}}
  \label{eq:ViscousDragFormulation}
\end{equation}

A quick note must be made concerning the volume average of the convection term in Equation 
\ref{eq:DA-MomentumEquation}. Using the double decomposition defined in 
Equation \ref{eq:doubleDecomposition}, the volume-averaged velocity product may be rewritten as:
\begin{eqnarray}
  \langle \bar u_{i} \bar u_{j}\rangle                = 
  \langle (\langle\overline{u}_{i}\rangle + 
           \tilde{u}_{i})
          (\langle\overline{u}_{j}\rangle + 
           \tilde{u}_{j}) \rangle                 \nonumber \\
  \langle \bar u_{i} \bar u_{j}\rangle                = 
  \langle (\langle\overline{u}_{i}\rangle\langle\overline{u}_{j}\rangle + 
            \tilde{u}_{i}\tilde{u}_{j} + 
            \langle\overline{u}_{i}\rangle\tilde{u_{j}} +
            \langle\overline{u}_{j}\rangle\tilde{u_{i}}) \rangle
  \label{eq:formInducedCrossTerms}
\end{eqnarray}
In line with the assumptions made for the calculation of a time average, a separation of scales 
is assumed to exist for the spatial average, such that 
$\langle\langle\theta\rangle\phi\rangle = \langle\theta\rangle\langle\phi\rangle$. 
Such assumptions allow for the simplification of Equation \ref{eq:formInducedCrossTerms} as:
%
\begin{equation}
  \langle \bar u_{i} \bar u_{j}\rangle                \approx
  \langle \bar u_{i}\rangle \langle \bar u_{j}\rangle +
  \langle \tilde u_{i} \tilde u_{j} \rangle
  \label{eq:formInducedThreeTerms}
\end{equation}
where $\langle \tilde u_{i} \tilde u_{j} \rangle$ is the so-called form-induced stress \cite{Nikora2007a}.
It is important to note that because the difference in the length scale of the averaging volume and the extent 
over which geometric heterogeneity creates variation in the flow field is likely on the order of the number 
of streamwise or spanwise unit cells, this assumption is 
less valid for the spatial decomposition than the temporal decomposition. Currently, this assumption 
only affects the calculation of the form-induced stresses. Consequently, errors accrued by this simplification
will need to be rigorously assessed if these stresses prove to be a dominant influence in the momentum
transport equation, or if this assumption is later used to make different observations.

\subsubsection*{Simplification of the Double-Averaged Equations}
Considering the chosen averaging slab, which occupies the whole $x-y$ plane, and the 
only source term in Equation \ref{eq:MomentumEquation}, a constant body force $b_{i}$, 
all of the double-averaged terms shown in Equation \ref{eq:DA-MomentumEquation} are functions 
of $z$ alone, which indicates that the derivatives of spatially averaged quantities only exist 
in the $z$-dimension. Additionally, all temporal derivatives are zero, as the time-averaged quantities 
are steady in time and fluctuating terms are mean-zero by definition. As a consequence, Equation 
\ref{eq:DA-MomentumEquation} may be simplified as:
%
\begin{equation}
  \frac{\rho}{\phi_{s}}\frac{\mathrm{d}\phi_{s}\langle \bar u_{i} \bar u_{3} \rangle}{\mathrm{d} x_{3}}    =
  -\frac{1}{\phi_{s}}\frac{\mathrm{d}\phi_{s}\langle \bar p \rangle}{\mathrm{d} x_{\overline{i}}} \delta_{\overline{i}3}         +
  \frac{1}{\phi_{s}}\frac{\mathrm{d}\phi_{s}\langle \overline {\tau_{i3}} \rangle}{\mathrm{d} x_{3}}       +
  \langle \overline {\tau_{ij,j}^{SGS}} \rangle                                                            +
  \frac{1}{\phi_{s}}\frac{\mathrm{d}\phi_{s}\langle \overline{\tau_{i3}^{R}} \rangle}{\mathrm{d} x_{3}}    -
  f_{i}^{p}                                                                                                +
  f_{i}^{v}                                                                                                +
  b_{i}
  \label{eq:DA-MomentumEquation-Simplification}
\end{equation}
where $\overline{i}$ denotes a lack of summation over the repeated index. Furthermore, to analyze 
functions of the streamwise velocity, the $i=1$ case is examined:
%
\begin{equation}
  \frac{\rho}{\phi_{s}}\frac{\mathrm{d}\phi_{s}\langle \bar u \bar w \rangle}{\mathrm{d} z}            =
  \frac{1}{\phi_{s}}\frac{\mathrm{d}\phi_{s}\langle \overline {\tau_{13}} \rangle}{\mathrm{d} z}       +
  \langle \overline {\tau_{1j,j}^{SGS}}\rangle                                                         +
  \frac{1}{\phi_{s}}\frac{\mathrm{d}\phi_{s}\langle \overline{\tau_{13}^{R}} \rangle}{\mathrm{d} z}    -
  f_{1}^{p,s}                                                                                          +
  f_{1}^{v,s}                                                                                          +
  b_{1}
  \label{eq:dansIeq1}
\end{equation}

A simplified version of the resolved viscous shear stress may be obtained using the incompressiblity 
assumption ($u_{i,i} = 0$), which will prove useful in the following analysis.
\begin{eqnarray}
  \langle \overline {\tau_{ij,j}} \rangle = 2\mu\langle\overline{S_{ij,j}}\rangle \\ \nonumber
                                          = \mu\langle\overline{u_{i,jj}} + \overline{u_{j,ji}}\rangle  \\ \nonumber
                                          = \mu\langle\overline{u_{i,jj}}\rangle 
  \label{eq:simpleViscShearStress}
\end{eqnarray}
Additionally, using the periodic and no-slip boundary conditions in concert with the dimensions of
the averaging volume, a simple argument can be constructed to show that $\langle\overline{w}\rangle =0$.
Thus, for the $i=1$ DANS momentum equation, Equation \ref{eq:formInducedThreeTerms} may be simplified as:
\begin{equation}
  \langle \bar u_{i} \bar u_{j}\rangle                \approx
  \langle \tilde u_{i} \tilde u_{j} \rangle
  \label{eq:formInducedApprox}
\end{equation}
which provides a convenient method for computing the form-induced stresses, given the time-averaged velocity
data. 

\subsection*{Integral form of Double-Averaged Momentum Equation}

A driving goal in creating a detailed model for flow over permeable beds is to determine which physics 
take a governing role in defining the double-averaged velocity profile, 
$\langle\overline{u}\rangle(z)$, within the transition region shown in Figures 
\ref{fig:porosity}. One way to better understand 
this dependence is to study the derivative of $\langle\bar{u}\rangle$ with respect to $z$ 
as a function of the other hydrodynamic quantities, obtained by integrating the DANS momentum equation.

To perform this analysis, Equation \ref{eq:dansIeq1} is first multiplied
by $\phi_{s}$ and then integrated from some datum, $z_{0}$, to some point of interest, $z$:
%
\begin{eqnarray}
  \int_{z_{0}}^{z} \rho\frac{\mathrm{d}\phi_{s}\langle\bar{u}\bar{w}\rangle}
                                       {\mathrm{d}z^{\prime}} \mathrm{d}z^{\prime}                  & = &
  \int_{z_{0}}^{z} \frac{\mathrm{d}\phi_{s}\langle\overline{\tau_{13}}\rangle} 
                                       {\mathrm{d} z^{\prime}} \mathrm{d}z^{\prime}         +
  \int_{z_{0}}^{z} \frac{\mathrm{d}\phi_{s}\langle\overline{\tau_{13}^{R}}\rangle}
                                       {\mathrm{d} z^{\prime}}   \mathrm{d}z^{\prime}   \nonumber     -
  \int_{z_{0}}^{z} \phi_{s}f_{1}^{p}\mathrm{d}z^{\prime}                                  \\ &&       +
  \int_{z_{0}}^{z} \phi_{s}f_{1}^{v}  \mathrm{d}z^{\prime}                                            +
  \int_{z_{0}}^{z} \phi_{s}\langle\overline{\tau_{13}^{SGS}}\rangle \mathrm{d}z^{\prime}              +
  b_{i}\int_{z_{0}}^{z} \phi_{s}\mathrm{d}z^{\prime}
  \label{eq:DA-MomentumEquation-IntegralForm}
\end{eqnarray}
By applying the fundamental theorem of calculus, several terms may be rewritten as a function evaluation
at the limits of integration:
%
\begin{eqnarray}
  \rho\phi_{s}(z)\langle\bar{u}\bar{w}\rangle(z)                      & = &
  \phi_{s}(z)\langle\overline{\tau_{13}}\rangle(z)                      +
  \phi_{s}(z)\langle\overline{\tau_{13}^{R}}\rangle(z)    -
  \int_{z_{0}}^{z} \phi_{s}f_{1}^{p}\mathrm{d}z^{\prime}  \nonumber\\ && +
  \int_{z_{0}}^{z} \phi_{s}f_{1}^{v}  \mathrm{d}z^{\prime}              +
  \int_{z_{0}}^{z} \phi_{s}\langle\overline{\tau_{13}^{SGS}}\rangle \mathrm{d}z^{\prime}              +
  b_{i}\int_{z_{0}}^{z} \phi_{s}\mathrm{d}z^{\prime}                    +
  C(z_0)
  \label{eq:DA-MomentumEquation-IntegralForm2}
\end{eqnarray}
where the constant $C(z_0)$ is defined as:
\begin{equation}
  C(z_0) = 
  \rho\phi_{s}(z_0)\langle\bar{u}\bar{w}\rangle(z_0) 
  -\phi_{s}(z_0)\langle\overline{\tau_{13}}\rangle(z_0)
  -\phi_{s}(z_0)\langle\overline{\tau_{13}^{R}}\rangle(z_0)
\end{equation}

The resolved shear stress may be simplified significantly using
the Equations \ref{eq:simpleViscShearStress}, \ref{eq:spatialAveTheorem} and the no-slip boundary conditions to provide:
%
\begin{equation}
  \phi_{s}\langle\overline{\tau_{13}}\rangle = 
  \mu\frac{\mathrm{d}\phi_{s}\langle\overline{u}\rangle}{\mathrm{d}z}
  \label{eq:simpleShearStress}
\end{equation}

Finally, substitution of Equation \ref{eq:simpleShearStress} into Equation \ref{eq:DA-MomentumEquation-IntegralForm2},
allows one to isolate $\phi_{s}\langle\overline{u}\rangle$ with respect to $z$:
%
\begin{eqnarray}
  \frac{\mathrm{d}\phi_{s}\langle\overline{u}\rangle(z)}{\mathrm{d}z}                       & = & 
  \frac{1}{\mu}\left[
    \rho\phi_{s}(z)\langle\bar{u}\bar{w}\rangle(z)                                            -
    \phi_{s}(z)\langle\overline{\tau_{13}^{R}}\rangle(z) \right]\nonumber               \\ && +
  \frac{1}{\mu}\left[
    \int_{z_{0}}^{z} \phi_{s}f_{1}^{p}\mathrm{d}z^{\prime}                                    -
    \int_{z_{0}}^{z} \phi_{s}f_{1}^{v}  \mathrm{d}z^{\prime}                                  -
    \int_{z_{0}}^{z} \phi_{s}\langle\overline{\tau_{13}^{SGS}}\rangle \mathrm{d}z^{\prime}    -
    b_{i}\int_{z_{0}}^{z} \phi_{s}\mathrm{d}z^{\prime} \right]\nonumber                 \\    -
    \frac{C(z_0)}{\mu}
  \label{eq:integralDANS}
\end{eqnarray}
which links the contribution of each term on RHS of above equation to
the variation of $\phi_{s}\langle\bar{u}\rangle$ along the vertical direction.
It is worthy to note that anytime the averaged porosity is constant 
(i.e. $\phi_{s} = \epsilon_{S}$ or $\phi_{s} = \epsilon_{B}$), the geometry 
indicator may be removed from the derivative and the change in $\langle\overline{u}\rangle$ 
may be examined explicitly.

%% file: texFiles/mainMatter/chapter1/initialStudiesAndResults.tex
\section{Initial Studies and Results}\label{sec:initResults}

\begin{figure}[!tbp]
    \centering  
    \subfloat[]
        {\includegraphics[width=0.5\textwidth]{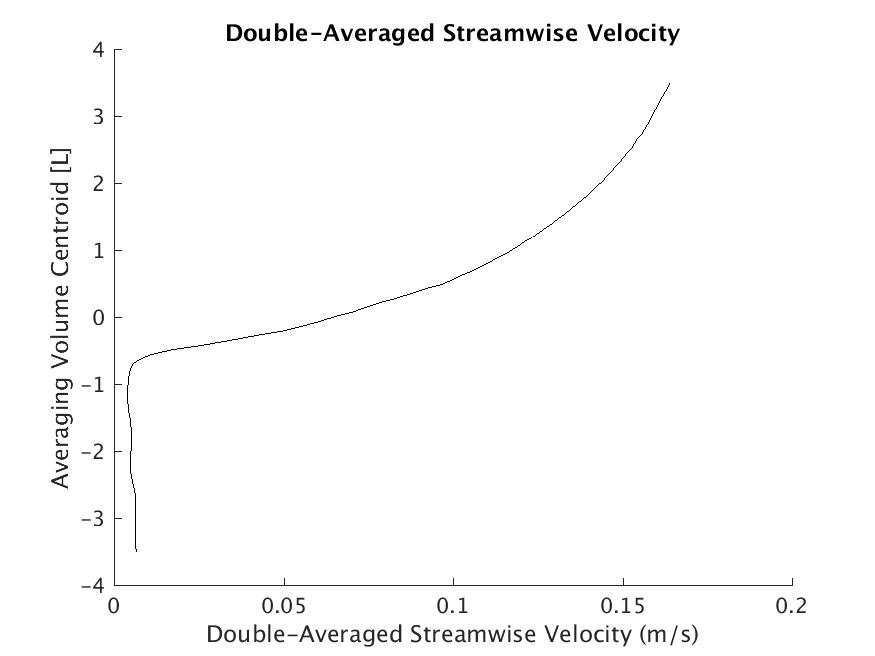}
         \label{fig:daSmag}}
    \subfloat[]
        {\includegraphics[width=0.5\textwidth]{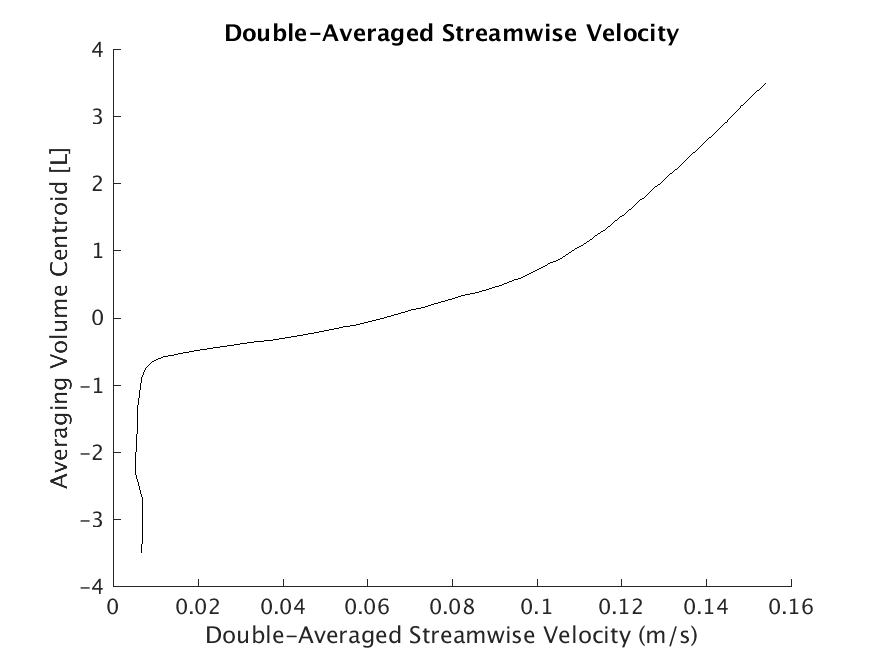}
         \label{fig:daWale}}
    \caption{Double-averaged streamwise velocity profiles computed by
      \protect\subref{fig:daSmag} the Smagorinsky model with a 15x5x4 bed and  
      \protect\subref{fig:daWale} the WALE model with a 10x5x4 bed.}
    \label{fig:initialDaConflictDa}
\end{figure}

Early on in this work, two test cases with different turbulence models were examined. Simulations 
were conducted on a 15x5x4 bed using the Smagorinsky model and on a 10x5x4 bed using the 
WALE model. Figure \ref{fig:initialDaConflictDa} shows the double-averaged streamwise 
velocity $\langle\overline{u}\rangle$, for each set of results. One need not examine a thorough set of 
turbulent statistics to identify several differences between 
the two curves. Looking at the surface flow profile, one might expect to see $\langle\overline{u}\rangle$ 
approach a constant value, indicating that the domain's surface flow height, H, is larger than that of 
a fully developed turbulent boundary layer, $\delta$, as discussed in \cite{Kundu2016}. 
While the Smagorinsky model results begins to approach this expected behavior at the top of the domain, 
neither model attains a constant velocity in the surface flow, indicating $H/\delta < 1$, which will likely
provide challenges in determining the locations of the inner, logarithmic and outer regions of the flow 
\cite{Manes2011}. Although both the model and streamwise domain size vary between 
the two simulations, making it difficult to understand the dependence of this difference exactly, it is 
clear that the treatment of subgrid stress and/or the scale of the resolved turbulent structures is the 
culprit.

While the simulations agree well in the homogeneous fluid and transition regions, the results are obviously 
different in the subsurface flow. The WALE model, and smaller geometry, predicts a minimum in 
$\langle\overline{u}\rangle$ at $z^{*} = -2.24$, while the Smagorinsky model predicts a minimum value at $z^{*} = -1.21$. 
Given the experimental confirmation of this minimum existing around the first 
layer of pores \cite{Pokrajac2007}, understanding this discrepancy is critical in performing model validation.

\begin{figure}[!tbp]
    \centering  
    \subfloat[]
        {\includegraphics[width=0.5\textwidth]{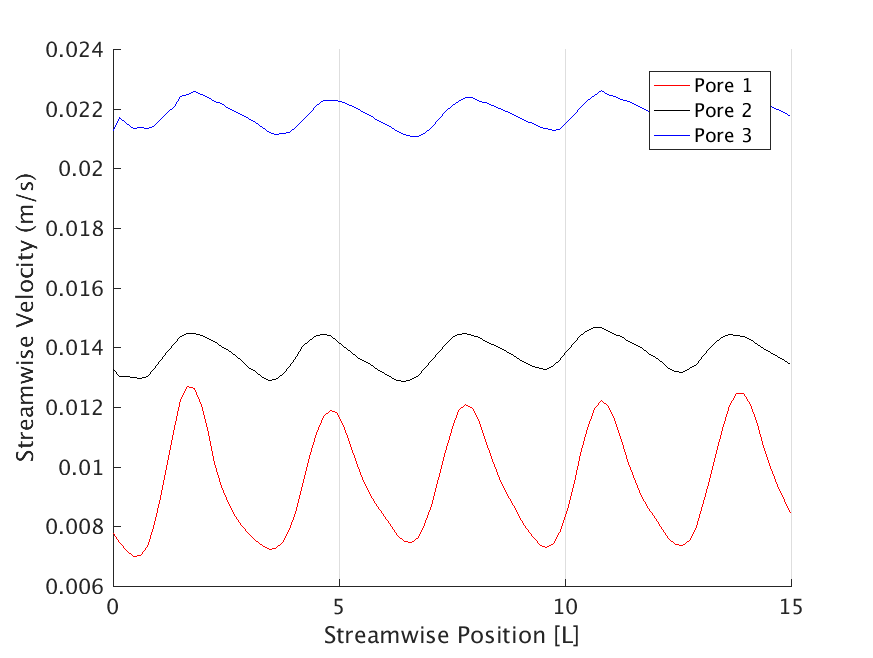}
         \label{fig:maxProbeSmag}}
    \subfloat[]
        {\includegraphics[width=0.5\textwidth]{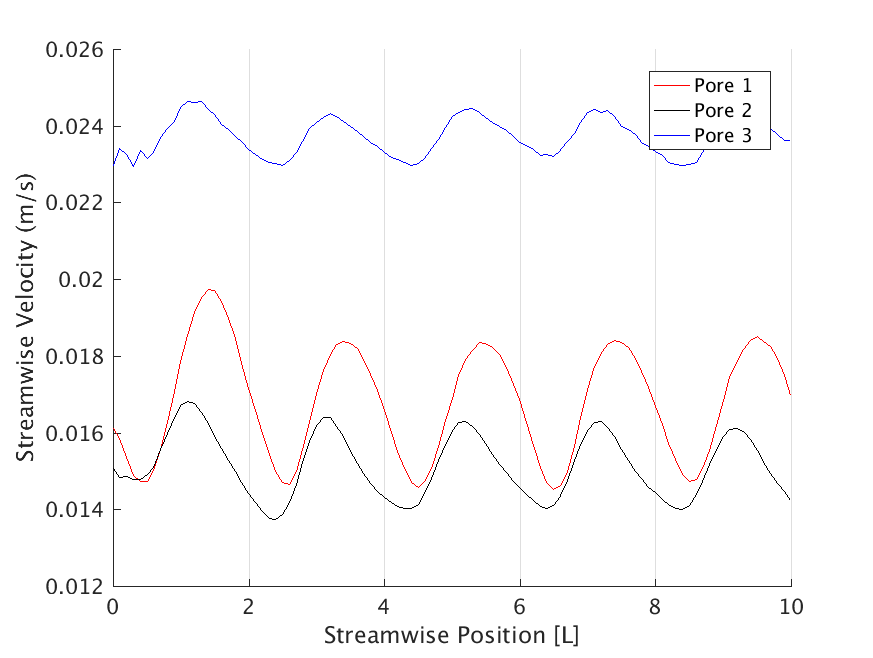}
         \label{fig:maxProbeWale}}
    \caption{Time-averaged streamwise velocity in the plane of maximum porosity along pore throats computed by
      \protect\subref{fig:maxProbeSmag} the Smagorinsky model and  
      \protect\subref{fig:maxProbeWale} the WALE model.}
    \label{fig:initialDaConflictMax}
\end{figure}

\begin{figure}[!tbp]
    \centering  
    \subfloat[]
        {\includegraphics[width=0.5\textwidth]{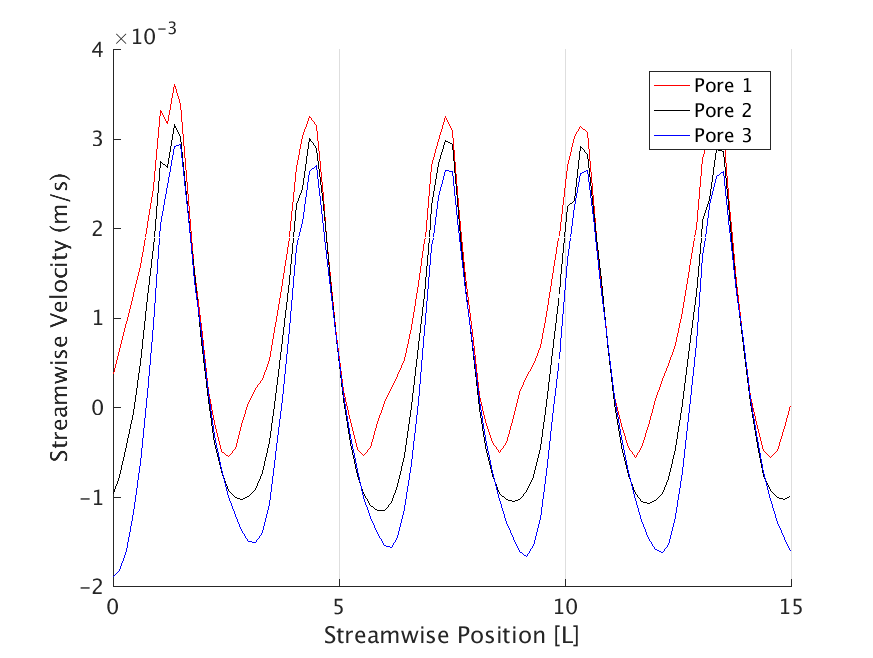}
         \label{fig:minProbeSmag}}
    \subfloat[]
        {\includegraphics[width=0.5\textwidth]{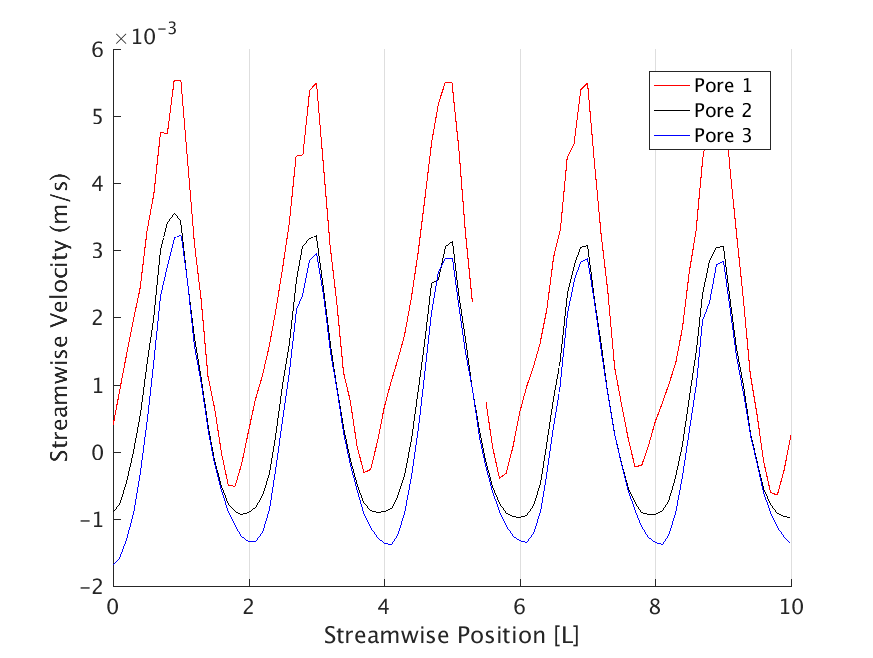}
         \label{fig:minProbeWale}}
    \caption{Time-averaged streamwise velocity in the plane of minimum porosity along pore throats computed by
      \protect\subref{fig:minProbeSmag} the Smagorinsky model and  
      \protect\subref{fig:minProbeWale} the WALE model.}
    \label{fig:initialDaConflictMin}
\end{figure}

To ensure that the observed disagreement is not simply a function of the double-averaging methodology, 
Figures \ref{fig:initialDaConflictMax} and \ref{fig:initialDaConflictMin} present the Reynolds-averaged 
velocities in the planes of maximum and minimum porosity, respectively. To obtain each curve,
200 points were distributed along the entire extent of the streamwise domain with equivalent $(y,z^{*})$ 
positions and velocities were interpolated at each point. The pore 1 $(y,z^{*})$ coordinates for the Smagorinsky 
and WALE models in the plane of maximum porosity are both $(L,-1)$. Pores 2 and 3 are
taken at $z^{*} = -2$ and $z^{*} = -3$, respectively, and measurements in the plane of 
minimum porosity are shifted by $-(L/2)$ in the spanwise direction.

By observing $\overline{u}$ in Figures \ref{fig:initialDaConflictMax} and \ref{fig:initialDaConflictMin}, 
it is clear that the resultant flow fields from these simulations are fundamentally 
different and that the locations of minimum $\langle\overline{u}\rangle$ are not misrepresenting the 
time-averaged velocities. Experimental plots similar to Figure \ref{fig:initialDaConflictMax} are seen in 
\cite{Pokrajac2007}, suggesting that the Smagorinsky model may be more reliably representing the physics at play. 
However, given that the Smagorinsky model is known to have trouble accurately predicting near-wall dynamics, 
and that the WALE model does a particularly good job capturing this behavior \cite{Temmerman2003}, 
this conclusion must be met with some skepticism.

Before moving forward with further analysis, the decision was made to rigorously assess the 
quality of the mesh (note the inconsistent detail along the boundaries and surface in Figure \ref{fig:initialMesh}), 
the impact of each turbulence model on the double-averaged profile and the effect of domain size 
on the observed hydrodynamics. Additionally, the computation of several terms important to the 
DANS momentum balance was added to the Nalu code base to support such analysis. 

%% file: texFiles/mainMatter/chapter2/introduction.tex
\section{Introduction}\label{sec:lesStudyIntro}
This chapter explores the different decisions made in creating the numerical model
proposed in Section \ref{sec:numericalModel}. In particular, the effects of mesh refinement,
domain size and turbulence model on the predicted flow field are examined. In addition,
analysis of both instantaneous and double-averaged flow data is used to elucidate the 
governing mechanisms underlying momentum transport between the surface and subsurface flows.

%% file: texFiles/mainMatter/chapter2/redoGeometryMeshModel.tex
\section{Mesh Refinement and Time Averaging Study}\label{sec:redoMesh}

\subsection*{Mesh Refinement Study on a Sample Geometry}

\begin{figure}[!tbp]
    \centering  
    \subfloat[Coarse: Min.]{\includegraphics[width=0.22\textwidth]{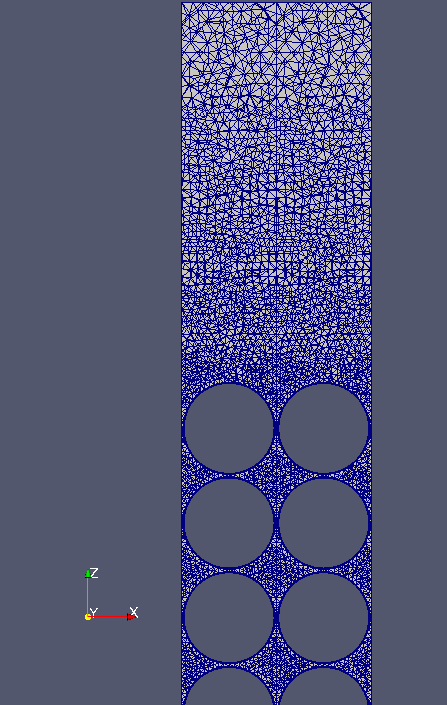}}
    \hfill
    \subfloat[H-Coarse: Min.]{\includegraphics[width=0.22\textwidth]{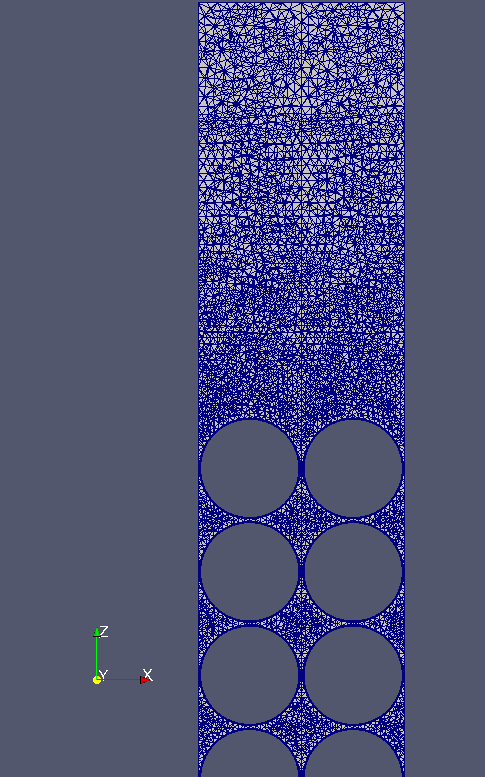}}
    \hfill
    \subfloat[Medium: Min.]{\includegraphics[width=0.22\textwidth]{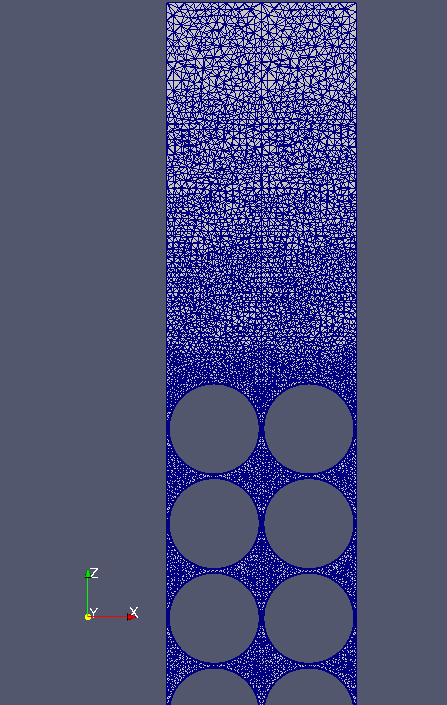}}
    \hfill
    \subfloat[Fine: Min.]{\includegraphics[width=0.22\textwidth]{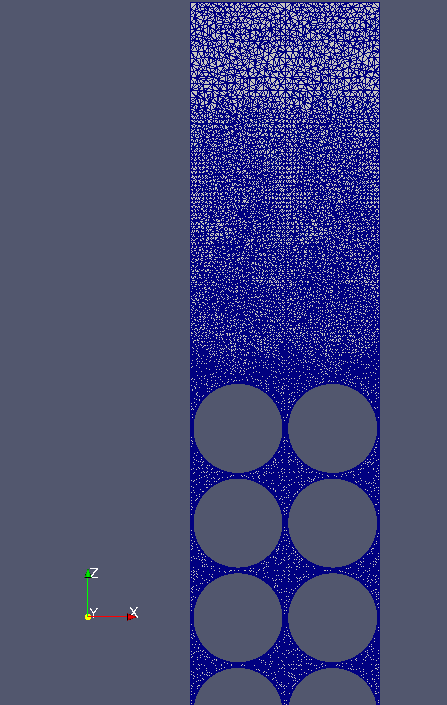}}
    \hfill
    \subfloat[Coarse: Max.]{\includegraphics[width=0.22\textwidth]{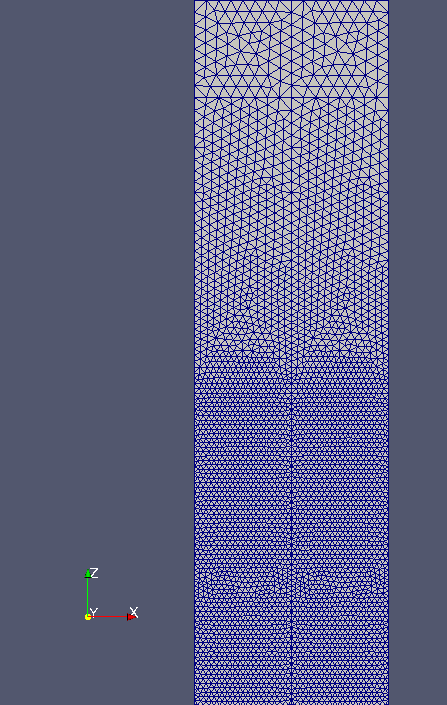}}
    \hfill
    \subfloat[H-Coarse: Max.]{\includegraphics[width=0.22\textwidth]{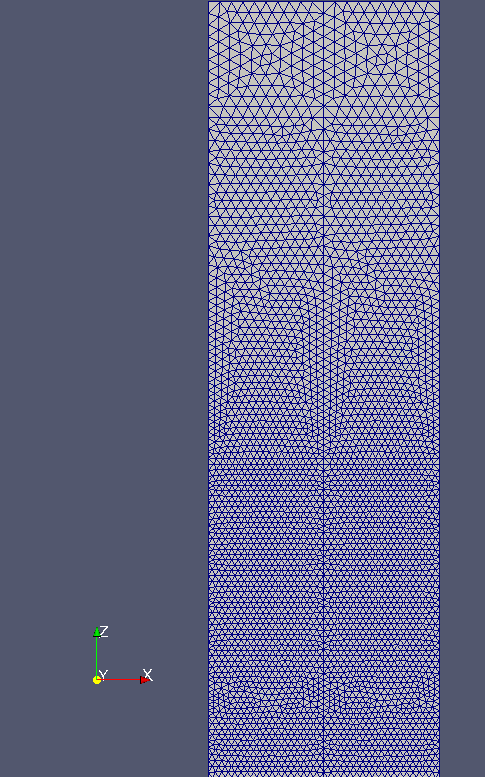}}
    \hfill
    \subfloat[Medium: Max.]{\includegraphics[width=0.22\textwidth]{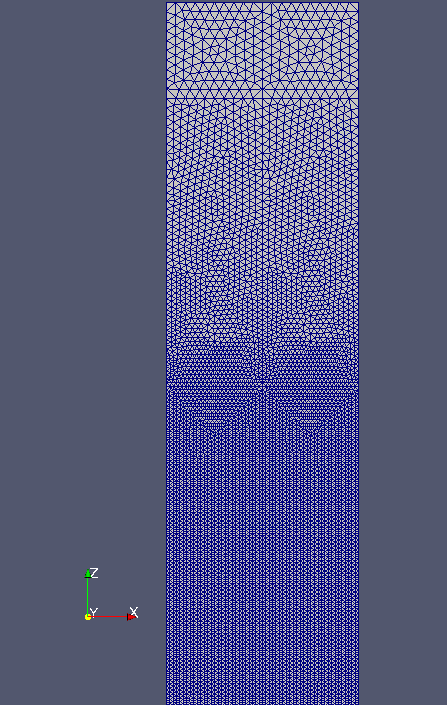}}
    \hfill
    \subfloat[Fine: Max.]{\includegraphics[width=0.22\textwidth]{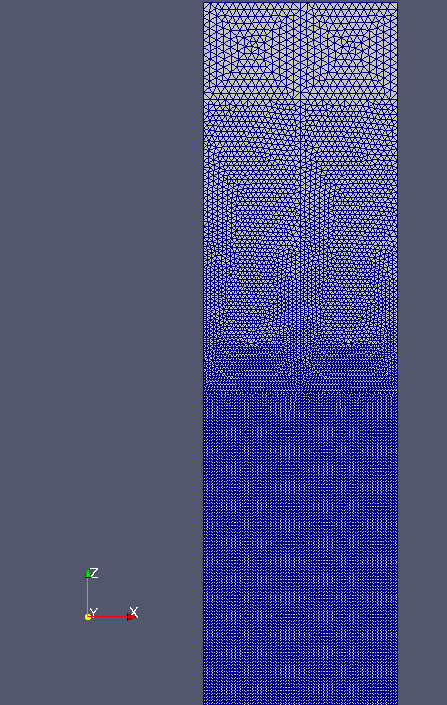}}
    \caption{Cross-section of each mesh used in refinement study.}
    \label{fig:meshes}
\end{figure}

In order to determine a suitable mesh for the detailed modeling of the geometries in Figure 
\ref{fig:geometries}, the 2x2x4 case with the WALE turbulence model was used to test several different levels 
of mesh refinement. 
Cross-sections of each mesh in the $x-z$ planes of maximum and minimum porosity are displayed in Figure 
\ref{fig:meshes}. To minimize computational cost, elements within $z^{*} \in [-4,0]$
are given a characteristic size $h_{B}$, deemed the bed-scale, which is smaller than the characteristic 
surface-scale element size of $h_{S}$, applied to elements within $z^{*} \in [3,4]$. Consequently, the 
elements in-between these regions have a length defined by the gradation from the bed-scale to the 
surface-scale. Such gradation in element size allows for appropriate resolution of the 
boundary layers surrounding the spherical voids, while focusing computational efforts on resolving larger 
turbulent structures within the homogeneous fluid region.

Table \ref{tab:mesh} details the $h_{B}$, $h_{S}$, and number of nodes and elements associated with each 
mesh used in this study. The meshes are labeled according to their relative level of refinement, where 
H-Coarse stands for Hybrid-Coarse, using Coarse's $h_{B}$ and Medium's $h_{S}$.

\begin{table}[h!]
  \centering
  \caption{Mesh characteristics used in refinement study.}
  \label{tab:particleClasses}
  \begin{tabular}{|l|l|l|l|l|}  
    \toprule
    Label     & Bed-scale (mm) & Surface-scale (mm) & Num. Elements      & Num. Nodes\\
    \midrule
    Coarse    & 2.133          & 5.330              & $8.963\cdot 10^5$  & $2.181\cdot 10^5$ \\
    H-Coarse  & 2.133          & 4.000              & $1.026\cdot 10^6$  & $2.405\cdot 10^5$ \\
    Medium    & 1.600          & 4.000              & $1.970\cdot 10^6$  & $4.487\cdot 10^5$ \\
    Fine      & 1.200          & 3.000              & $4.554\cdot 10^6$  & $9.762\cdot 10^5$ \\
    \bottomrule
  \end{tabular}
  \label{tab:mesh}
\end{table}

\begin{figure}[!tbp]
    \centering  
    \subfloat[]
        {\includegraphics[width=0.5\textwidth]{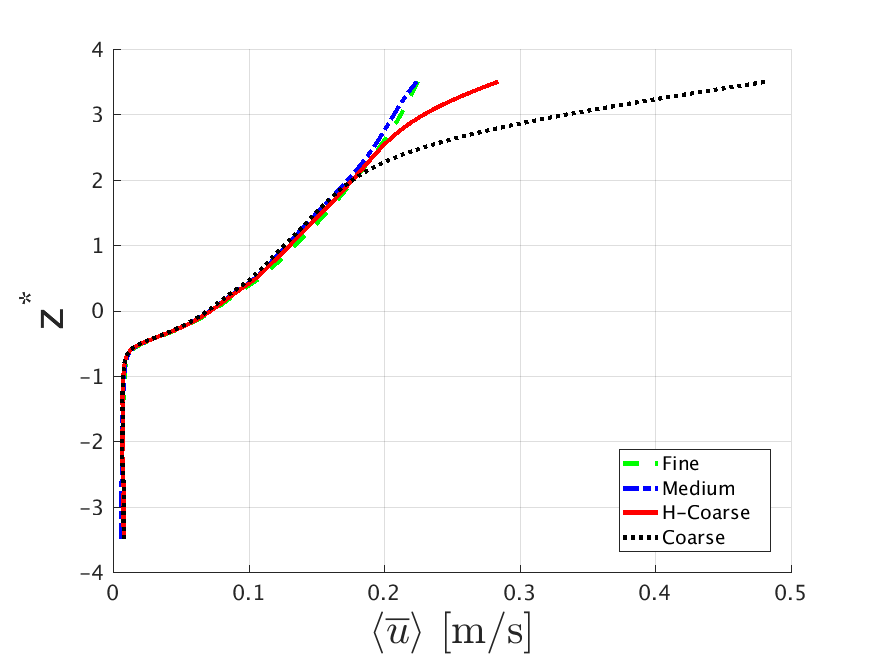}
         \label{fig:daTestMeshFull}}
    \subfloat[]
        {\includegraphics[width=0.5\textwidth]{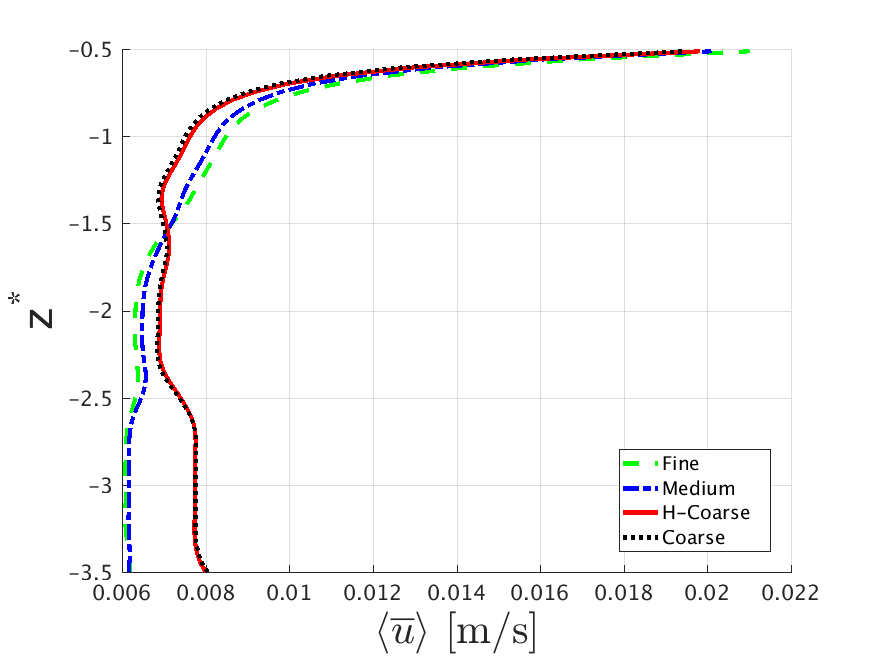}
         \label{fig:daTestMeshSub}}
    \caption{$\langle\overline{u}\rangle(z^{*})$ as predicted by each test mesh. 
      \protect\subref{fig:daTestMeshFull} $z^{*} \in [-3.5,3.5].$
      \protect\subref{fig:daTestMeshSub}  $z^{*} \in [-3.5,-0.5]$.}
    \label{fig:meshTestProfile}
\end{figure}

Each mesh was used to simulate the model described in Section \ref{sec:numericalModel}. The resultant double-averaged 
streamwise velocity profiles for the entire domain and the subsurface flow are shown in Figure 
\ref{fig:daTestMeshFull} and \ref{fig:daTestMeshSub}, respectively. As seen in Figure \ref{fig:daTestMeshFull}, 
the surface flow profiles vary greatly near the top of the domain. While Fine and Medium show 
reasonable agreement, Coarse and H-Coarse produce diverging surface profiles. Interfacial and subsurface 
profiles are more clearly observed in Figure \ref{fig:daTestMeshSub}, where we see bi-modal agreement between the  
coarse and fine meshes. 

The divergence seen near the top of the domain grows worse in time, and, given that the level of mesh refinement 
is the only variable among simulation parameters, clearly shows that some instability is present in the coarser 
meshes. One possible explanation is due to an abuse of the LES closure model. As the mesh grows coarse in 
the transition region, fewer details of the turbulence are resolved. This 
places an increased demand on the closure model to dissipate energy in the domain, as these energetic losses 
to viscosity are not directly computed. With insufficient resolution, it appears the numerical closure model 
cannot dissipate energy quickly enough to comply with the physics of the flow, and energy erroneously accumulates. 

Noting the relative convergence of Fine and Medium as support that a nearly mesh-independent solution has been computed, 
these two cases were run once more to study how the quality of the Reynolds-averaged value 
(i.e. number of samples used in computing the time average) influenced the double-averaged profile.
In each case, the simulation advanced with no data collection until it achieved a turbulent steady state.
Once steady, evolving with a time step of $\Delta{t}$ = 0.02 s, the current value of the Reynolds-averaged 
velocities were output every 180 seconds of simulation time for 900 seconds. 

\begin{figure}[!tbp]
    \centering  
    \subfloat[]
        {\includegraphics[width=0.5\textwidth]{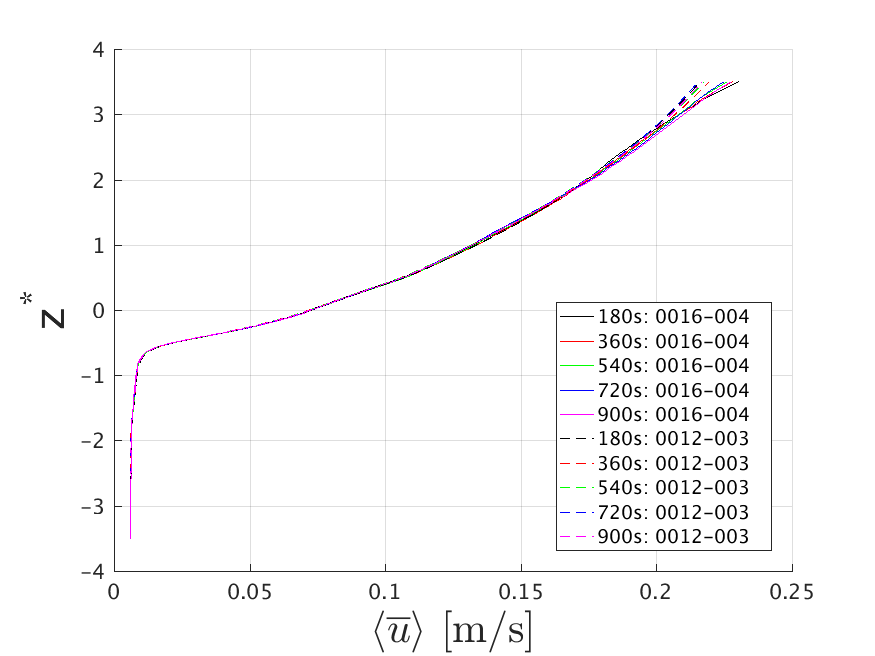}
         \label{fig:timeFilterFull}}
    \subfloat[]
        {\includegraphics[width=0.5\textwidth]{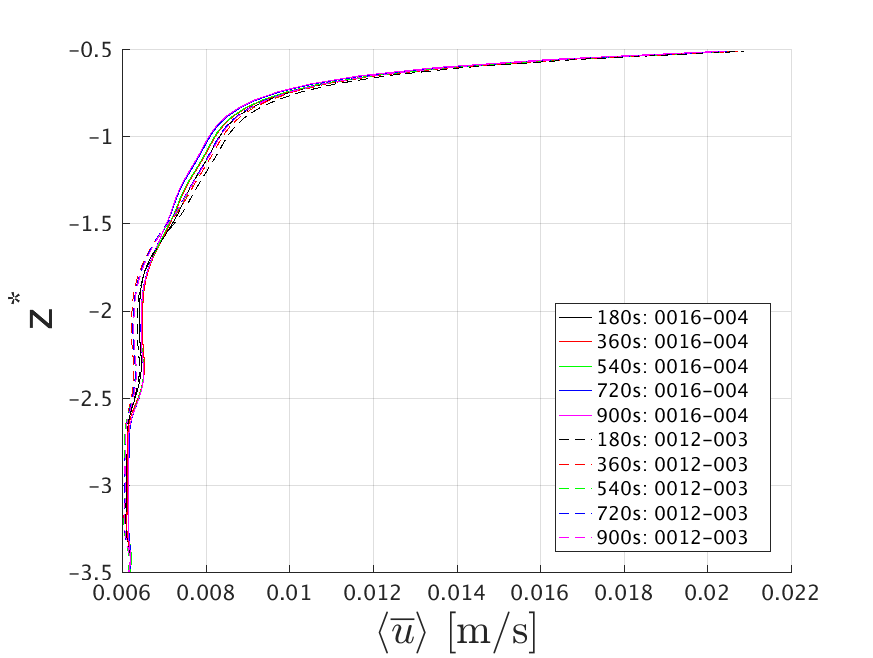}
         \label{fig:timeFilterSub}}
    \caption{$\langle\overline{u}\rangle(z^{*})$ as predicted using several time-averaging filters for both Fine and Medium.
      \protect\subref{fig:daTestMeshFull} $z^{*} \in [-3.5,3.5].$ 
      \protect\subref{fig:daTestMeshSub}  $z^{*} \in [-3.5,-0.5]$. The legend numbering refers to the length of time-averaging 
      and the mesh size used, specified in Table \ref{tab:mesh}.}
    \label{fig:timeFilterTest}
\end{figure}

The results of these tests are shown in Figure \ref{fig:timeFilterTest}, where dashed and solid lines correspond 
to the fine and medium meshes, respectively. Results throughout the domain are in generally good agreement, 
with a maximum difference of $6.7\%$ for any equivalent time interval measure in the maximum surface velocity. 
Moving away from the top of the domain where the flow rate is highest, inspection of the subsurface 
flow shows a desirable overlap in the profiles near the interface, and the maximum difference for 
any equivalent time measure drops to $4.2\%$ within the bed. Furthermore, for any equivalent time measure, 
the mean difference within the region $z^{*} \in [-2,2]$ is no more than $0.83\%$, indicating excellent 
agreement within the transition region. As Fine and Medium produce nearly identical 
results throughout the interface region and differences between the two results are shown to be small across 
the entire domain, all following simulations use Medium with intention to study larger geometries at the expense of 
minimal additional resolution.

%% file: texFiles/mainMatter/chapter2/domainSize.tex
\section{The Influence of Domain Size on the Flow Field}\label{sec:domainSize}

When performing LES, periodic boundary conditions enable one to obtain a fully developed
turbulent inflow condition by recycling the fluctuating velocity field exiting the domain at the
inlet. One must be careful using these conditions, however, as the constraint of perfect correlations at
the periodic domain boundaries may produce nonphysical influences on the flow field if the simulation box
is too small \cite{Meyers2016}. In particular, the domain size directly limits the size of 
the largest turbulent structures which may be resolved \cite{Frohlich2005,Mendez2008}. As the larger
structures are expected to play an important role in interfacial transport, the fidelity of the simulation
depends on using a sufficiently large domain.

\begin{figure}[!tbp]
    \centering  
    \subfloat[]
        {\includegraphics[width=0.5\textwidth]{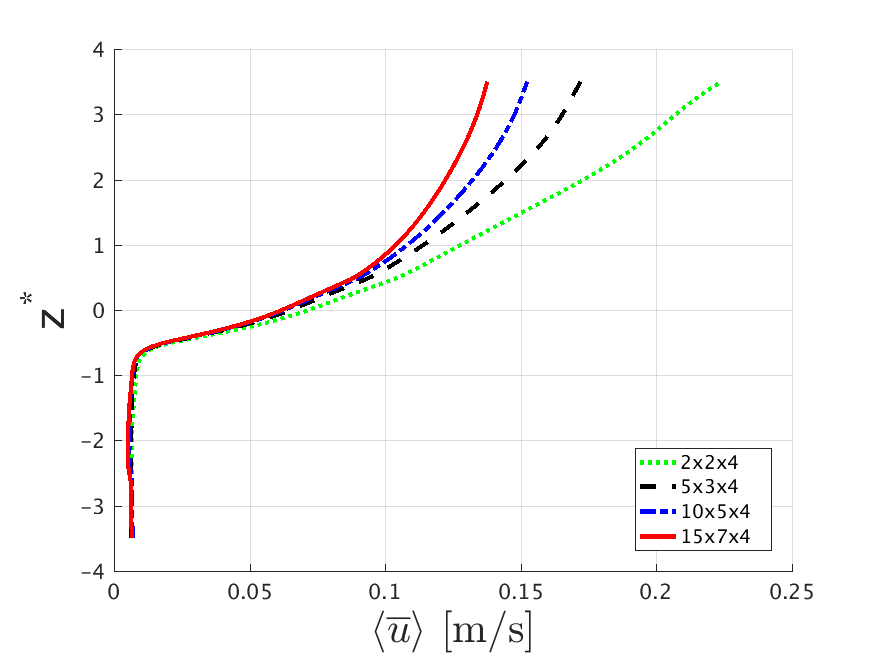}
         \label{fig:domainFull}}
    \subfloat[]
        {\includegraphics[width=0.5\textwidth]{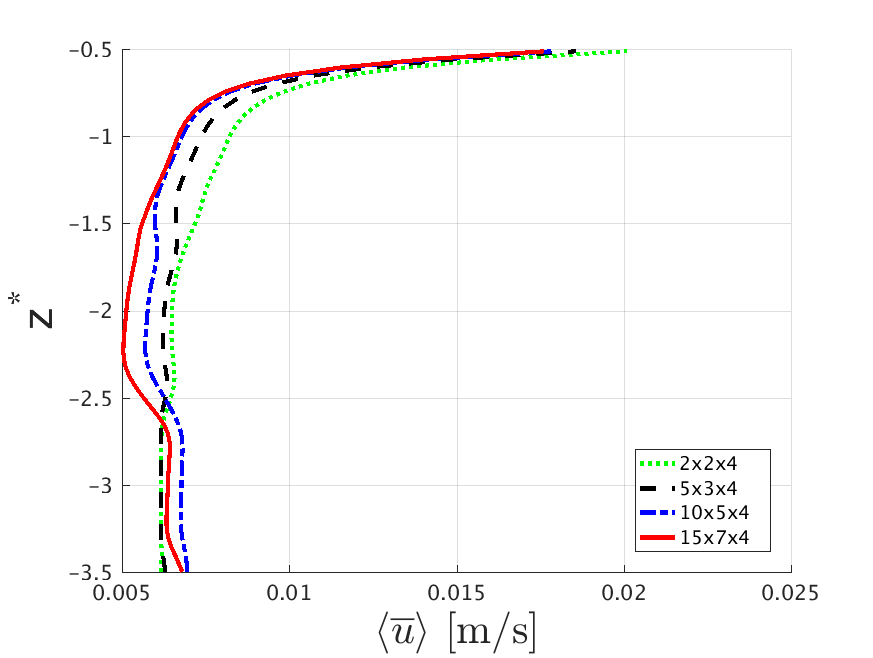}
         \label{fig:domainSub}}
    \caption{$\langle\overline{u}\rangle(z^{*})$ as predicted by each domain.
      \protect\subref{fig:domainFull} $z^{*} \in [-3.5,3.5].$ 
      \protect\subref{fig:domainSub}  $z^{*} \in [-3.5,-0.5]$. The 2x2x4 curve uses the same domain and meshing
      scheme as the Medium scheme seen in the mesh refinement study.}
    \label{fig:domainVelocity}
\end{figure}

To clarify the impact of domain size on the LES results, all four geometric models presented in Figure \ref{fig:geometries} 
were studied subject to the problem constraints specified in Section \ref{sec:numericalModel}, using the meshing
scheme Medium specified in Section \ref{sec:redoMesh} and the WALE turbulence model. Double-averaged streamwise 
velocity profiles are shown 
in Figures \ref{fig:domainFull} and \ref{fig:domainSub}, respectively. Figure \ref{fig:domainFull} 
demonstrates a clear size dependence in the magnitude of the surface flow, which decreases
with increasing domain size. With increasing distance from the bed, velocities more rapidly tend 
toward a constant as the bed size increases, in line with the profile one might expect
when recalling the velocity profile for the standard smooth wall-bounded case \cite{Kundu2016}.
A bulk Reynolds number, $Re = \langle\overline{u_1}\rangle{H}/\nu$, where the double average is taken
over the surface flow of depth H = 0.161 m, may be computed for each simulation as: 
$Re_{15x7} = 19,297$, $Re_{10x5} = 20,246$, $Re_{5x3} = 22,162 $, and $Re_{2x2} = 26,904$. 

One possible reason for the inverse relationship between surface velocity magnitude and domain size
is that the smaller domains restrict the development of large turbulent structures, squeezing the size
of the largest eddies. Wall permeability is known to increase the presence of large-scale vortical motions, and 
these motions encourage exchange between high momentum fluid in the surface flow and low momentum fluid between 
the roughness elements and within the bed \cite{Breugem2006}. If turbulence is being constrained by the 
size of the domain and then this domain size is increased while the 
driving force of the flow is held constant, the mean flow in the homogeneous fluid region will slow down as
high momentum fluid is moved towards the bed.

Compared to the surface flow, trends in subsurface flow behavior with increasing domain size are less consistent. 
In Figure \ref{fig:domainSub} the velocities in $z^{*} \in [-2.5, -0.5]$ are seen to decrease as the 
bed grows, while there appears to be no clear trend for velocities below $z^{*}=-2.5$. However, 
a minimum in $\langle{\overline{u}}\rangle(z^{*})$ is obtained at $z^{*}=-2.24$ in the two largest domains. In addition,
neither of these domains tend toward a constant subsurface velocity within the bed. These trends are more pronounced 
in the 15x7x4 bed, suggesting that the presence of turbulent flow is being felt
deeper within the bed. 

Given that eddies carry correlated fluid motion with them, another approach to assess the domain 
size dependence of the largest resolved vortical structures is to compute
the correlation coefficient for the fluctuating velocity field along some path. The size of the largest eddies 
can be understood from the length over which fluid motion remains correlated, and thus the correlation 
coefficient provides a metric to assess whether periodic boundary conditions are artificially constraining the 
size of turbulent structures. To be confident that structures are not being unphysically squeezed by the domain,
fluctuations should become decorrelated over a length less than half of the domain size \cite{Frohlich2005,Mendez2008}. 

The correlation coefficient, $R_{ij}$, may be computed as a function of some time lag, $\Delta{t}$ and 
some space lag $\Delta{x}$, as described in Equation \ref{eq:generalCC}. 
%
\begin{equation}
  R_{ij}(\Delta{x},\Delta{t}) = 
              \frac{\overline{u_{i}(\textbf{x}_1,t_1) u_{j}(\textbf{x}_1+\Delta{x},t_1+\Delta{t}) }}
              {  \sqrt{\overline{ u_{i}^{2}(\textbf{x}_1,t_1)}}
                 \sqrt{\overline{ u_{j}^{2}(\textbf{x}_1+\Delta{x},t_1+\Delta{t})}} }
  \label{eq:generalCC}
\end{equation}
Because the correlation coefficient
is the covariance of two random variables normalized by the root mean square of each variable, 
the correlation coefficient may hold values between $-1$ and $1$, where the former denotes perfect
anticorrelation and the latter denotes perfect correlation \cite{Kundu2016}.

\begin{figure}[!tbp]
    \centering  
    \subfloat[]
        {\includegraphics[width=0.35\textwidth]{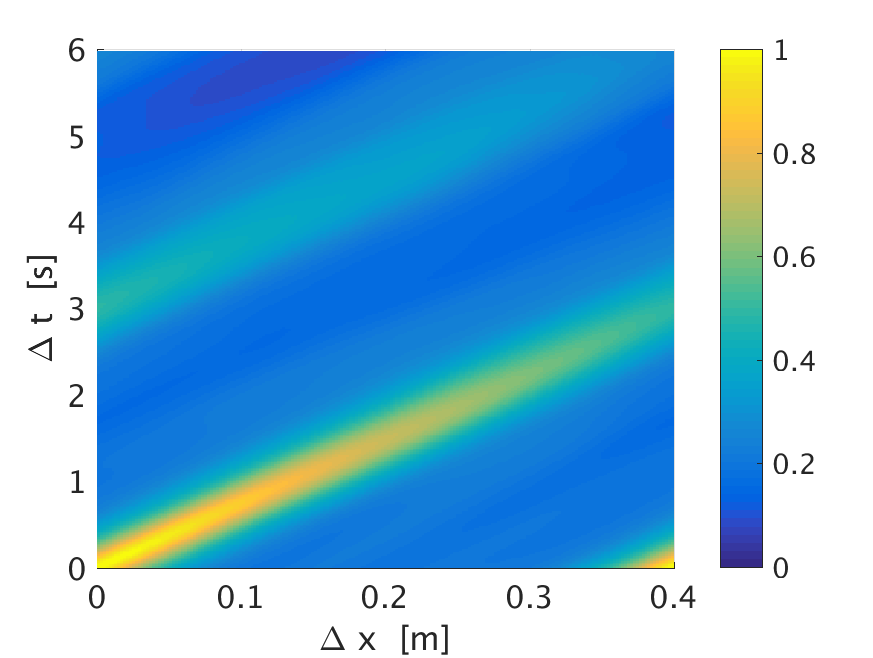}
         \label{fig:corrDomainCompare10}}
    \subfloat[]
        {\includegraphics[width=0.35\textwidth]{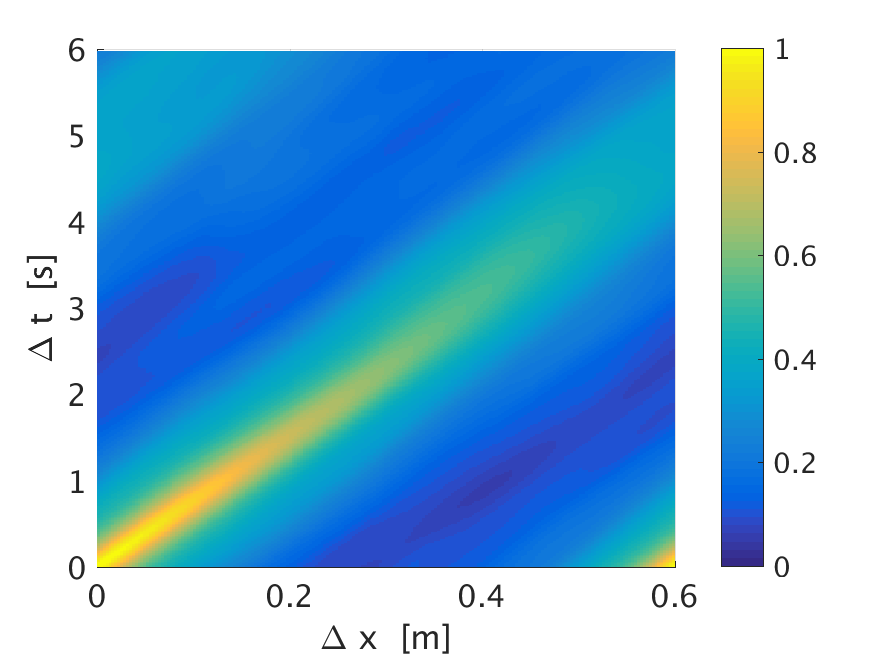}
         \label{fig:corrDomainCompare15}}
    \subfloat[]
        {\includegraphics[width=0.35\textwidth]{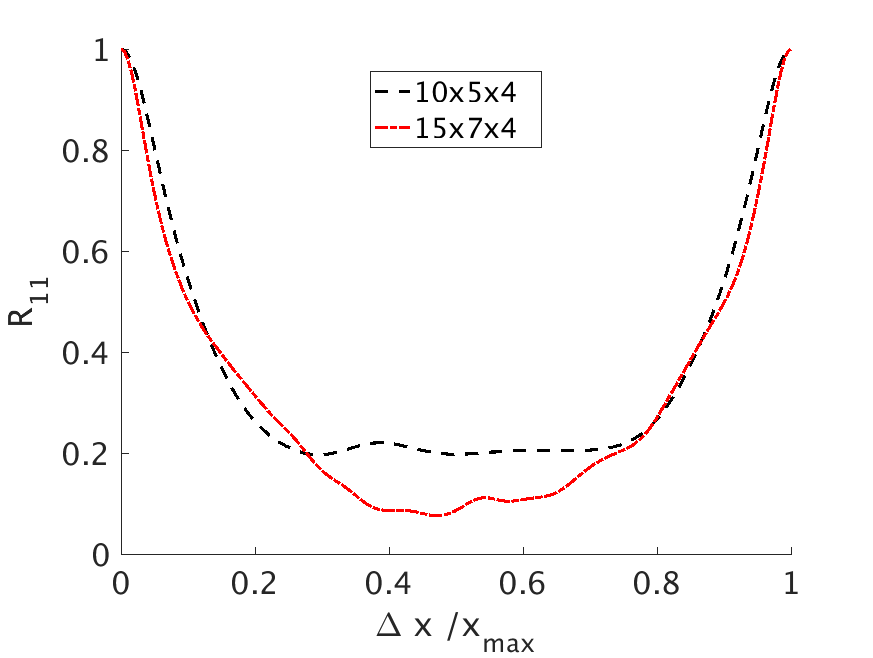}
         \label{fig:corrDomainCompare1510}}
    \caption{Autocorrelation plots of $R_{11}(\Delta{x},\Delta{t})$ for  
      \protect\subref{fig:corrDomainCompare10} the 10x5x4 domain and  
      \protect\subref{fig:corrDomainCompare15} the 15x7x4 domain.
      \protect\subref{fig:corrDomainCompare1510} A comparison $R_{11}(\Delta{x},0)$ for both domains.}
    \label{fig:corrDomainCompare}
\end{figure}

Figures \ref{fig:corrDomainCompare10} and \ref{fig:corrDomainCompare15} display $R_{11}(\Delta{x},\Delta{t})$
computed over a line with coordinates $(\Delta{x}, 0.04, 0.24)$ m, where $\Delta{x} \in [0, L_{x}]$m and 
$\Delta{t} \in [0, 6]$ s, for the 10x5x4 and 15x7x4 beds, respectively. 
For each domain, a line of strong correlation is seen moving with a velocity $\Delta{x}/\Delta{t}$ which
roughly matches the double-averaged velocity specified in Figure \ref{fig:domainFull}. This indicates that 
eddies in the surface are moving with the mean flow, and provides a sense of how long structures in the homogeneous
fluid region maintain coherency.

To inspect the correlations strictly as a function of space, Figure \ref{fig:corrDomainCompare1510}
shows $R_{11}(\Delta{x},0)$ for both domains. As expected when using periodic boundary conditions, 
$R_{11}(0,0) = R_{11}(L_{x},0) = 1$, as the flow sees these locations as the same position. Neither geometry 
achieves complete decorrelation of streamwise velocity fluctuations at a lag $\Delta{x} = \frac{L_{x}}{2}$, 
with the 10x5x4 and 15x7x4 beds producing $R_{11}(L_{x}/2,0) = 0.197$ and  $R_{11}(L_{x}/2,0) = 0.086$, 
respectively. It is evident that the growth of vortical structures is suppressed by the domain size in
both cases, although the larger domain is relatively close to meeting the specified criterion. 

This investigation into the impact of domain size on the double-averaged streamwise velocity profile provides
multiple indications that the expansion of the domain leads to an increase in penetration depth of turbulent structures
into the bed, and possibly an increase in the strength of those structures. Increased penetration can occur due to
larger turbulent structures, arising from the presence of wall permeability \cite{Breugem2006}, and therefore 
corroborates the argument that adequate resolution of large turbulent scales, and consequently the use of a sufficiently 
large domain, is critical in performing high-fidelity LES on this type of system.

%% file: texFiles/mainMatter/chapter2/turbulenceModel.tex
\section{Comparison of The Smagorinsky and WALE Turbulence Models}
Within this section, data sets resulting from simulations using the WALE and Smagorinsky
turbulence models are compared. Both simulations were run on the 10x5x4 geometry discussed
in Section \ref{sec:domainSize} using the Medium mesh, detailed in Section \ref{sec:redoMesh}.
Analysis is split into two parts. First, double-averaged data is presented to look at 
large-scale persistent hydrodynamic information. These results are followed by instantaneous
flow data, providing a more detailed look at the presence of turbulence near and within the 
bed.
\subsection*{Double-Averaged Flow Measurements}
\begin{figure}[!tbp]
    \centering  
    \subfloat[]
        {\includegraphics[width=0.5\textwidth]{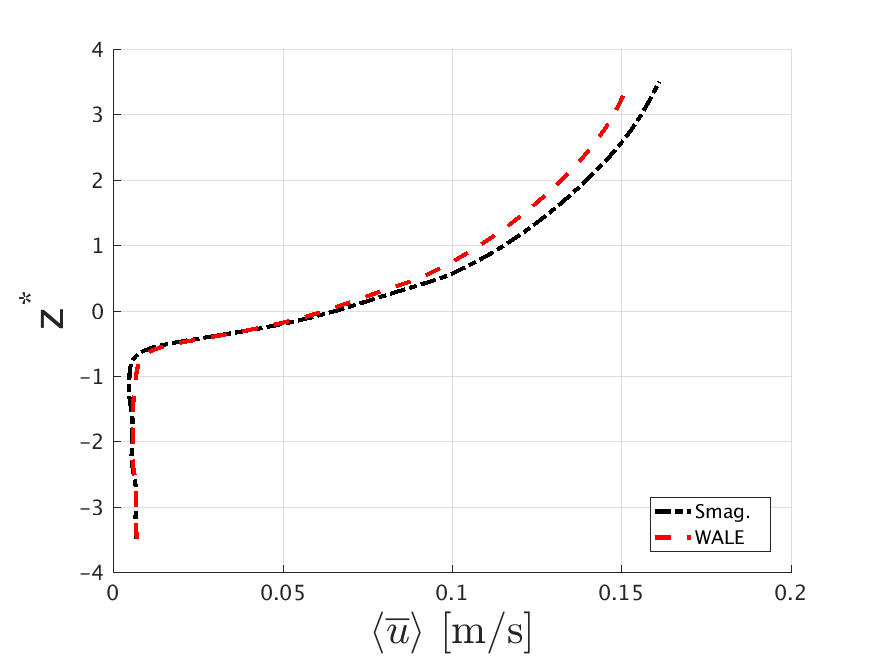}
         \label{fig:daVelTurbFull}}
    \subfloat[]
        {\includegraphics[width=0.5\textwidth]{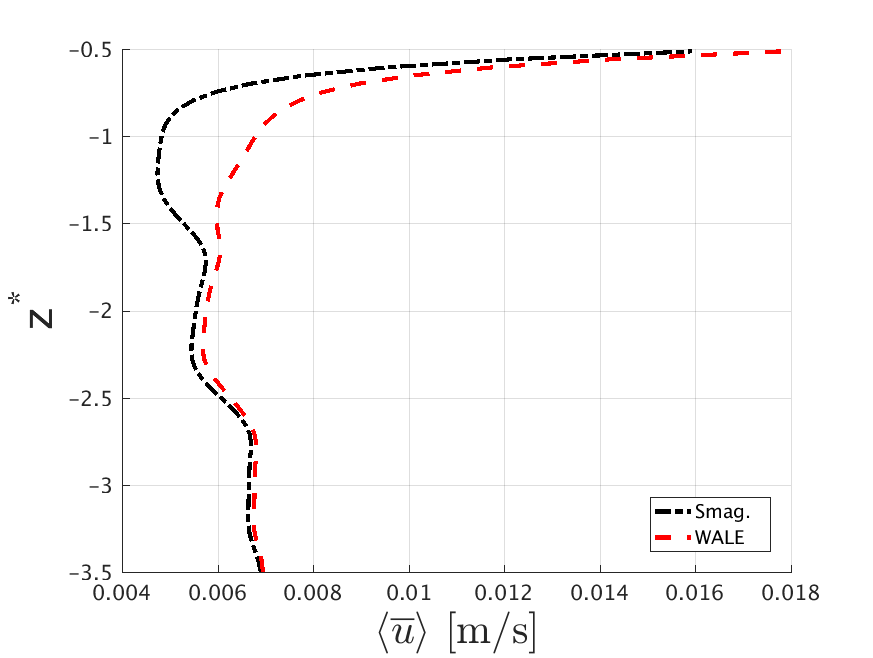}
         \label{fig:daVelTurbSub}}
    \caption{$\langle\overline{u}\rangle(z^{*})$ as predicted by the WALE and Smagorinsky turbulence models on the
             10x5x4 domain.
      \protect\subref{fig:daVelTurbFull} $z^{*}\in [-3.5,3.5].$
      \protect\subref{fig:daVelTurbSub}  $z^{*}\in [-3.5,-0.5].$}
    \label{fig:velocityModCompare}
\end{figure}

Double-averaged streamwise velocity profiles for simulations using the WALE and Smagorinsky closure
models are shown in Figure \ref{fig:velocityModCompare}. Given the well documented differences in the 
treatment of sub-grid stresses between the models (e.g. \cite{Temmerman2003,Ducros1998}), it is
no surprise that there exists discrepancy between the velocity profiles, even when the simulations
share identical domains and meshes. However, it is clear that the concerns discussed in Section \ref{sec:initResults} were not due
to the mesh or geometric model, as velocity minima for the WALE and Smagorinsky models are found at 
nearly the same locations ($z^{*} = -2.19$ and $z^{*} = -1.21$, respectively).

\begin{figure}[!tbp]
    \centering  
    \subfloat[]
        {\includegraphics[width=0.5\textwidth]{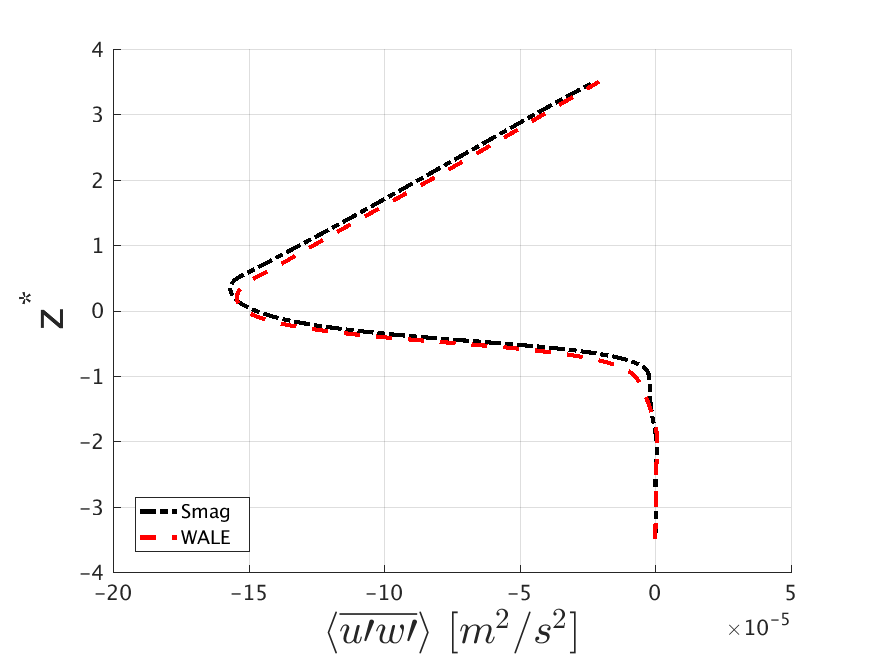}
         \label{fig:daRsTurbFull}}
    \subfloat[]
        {\includegraphics[width=0.5\textwidth]{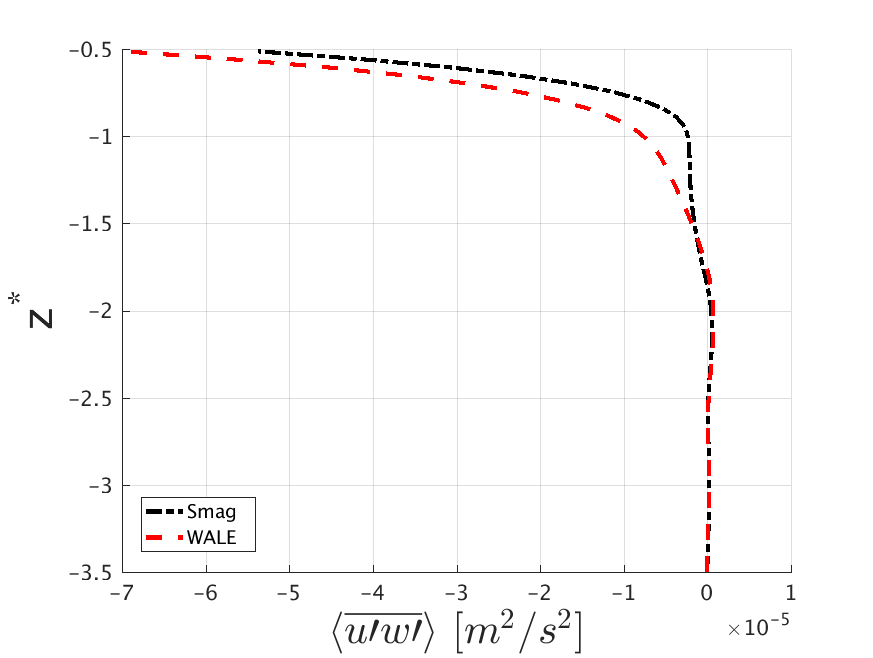}
         \label{fig:daRsTurbSub}}
    \caption{$\langle\overline{u^{'}w^{'}}\rangle(z^{*})$ as predicted by each turbulence model.
      \protect\subref{fig:daRsTurbFull} $z^{*}\in [-3.5,3.5].$
      \protect\subref{fig:daRsTurbSub}  $z^{*}\in [-3.5,-0.5].$}
    \label{fig:daRsTurbCompare}
\end{figure}

The corresponding double-averaged Reynolds shear stresses (1,3) for the two closures are presented in
Figure \ref{fig:daRsTurbCompare}. Both curves assume nearly the same profile, growing linearly in magnitude
when approaching the permeable wall from the surface flow and then decreasing rapidly with penetration into
the bed. Although similar, the WALE model predicts a more mild decay with depth while the Smagorinsky model 
predicts nearly laminar flow within the first pore, indicating that
vortical motions produced by the WALE model are able to penetrate further beyond the permeable wall and 
have a more pronounced role in transport. 

One notable feature belonging to these profiles is the rapid decrease 
in magnitude near the interface, which makes clear a challenge in linking surface and subsurface models
via a boundary condition when studying this system. Knowing that the double-average effectively smears the 
presence of highly localized events across an entire averaging volume, one can expect both the peak and rate 
of decay of the stress to increase when examining only the time-averaged quantity (note the decay of the 
TKE below the permeable wall in Figure \ref{fig:tkeCompare}). Thus, this double-averaged profile may prove useful
in the parameterization of such reduced-order models.

\begin{figure}[!tbp]
    \centering  
    \subfloat[]
        {\includegraphics[width=0.5\textwidth]{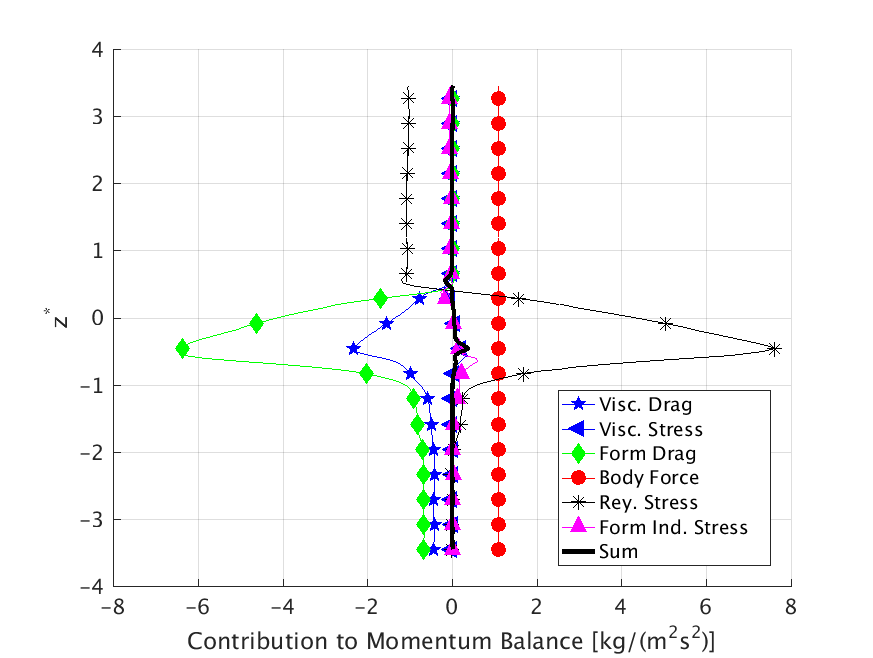}
         \label{fig:diffMomWale}}
    \subfloat[]
        {\includegraphics[width=0.5\textwidth]{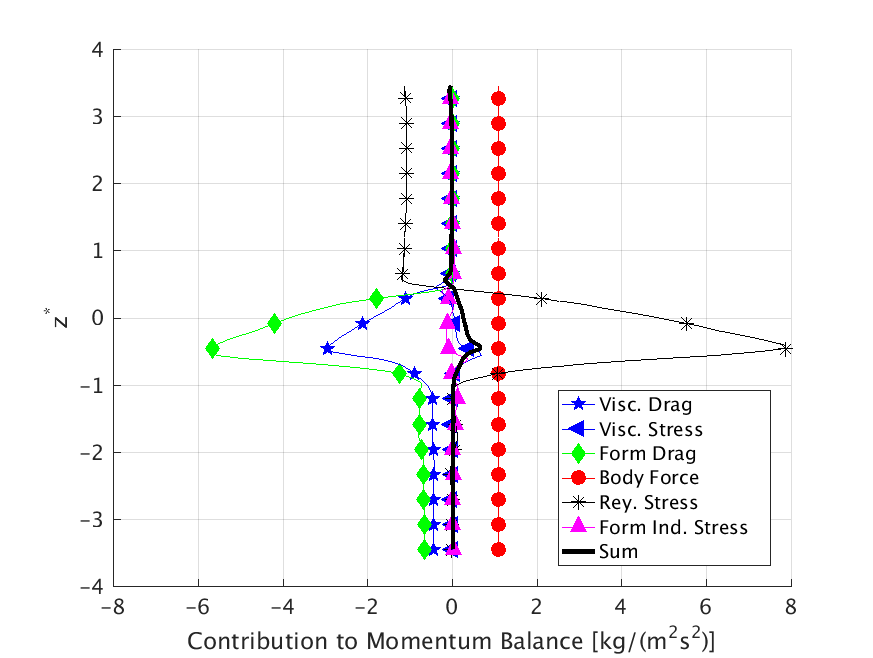}
         \label{fig:diffMomSmag}}
    \caption{Terms of the DANS momentum balance as predicted by 
      \protect\subref{fig:diffMomWale} the WALE model and 
      \protect\subref{fig:diffMomSmag} the Smagorinsky model. 
      Markers are used only for distinguishing curves
      and do not reflect resolution of the data.}
    \label{fig:diffMomCompare}
\end{figure}

Recalling the formulation of the DANS equations presented in Section \ref{sec:dansFormulation},
the differential form of the DANS momentum balance for the $i=1$ case is restated in Equation
\ref{eq:diffDANS} for reference. 
%
\begin{equation}
  0 =                                                                                                  -
  \frac{\rho}{\phi_{s}}\frac{\mathrm{d}\phi_{s}\langle \bar u \bar w \rangle}{\mathrm{d} z}            +
  \frac{1}{\phi_{s}}\frac{\mathrm{d}\phi_{s}\langle \overline{\tau_{13}^{R}} \rangle}{\mathrm{d} z}    +
  \frac{1}{\phi_{s}}\frac{\mathrm{d}\phi_{s}\langle \overline {\tau_{13}} \rangle}{\mathrm{d} z}       +        
  \langle \overline {\tau_{1j,j}^{SGS}}\rangle                                                         -
  f_{1}^{p,s}                                                                                          +
  f_{1}^{v,s}                                                                                          +
  b_{1} 
  \label{eq:diffDANS}
\end{equation}
The terms of this equation are shown in Figure \ref{fig:diffMomCompare}.
Simulations were performed on the 10x5x4 geometry with both the WALE (\ref{fig:diffMomWale}) and 
Smagorinsky (\ref{fig:diffMomSmag}) turbulence models.

Simulations with the different models show agreement in regions of constant porosity. 
Within the homogeneous fluid region, the body force, which drives the flow, is balanced by the derivative of the Reynolds 
shear stress. Additionally, viscous forces are seen to have little relevance away from the wall, as 
noted in \cite{Breugem2006} for walls with appreciable permeability.
Throughout the homogeneous porous region, the form and viscous drag forces act to balance the 
body force. According to arguments in \cite{Pokrajac2007}, one should expect to see the driving force
of the flow primarily balanced by form drag in the upper pores and viscous drag in the lower pores, but both
turbulence models predict that form drag does a better job than viscous drag at extracting fluid
momentum at all depths. Interestingly, the WALE model predicts the Reynolds shear stress becomes negligible
roughly a full unit cell diameter below the Smagorinsky model. In each case, the point at which the Reynolds
shear stress has nearly completely decayed roughly corresponds to the location of the minimum double-averaged
streamwise velocity. This suggests that the well noted minimum in the double-averaged velocity, 
which has been found to appear in the first layer of pores \cite{Pokrajac2007,Manes2009}, is 
related to the penetration depth of turbulence and not solely described by the form and viscous 
drag terms.

\begin{figure}[!tbp]
    \centering  
    \subfloat[]
        {\includegraphics[width=0.5\textwidth]{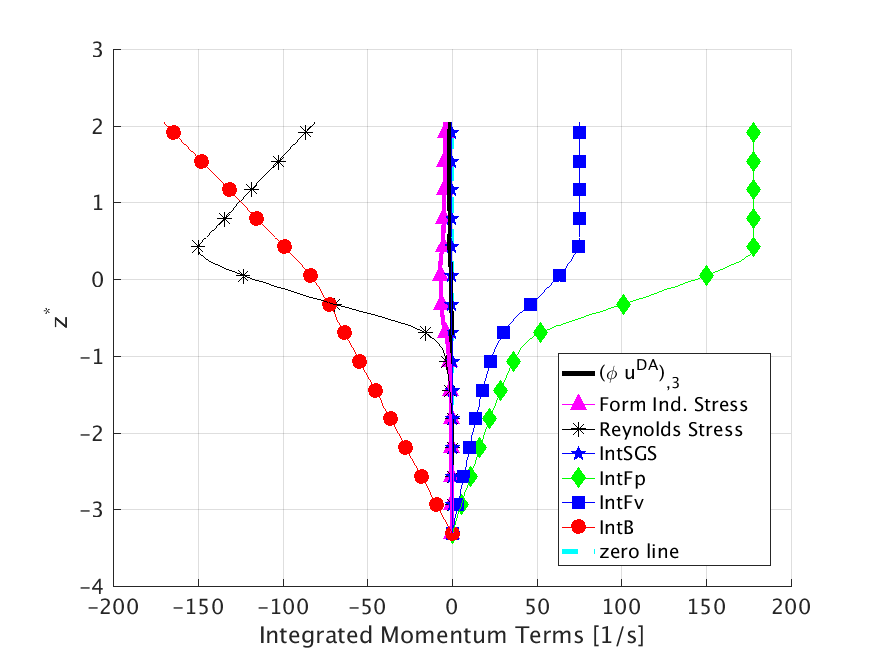}
         \label{fig:intMomWale}}
    \subfloat[]
        {\includegraphics[width=0.5\textwidth]{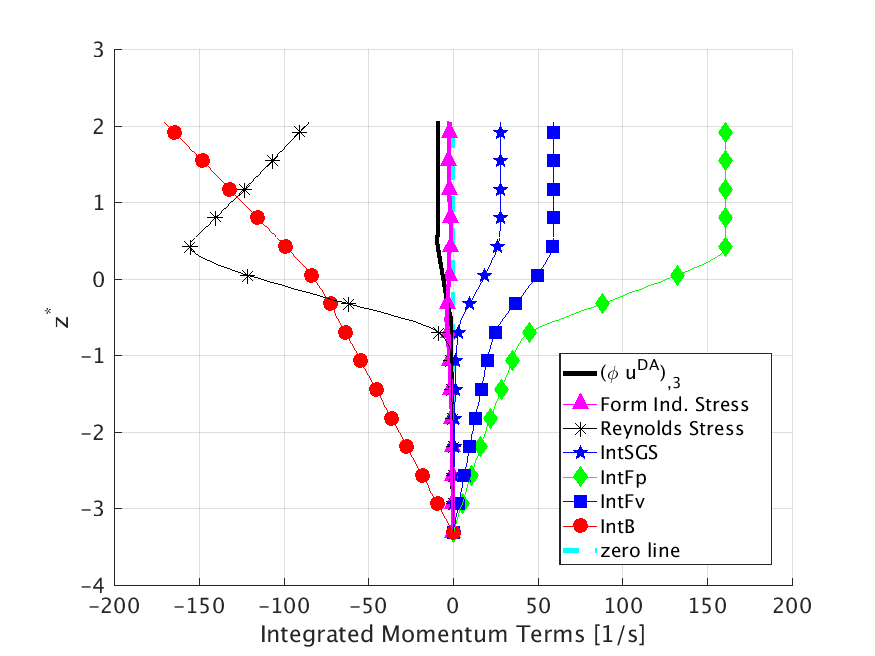}
         \label{fig:intMomSmag}}
    \caption{Terms of the integrated DANS momentum balance as predicted by 
      \protect\subref{fig:intMomWale} the WALE model and 
      \protect\subref{fig:intMomSmag} the Smagorinsky model. 
       The use of 'Int' in the legend denotes an integral. Markers are used only for distinguishing curves 
       and do not reflect resolution of the data.}
    \label{fig:intMomCompare}
\end{figure}

Once again recalling the formulations laid out in Section \ref{sec:dansFormulation}, the integrated
DANS momentum equation is given in Equation \ref{eq:integralDANS} for reference, and its constituent 
terms are plotted in Figure \ref{fig:intMomCompare} for the same simulations just discussed. 
%
\begin{eqnarray}                                          
  \frac{\mathrm{d}\phi_{s}\langle\overline{u}\rangle(z)}{\mathrm{d}z}                       & = & 
  \frac{1}{\mu}\left[
    \rho\phi_{s}(z)\langle\bar{u}\bar{w}\rangle(z)                                            -
    \phi_{s}(z)\langle\overline{\tau_{13}^{R}}\rangle(z) \right]\nonumber               \\ && +
  \frac{1}{\mu}\left[
    \int_{z_{0}}^{z} \phi_{s}f_{1}^{p}\mathrm{d}z^{\prime}                                    -
    \int_{z_{0}}^{z} \phi_{s}f_{1}^{v}  \mathrm{d}z^{\prime}                                  -
    \int_{z_{0}}^{z} \phi_{s}\langle\overline{\tau_{13}^{SGS}}\rangle \mathrm{d}z^{\prime}    -
    b_{i}\int_{z_{0}}^{z} \phi_{s}\mathrm{d}z^{\prime} \right]\nonumber                 \\    -
    \frac{C(z_0)}{\mu}
  \label{eq:integralDANS}
\end{eqnarray}

As this formulation allows for the isolation of the derivative of $\langle{\overline{u}}\rangle$ within
the homogeneous porous region, where $\phi_{s} = \epsilon_{B}$, these plots are of interest
primarily in the region $z^{*} \in [-3.5,-0.5]$. In this region, results disagree with the conceptual
model put forth in
\cite{Pokrajac2007} and show that form drag extracts more momentum than viscous drag, even deep within the
bed. However, due to a small but nonzero value in the momentum residuals shown in Figure 
\ref{fig:diffMomCompare}, the sum of the right hand side terms in Equation \ref{eq:integralDANS} 
does not yield a critical point at the location of the minimum in the double-averaged streamwise velocity 
profile. Thus, the integrated DANS equation has limited use in the proposed analysis, and proves to 
be sensitive to even small numerical errors.

\begin{figure}[!tbp]
    \centering  
    \subfloat[]
        {\includegraphics[width=0.5\textwidth]{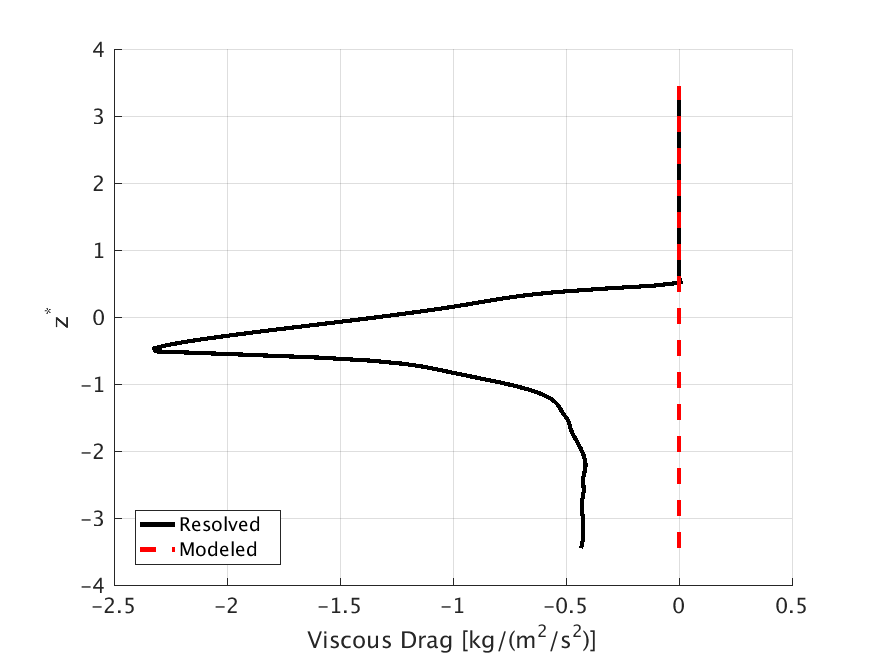}
         \label{fig:dragResolutionWale}}
    \subfloat[]
        {\includegraphics[width=0.5\textwidth]{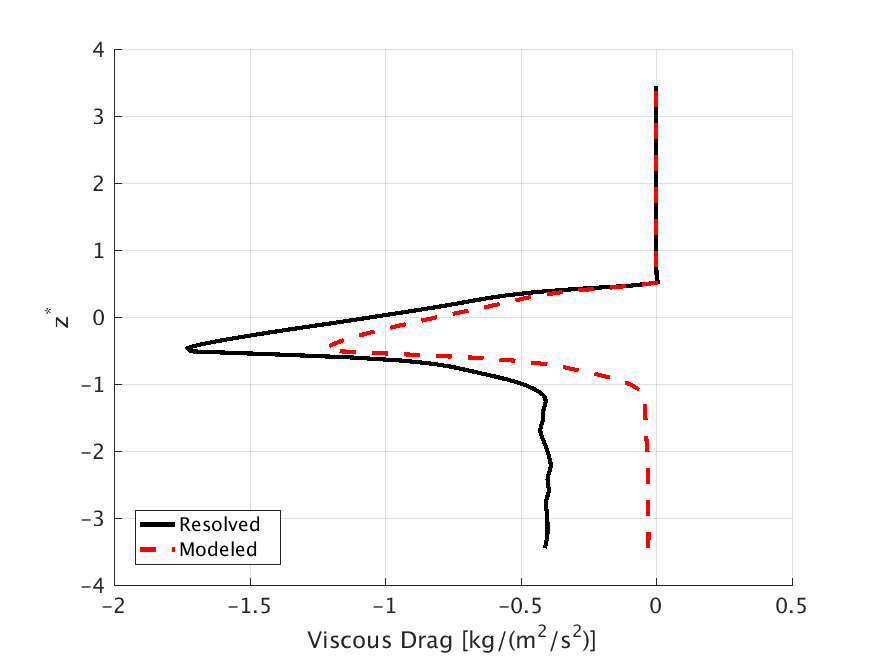}
         \label{fig:dragResolutionSmag}}
    \caption{Resolved and modeled contributions to the viscous drag force as predicted by
      \protect\subref{fig:diffMomWale} the WALE model and 
      \protect\subref{fig:diffMomSmag} the Smagorinsky model.}
    \label{fig:dragResolutionCompare}
\end{figure}

In an effort to better understand the contribution of modeled viscosity to the total viscous momentum 
transport, and thus measure the sensitivity of the flow to the turbulence model, the viscous drag force 
for each closure scheme has been separated into its resolved and modeled components, shown in 
Figure \ref{fig:dragResolutionCompare}. As expected, neither model produces any noticeable
measure of viscous drag far from the permeable wall. Given the high resolution boundary layer mesh surrounding 
each grain and the fine mesh within the bed, one might expect that sub-grid viscous contributions should only be a small 
fraction of the resolved viscous forces. The Smagorinsky model appears to meet expectations of 
overdamping in the near-wall limit, noted by the roughly constant, non-zero modeled viscous drag deep within
the homogeneous porous region and large contribution of modeled viscous drag in the transition layer. 
The negligible sub-grid contribution computed by the WALE model, however, suggests that the mesh is fine enough for
high quality LES. To more carefully investigate this difference, another mesh refinement study 
may be useful, executed with a focus on the full 
double-averaged momentum balance, or at least a more thorough study of the resolved and modeled
viscous forces.  

\subsection*{Verification of the Drag Force Calculations}
In an effort to verify the double-averaging methodology's ability to accurately recover
the form and viscous drag terms given the proposed averaging domain and geometric model, 
both viscous and pressure based contributions to the mean drag force per particle 
have been computed in two ways. Each drag force was computed indirectly via the DANS equations,
as described in Section \ref{sec:dansFormulation}, and directly via a surface integration 
within Nalu according to Equation \ref{eq:surfaceIntegration}. 
%
\begin{equation}
  F_{i} =  \int_{\Gamma} [\tau_{ij} - p\delta_{ij}]n_{j}dA 
  \label{eq:surfaceIntegration}
\end{equation}
where $\Gamma$ is the surface of each sphere and the viscous term, $\tau_{ij}$, uses the effective 
viscosity (i.e. $\mu_{e} = \mu + \mu_{t}$). This force may be decomposed into contributions from 
pressure and viscous terms, and then time-averaged to obtain the respective mean drag forces.

\begin{figure}[!tbp]
    \centering  
    \subfloat[]
        {\includegraphics[width=0.5\textwidth]{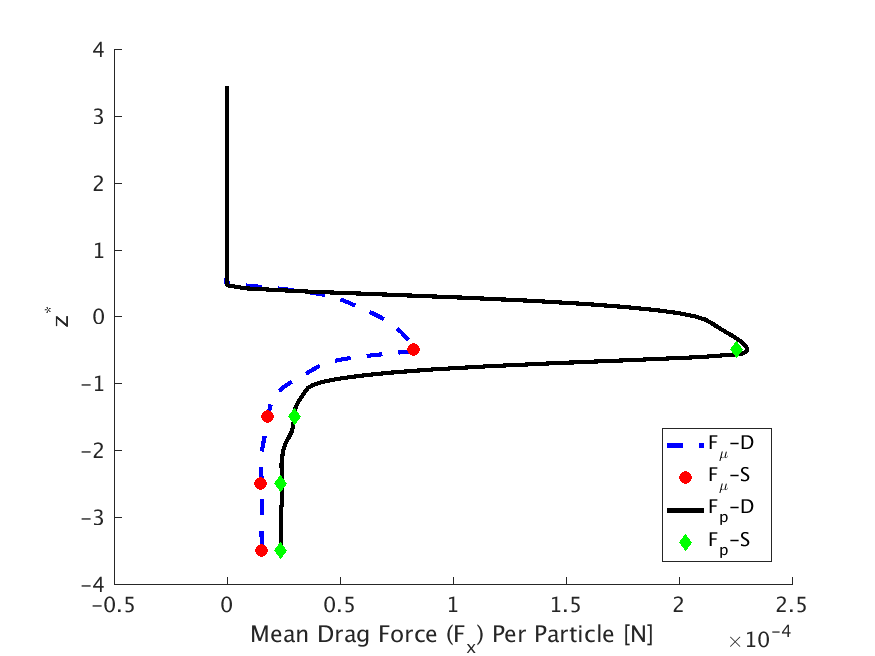}
         \label{fig:waleDragX}}
    \subfloat[]
        {\includegraphics[width=0.5\textwidth]{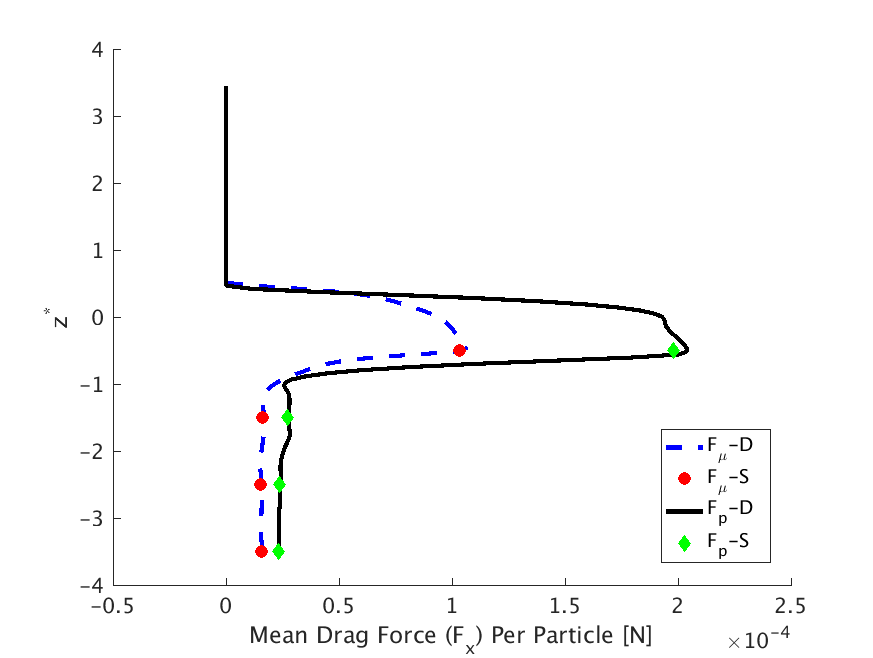}
         \label{fig:smagDragX}}
    \caption{Comparisons of surface integration and double-averaging for drag force calculations 
             (x-component) as predicted by
      \protect\subref{fig:waleDragX} the WALE model and
      \protect\subref{fig:smagDragX} the Smagorinsky model. The 'D' denotes double-averaging 
      and the 'S' denotes surface integration.}
    \label{fig:dragTermsX}
\end{figure}
\begin{figure}[!tbp]
    \centering  
    \subfloat[]
        {\includegraphics[width=0.5\textwidth]{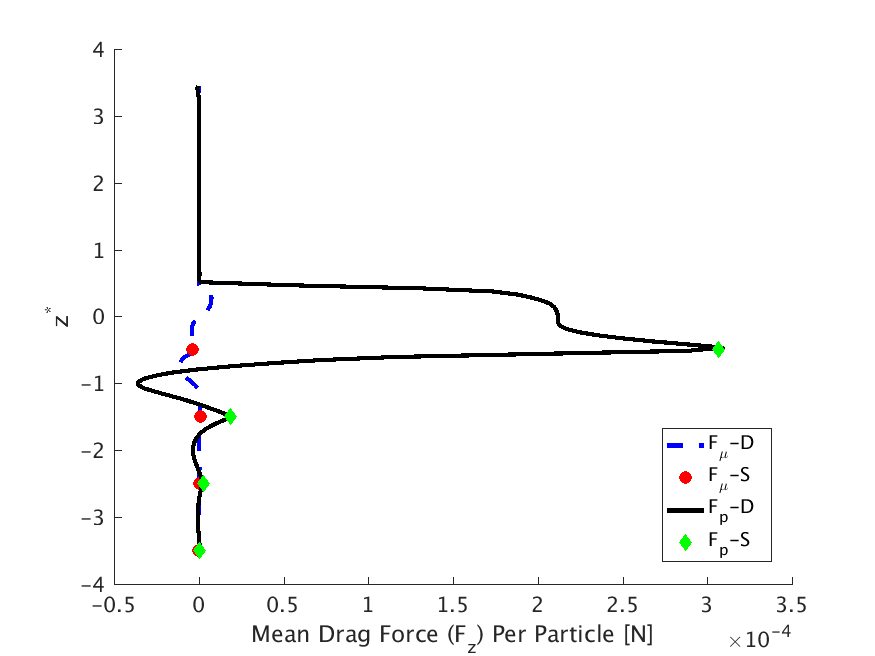}
         \label{fig:waleDragZ}}
    \subfloat[]
        {\includegraphics[width=0.5\textwidth]{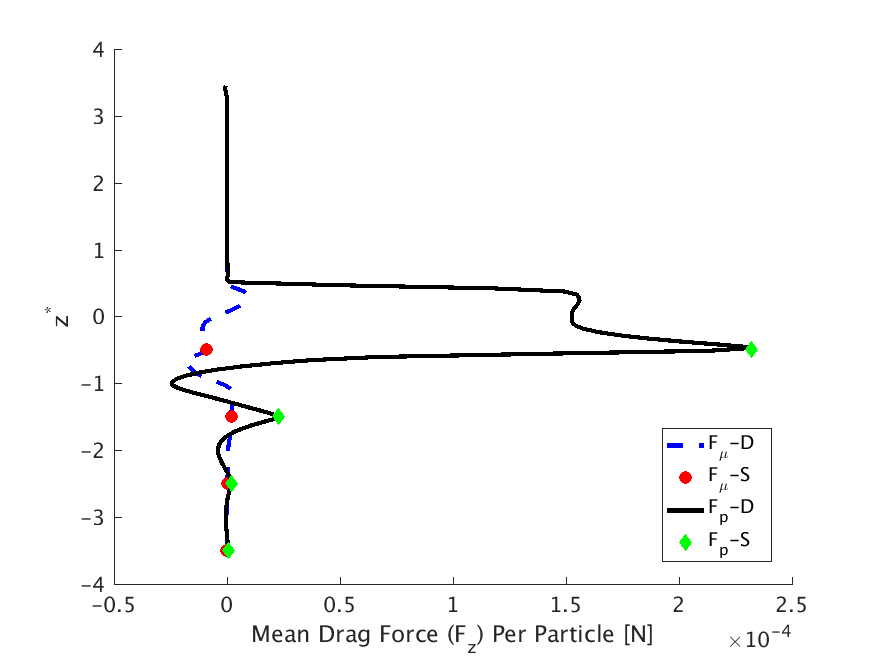}
         \label{fig:smagDragZ}}
    \caption{Comparisons of surface integration and double-averaging for drag force calculations
             (z-component) as predicted by
      \protect\subref{fig:waleDragX} the WALE model and
      \protect\subref{fig:smagDragX} the Smagorinsky model. The 'D' denotes double-averaging 
      and the 'S' denotes surface integration.}
    \label{fig:dragTermsZ}
\end{figure}

Figures \ref{fig:dragTermsX} and \ref{fig:dragTermsZ} compare the calculation of the form and viscous 
drag terms for both turbulence models components in the $x$ and  $z$ directions, respectively.
Excellent agreement is seen for all calculations, even in regions where the drag exhibits a large
rate of change. This agreement supports the use of double-averaging to obtain continuous profiles of 
the drag force, given an averaging domain with the same length scale as the grain diameter. 

\subsection*{Instantaneous Flow Measurements}
\begin{figure}[!tbp]
    \centering  
    \subfloat[]
        {\includegraphics[width=0.5\textwidth]{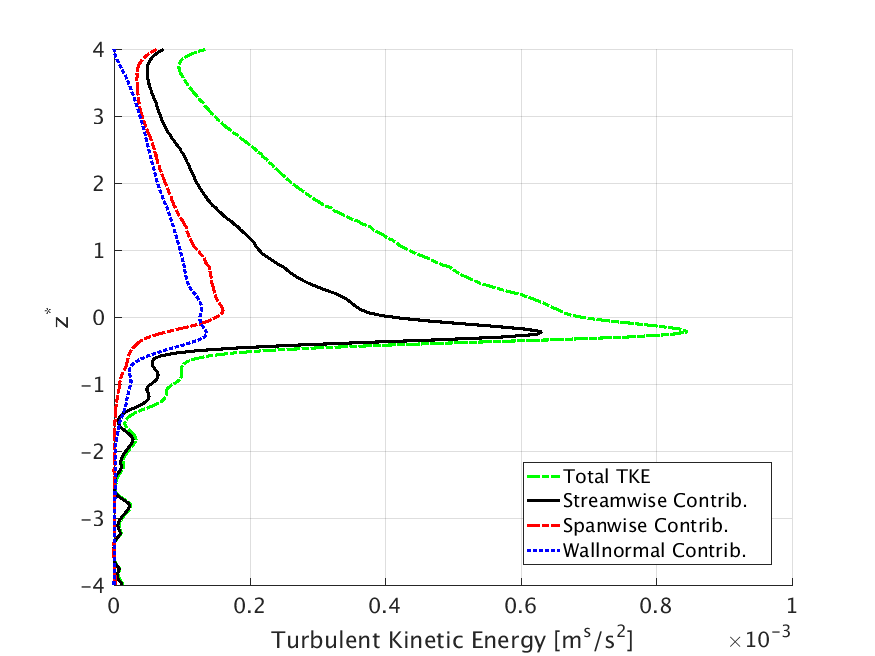}
         \label{fig:tkeWale}}
    \subfloat[]
        {\includegraphics[width=0.5\textwidth]{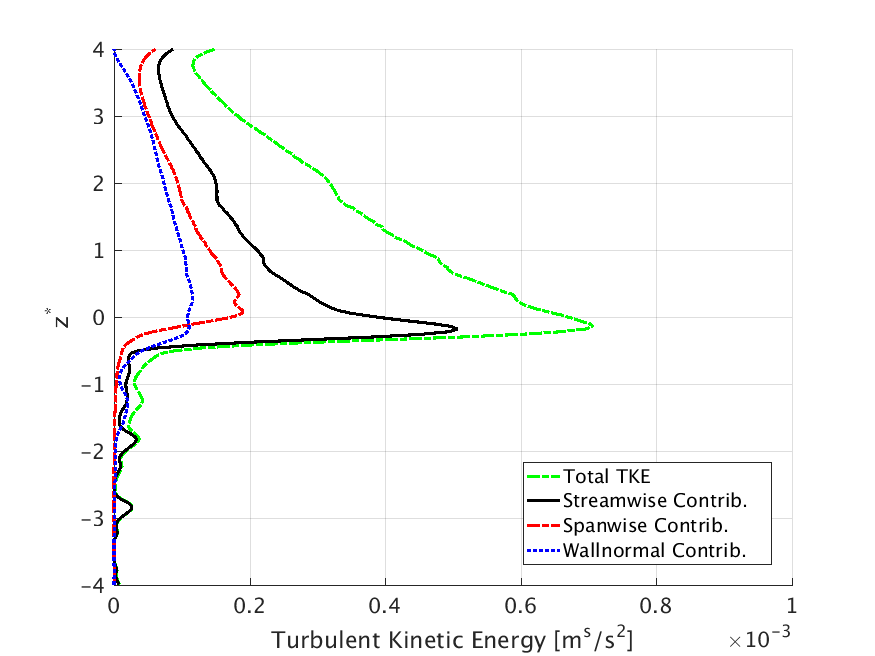}
         \label{fig:tkeSmag}}
    \caption{Turbulent kinetic energies, 
             $\frac{1}{2}(\overline{u^{\prime{2}}_{1}}+\overline{u^{\prime{2}}_{2}}+\overline{u^{\prime{2}}_{3}})$
, 
             along a vertical line through the pore throats as predicted by
      \protect\subref{fig:daSmag} the WALE model and
      \protect\subref{fig:daWale} the Smagorinsky model.}
    \label{fig:tkeCompare}
\end{figure}

In addition to the use of the double-averaging formulation, analysis of the instantaneous flow 
data can help to elucidate the role of turbulence in interfacial transport. Figure \ref{fig:tkeCompare}
displays the turbulent kinetic energy (TKE) 
$\frac{1}{2}(\overline{u^{\prime{2}}_{1}}+\overline{u^{\prime{2}}_{2}}+\overline{u^{\prime{2}}_{3}})$
as well as the TKE contribution from each dimension predicted by simulations with both turbulence models along
a wall-normal line with position ($nL$,$mL$,$z^{*}$), where $n$ and $m$ are integers.
Each model predicts a maximum in the TKE very near $z^{*} = -0.25$, just below the permeable wall, with streamwise 
fluctuations provided the majority contribution. High TKE in this region supports experimental findings 
\cite{Pokrajac2009} that strong shearing between the high momentum surface flow and low momentum flow
around the roughness elements is a primary driver of the interfacial turbulence. Noting that the Smagorinsky
model predicts a $16.5\%$ smaller peak TKE relative to the WALE model suggests that the Smagorinsky 
model is artificially restricting turbulence generation and 
thereby reducing the role of coherent vortical structures in interfacial momentum transport.

Common between the two turbulence models, a rapid decay of TKE is observed from the peak value with
depth inside the bed. Both spanwise and wall-normal fluctuations are negligible below 0.09 m, 
and smaller peaks in TKE are observed near the top of each layer of grains within the bed, suggesting
that the pore spacing between layers of grains may act as a nucleation site for turbulent eddies. 

Quadrant analysis provides another insightful tool for examining the role of turbulence in momentum
transport. By visualizing the instantaneous flow data in this way, one may determine what kind of 
temporary motions are common at different depths within the domain. Four types of events are described
in quadrant analysis: outward interactions (Q1), ejection events (Q2), inward interactions (Q3) and sweeps 
(Q4). The meaning of each event becomes clear when observing the sign of the fluctuations in each quadrant. 

\begin{figure}[!tbp]
    \centering  
    \subfloat[$z^{*} = 0.5$: WALE]
	{\includegraphics[width=0.25\textwidth]{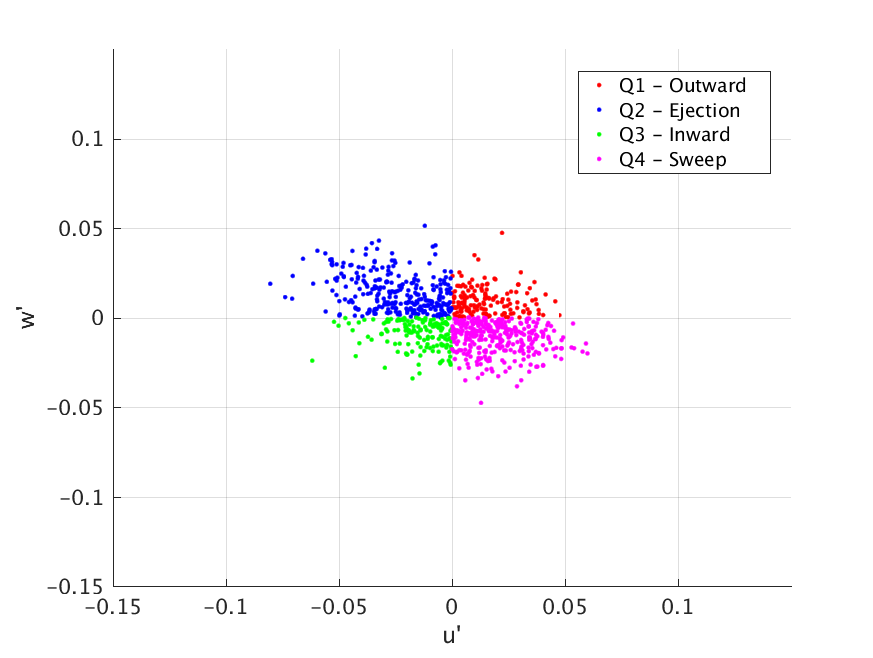}
	 \label{fig:qa24w}}
    \subfloat[$z^{*} = 0.5$: Smag.]
	{\includegraphics[width=0.25\textwidth]{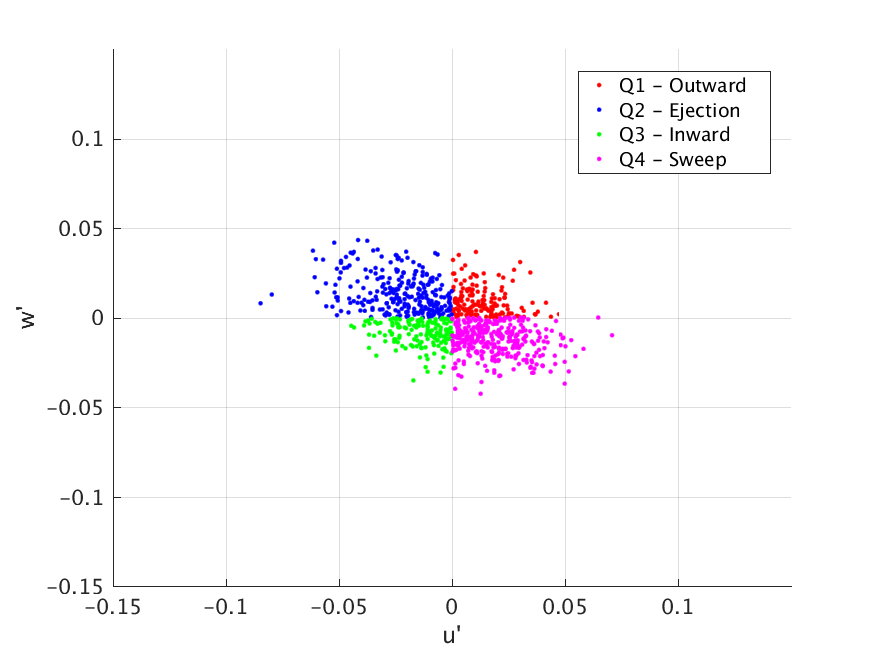}
	 \label{fig:qa24s}}
    \hfill
    \subfloat[$z^{*} = 0.0$: WALE]
	{\includegraphics[width=0.25\textwidth]{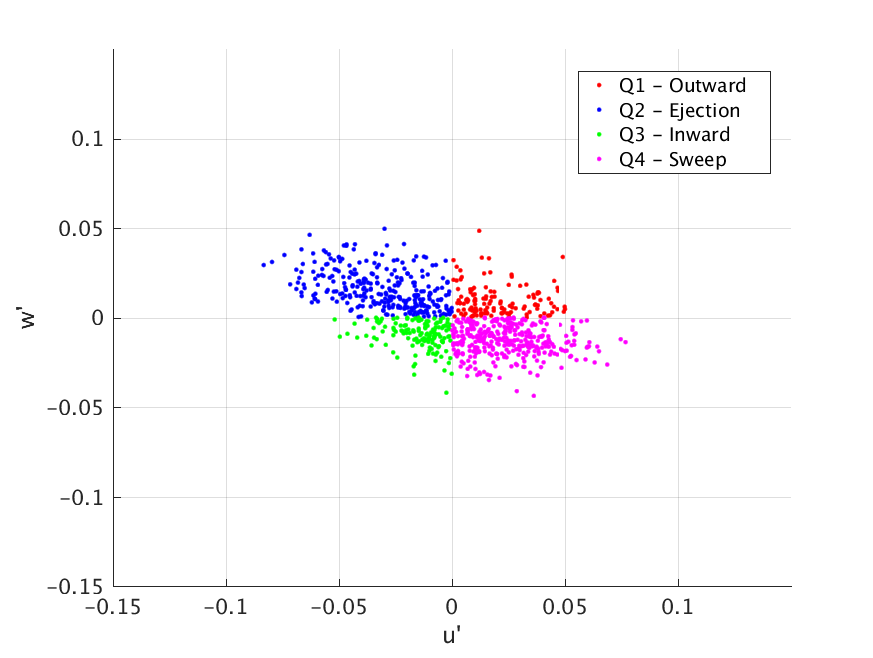}
	 \label{fig:qa16w}}
    \subfloat[$z^{*} = 0.0$: Smag.]
	{\includegraphics[width=0.25\textwidth]{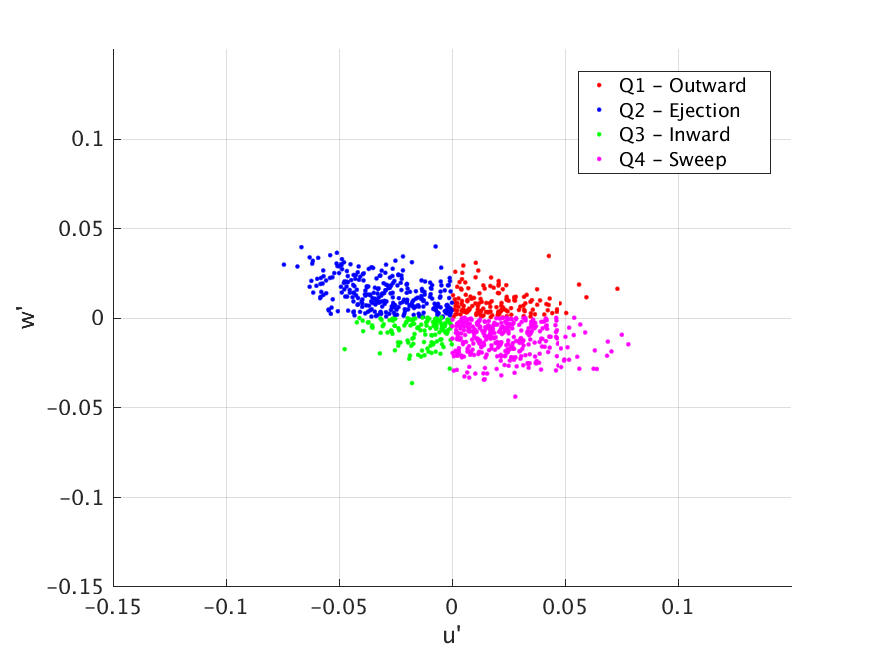}
	 \label{fig:qa16s}}
    \hfill
    \subfloat[$z^{*} = -0.25$: WALE]
	{\includegraphics[width=0.25\textwidth]{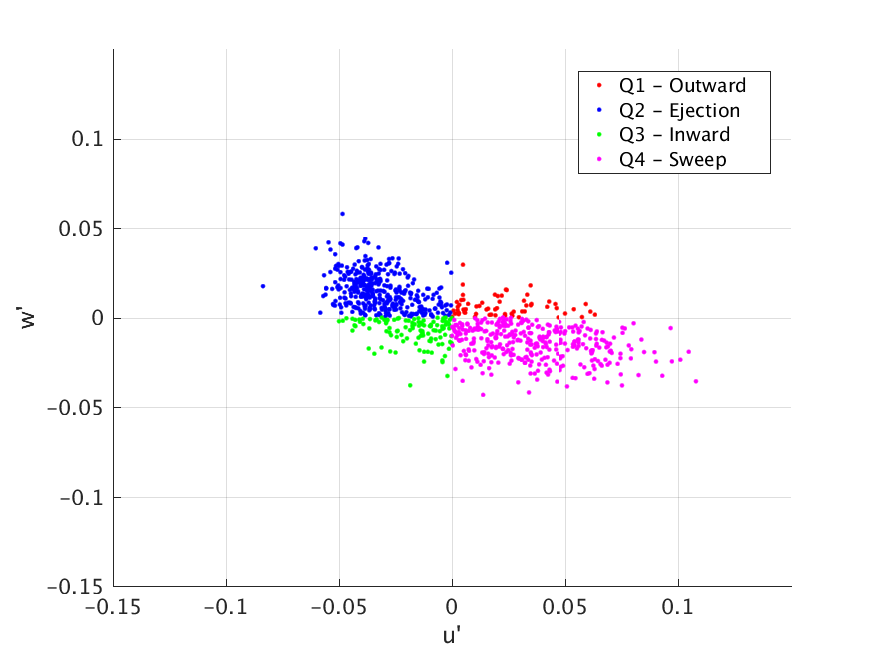}
	 \label{fig:qa14w}}
    \subfloat[$z^{*} = -0.25$: Smag.]
	{\includegraphics[width=0.25\textwidth]{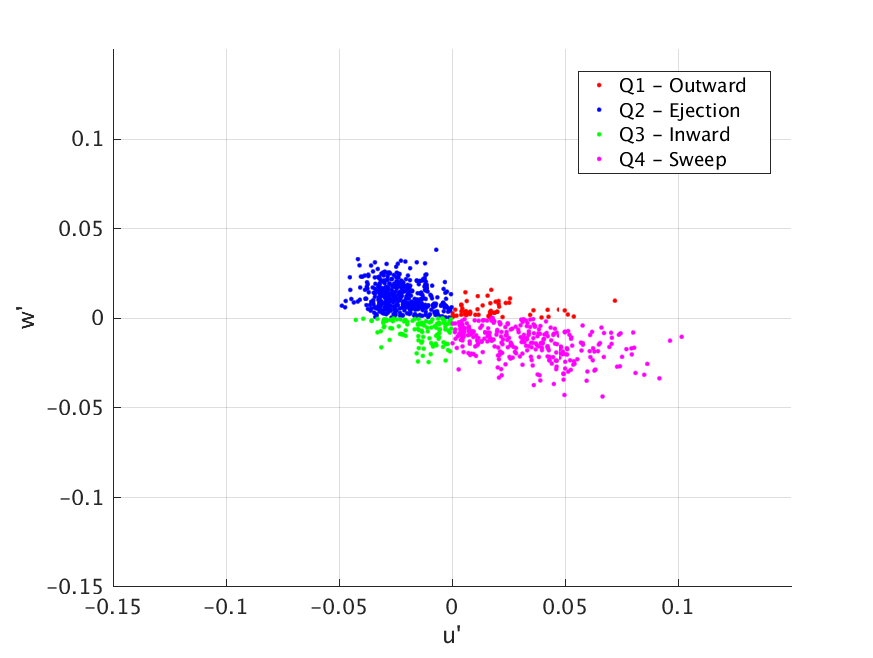}
	 \label{fig:qa14s}}
    \hfill
    \subfloat[$z^{*} = -0.5$: WALE]
	{\includegraphics[width=0.25\textwidth]{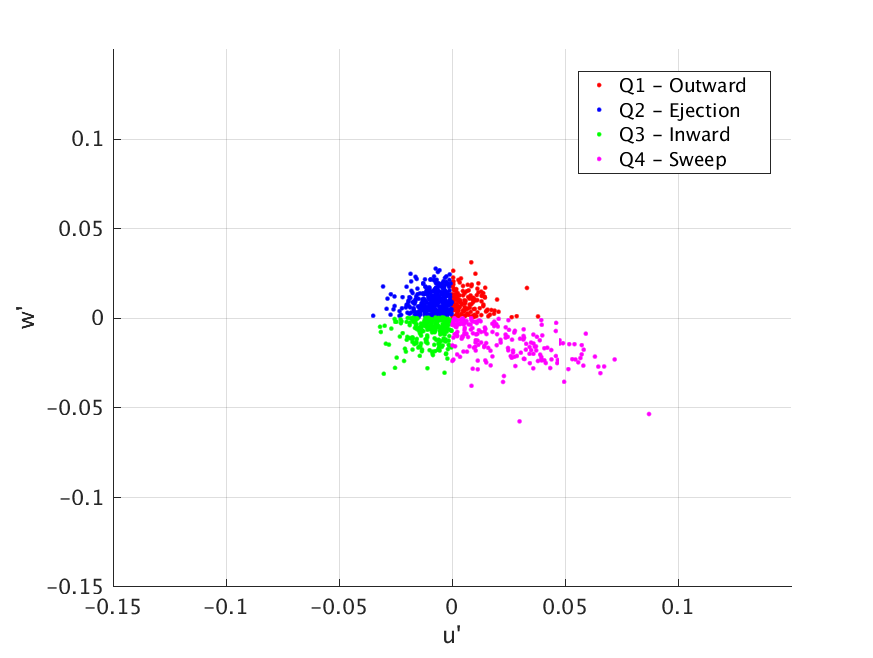}
	 \label{fig:qa12w}}
    \subfloat[$z^{*} = -0.5$: Smag.]
	{\includegraphics[width=0.25\textwidth]{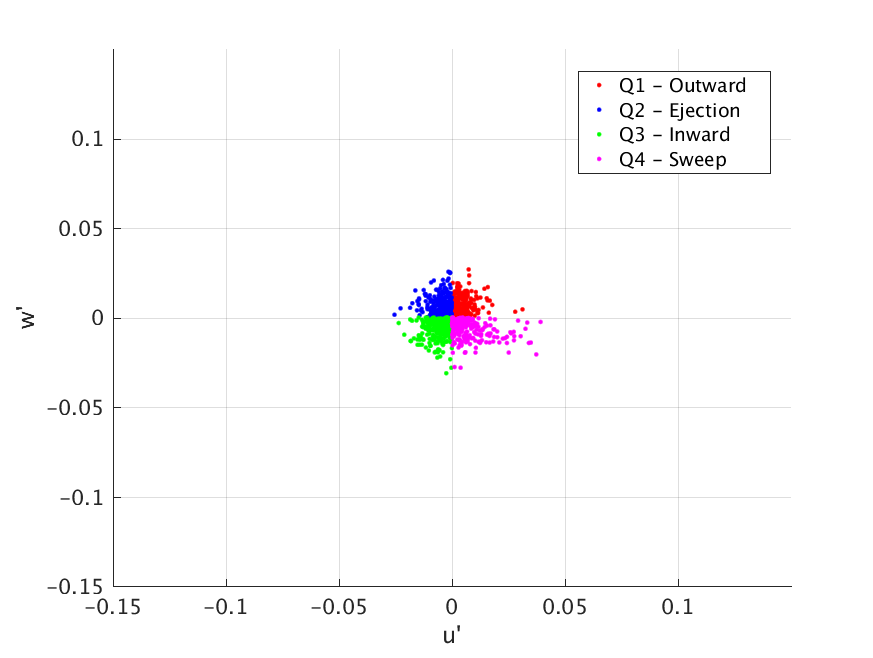}
	 \label{fig:qa12s}}
    \hfill
    \subfloat[$z^{*} = -1.0$: WALE]
	{\includegraphics[width=0.25\textwidth]{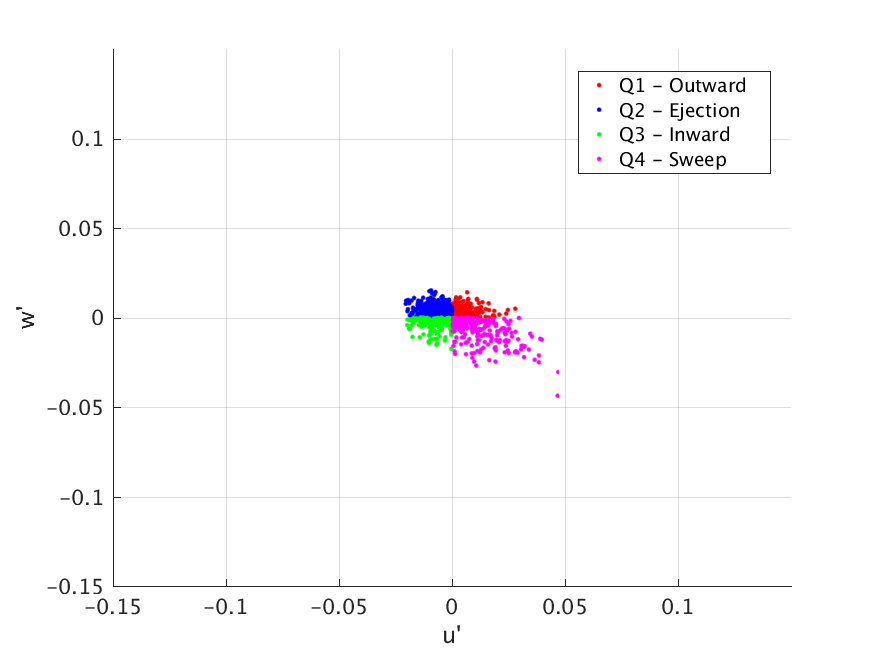}
	 \label{fig:qa10w}}
    \subfloat[$z^{*} = -1.0$: Smag.]
	{\includegraphics[width=0.25\textwidth]{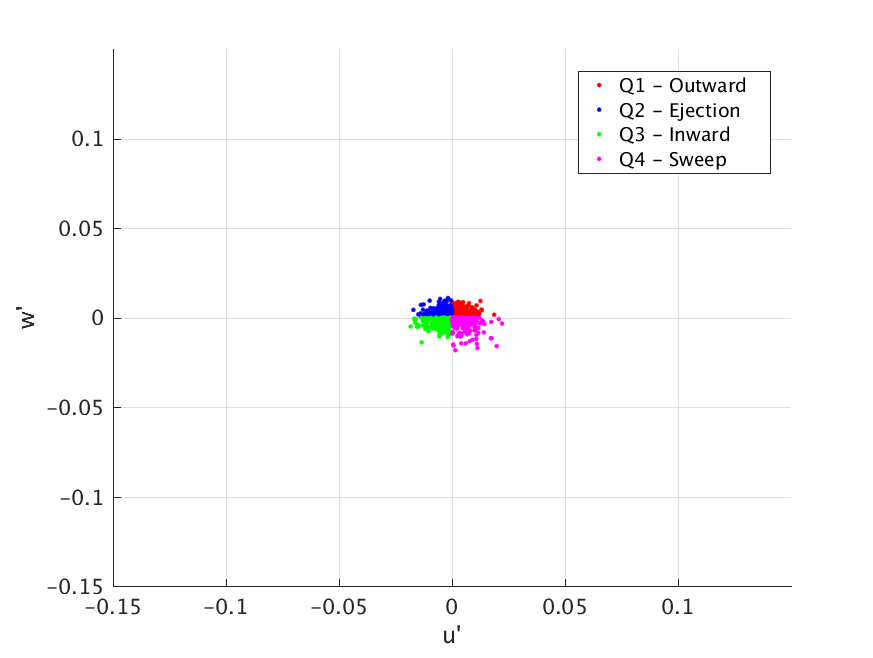}
	 \label{fig:qa10s}}
    \hfill
    \caption{Quadrant analysis at various heights along a vertical line passing through the pore throats for both
             the WALE and Smagorinsky models. All velocities have units $m/s$.}
    \label{fig:quadrantAnalysis}
\end{figure}

Figure \ref{fig:quadrantAnalysis} presents the history of turbulent events at different
depths along the center of a pore throat for the WALE (left) and Smagorinsky (right) models, respectively.
Every image contains 900 events, with each event separated by $20\Delta{t}$ to avoid recording the same event
several times. Approaching the permeable wall from the homogeneous fluid region, both closure schemes
show a growth in the presence of ejection and sweeps. This behavior is expected, as vortical motions
in this region will encourage exchange between the low momentum fluid of the bed and high momentum
fluid of the channel. Immediately below the permeable wall, however, a decrease in 
strength of Q1 and Q2 events occurs, accompanied by a significant increase in the strength of Q4 events. Such
behavior corroborates previous findings (e.g. \cite{Kuwata2016}) and highlights the importance of strong 
vortical surface motions on interfacial transport.

Moving deeper into the bed, the effects of the turbulence model are noticeable. Over the depth of a single
unit cell, the strength of Q4 events becomes insignificant according to the Smagorinsky model, while the 
WALE closure scheme exhibits relatively strong sweep events even at the bottom of the first layer of grains. 
This discrepancy agrees with the difference in TKE decay seen in Figure \ref{fig:tkeCompare}, and confirms 
that vortical structures predicted by the Smagorinsky model generally do not penetrate as deeply into the
bed as those predicted by the WALE model.

\begin{figure}[!tbp]
    \centering  
    \subfloat[]
        {\includegraphics[width=0.35\textwidth]{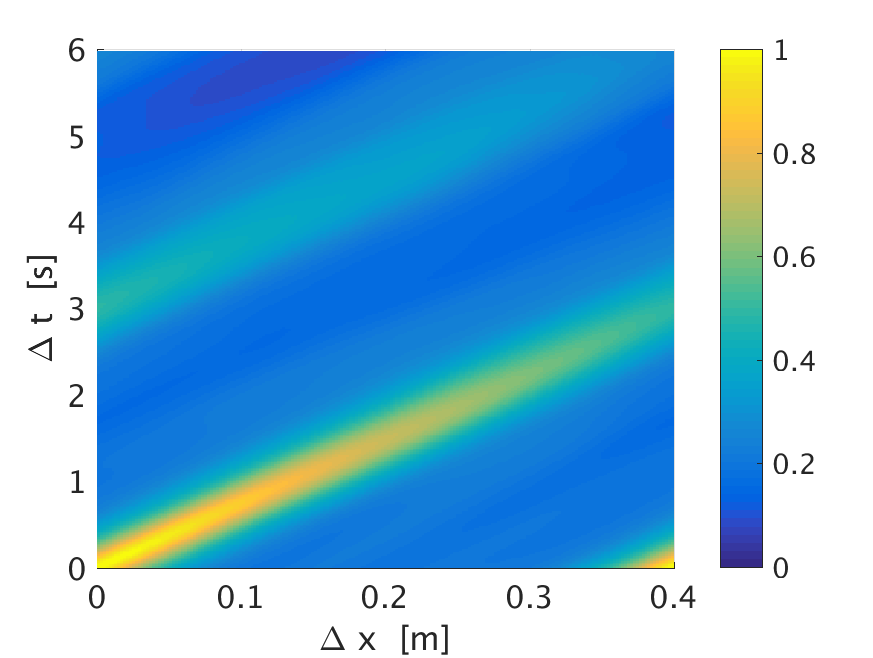}
         \label{fig:corrModelCompare10w}}
    \subfloat[]
        {\includegraphics[width=0.35\textwidth]{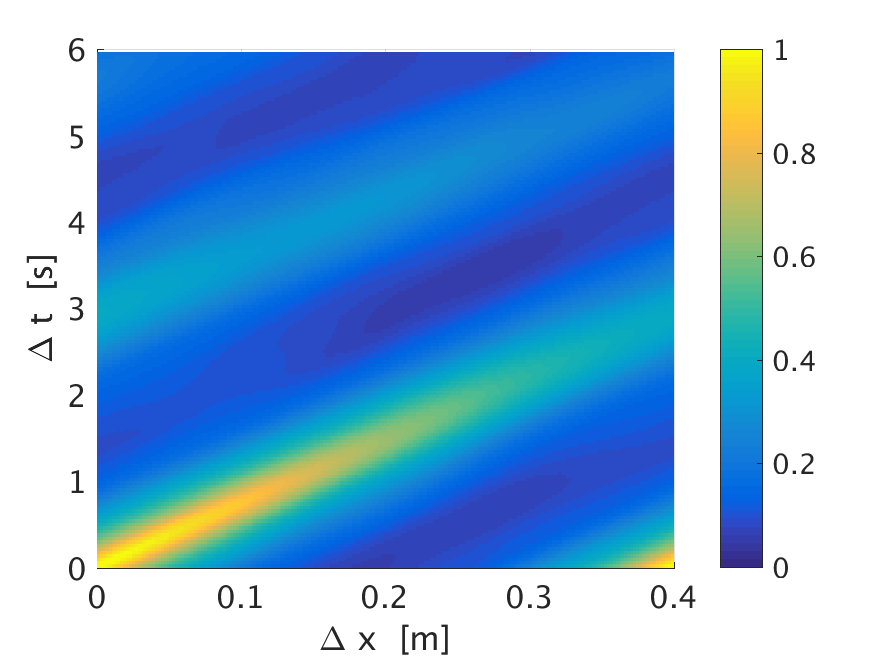}
         \label{fig:corrModelCompare10s}}
    \subfloat[]
        {\includegraphics[width=0.35\textwidth]{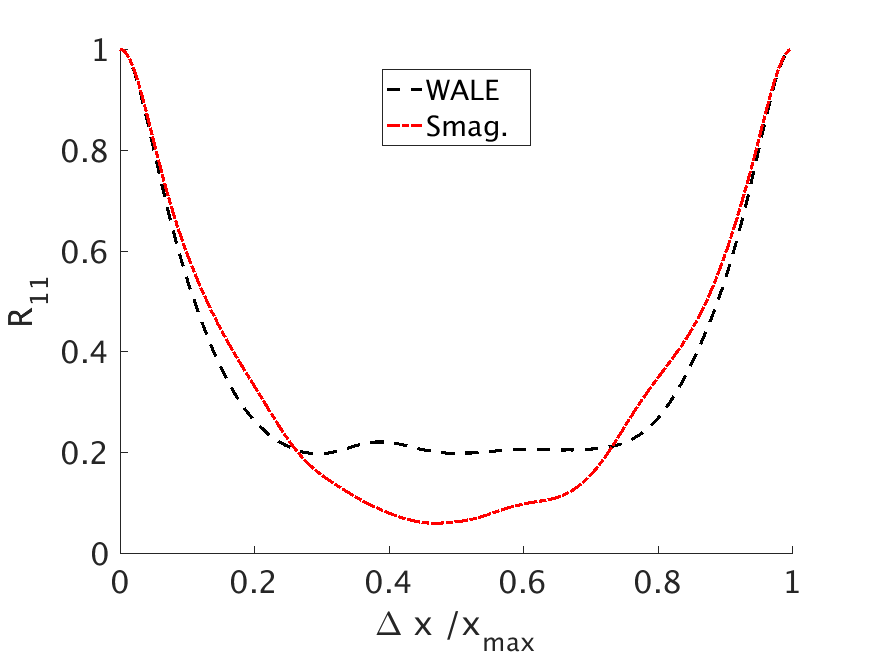}
         \label{fig:corrModelCompare10ws}}                                                                     
    \caption{Autocorrelation plots of $R_{11}(\Delta{x},\Delta{t})$ for 
      \protect\subref{fig:corrModelCompare10w} the WALE model and 
      \protect\subref{fig:corrModelCompare10s} the Smagorinsky model.
      \protect\subref{fig:corrModelCompare10ws} A comparison $R_{11}(\Delta{x},0)$ for both models.}
    \label{fig:corrModelCompare}
\end{figure}

As a final measure of comparison, the same correlations presented
in Figure \ref{fig:corrDomainCompare} have been computed for the WALE and Smagorinsky models, 
shown in Figure \ref{fig:corrModelCompare}. Again, the slope associated with the line of strong correlation
roughly equals the inverse of the velocity at the same depth ($z^{*} = 2$), shown in Figure 
\ref{fig:velocityModCompare}, highlighting that turbulent motions are generally moving with the mean flow. 
Additionally, the line of strong correlation attenuates across space and time slower for the
the WALE closure, indicating that its predicted structures moving with the mean 
flow are more successful at maintaining coherent motion when compared to those predicted by the Smagorinsky model.

A comparison of streamwise instantaneous spatial correlations demonstrates a stark difference between 
the two closure schemes. Although neither model achieves complete decorrelation, the streamwise fluctuations 
in the WALE and Smagorinsky models produce a correlation coefficient of $R_{11} = 0.197$ and $R_{11} = 0.060$, 
respectively, at a distance $L_{x}/2$ from the measurement origin. This difference in $R_{11}$ suggests
that the large vortical motions predicted by the Smagorinsky model are either weaker or less constrained
by the simulation box than those predicted by the WALE model, which corroborates the findings in 
Section \ref{sec:domainSize}. 

Given the identical domain dimensions used between these two simulations, 
however, the weaker correlation may be explained by looking at Figure \ref{fig:diffMomCompare}.
Within the transition region, the Smagorinsky model predicts an increased and decreased loss of momentum 
due to viscous stresses and form drag, respectively, when compared to the WALE model. The
Smagorinsky model is known to overcompensate for shear in the near-wall limit by producing an unphysical amount of 
turbulent viscosity \cite{Temmerman2003}, and consequently overdamping the fluctuating velocity field. One possible
explanation for this discrepancy is that increased interfacial viscous stress is hindering the development of larger structures
by entrainment or coalescence of smaller turbulent scales. Consequently, such coherent motions 
lack the strength to push recirculation regions and low-momentum fluid deep into the bed. The difference
in strength of Q3 interactions at $z^{*} = -0.5$ between the models supports this argument. It has been suggested that 
the bed geometry acts to transform Q4 events into Q3 events \cite{Pokrajac2009}. Given the increased strength of
sweep events within the first pore layer predicted by the WALE model, a more thorough study of quadrant events
within the pore spaces may help to elucidate the role of geometry in this proposed transition and clarify the
link between the penetration depth of turbulence and the profiles seen in Figure \ref{fig:velocityModCompare}.

%% file: texFiles/mainMatter/chapter2/conclusions.tex
\section{Concluding Remarks}\label{sec:lesStudyConclusion}

Within these last two chapters, the formulation of a detailed LES for studying turbulent
flow over a permeable bed has been examined. The double-averaging methodology has been implemented
to aid in the study of large-scale, persistent flow structure in the presence of temporal and 
spatial heterogeneity. Additionally, the influence of mesh size, domain size and turbulence model
on the predicted hydrodynamics has been discussed.

Notably, results of the mesh refinement study show that achieving a fine resolution in the
transition region is critical to avoid generating numerical instabilities. Moreover,
the double-averaging methodology reduces the number of temporal measurements needed to capture
trends in the first and second moments of the velocity. Simulations with various domain sizes show
that the size of the largest resolved turbulent structures is sensitive to the size of the simulation
box when using periodic boundary conditions, and that the artificial restriction of such structures
may reduce turbulence-induced momentum exchange within the transition region. Finally, two LES runs were
performed on the same domain using the WALE and Smagorinsky turbulence closures. Similar to the effect
of a smaller domain size, the increased interfacial turbulent viscosity produced by the Smagorinsky
model acts to restrict the size, and possibly strength, of large vortical motions, reducing
the influence of turbulence on the subsurface flow when compared to the WALE model.

It is clear from these studies that the development of a computational model well suited for performing
LES on a system with such a broad range of length and time scales is no trivial task. Many aspects of the 
model must be considered, and variations in any of them may profoundly affect the predicted 
hydrodynamic behavior. In addition to rigorous validation against experimental studies, more LES work is 
needed within the field to better understand the impact of the discussed features on the performance of 
the LES, as well as many other parameters left undiscussed (e.g. body force, turbulence model constant).
However, it is the author's hope that the explorations presented in this work inform further use of 
LES in this research area and lead to a more complete picture of turbulent momentum exchange in flows
over permeable beds.

%% file: texFiles/mainMatter/chapter3/introduction.tex
\section{Introduction}\label{sec:particleIntro}

As discussed in Section \ref{sec:thesisIntro}, a primary interest of those studying hyporheic
exchange, and more broadly, flows of permeable bed forms, is to understand how scalar quantities
are exchanged between the surface and subsurface flows. This process is often modeled by
lower-order stochastic particle tracking models or advection-dispersion equations 
\cite{Berkowitz2006,Schumer2009,Chakraborty2009}, which 
could likely be made more accurate with the adoption of parameterizations derived from pore scale information \cite{Sund2015}.
Given the wealth of hydrodynamic detail available at this length scale when using Large
Eddy Simulation (LES), provided the necessary information can be extracted,
the detailed flow model proposed and evaluated in the previous chapters is a viable candidate for producing such parameters.

The primary challenge in implementing a Lagrangian particle tracking model within an 
Eulerian system is the continual need to locate the particles' host elements, which are necessary for 
accurate interpolation of the fluid fields to the particles' locations. Several methods have been proposed
for handling these problems of search and interpolation. Notably, algorithms have been developed to 
efficiently handle particle tracking on unstructured and mixed element meshes \cite{Martin2009,Cheng1996,Lohner1990}. 
Efforts have also been focused on reducing interpolation errors 
for particles crossing element boundaries \cite{Pokrajac2002}.

However, these methods generally don't address the influence of domain decomposition and distributed 
memory on the efficiency/applicability of the proposed algorithms. As the implementation of this 
particle tracking module will be within an existing open source software, designed for efficient computation
of fluid fields across unstructured meshes on massively parallel machines, host-element determination 
for each particle becomes a critical matter of computational efficiency. Additionally, the treatment of 
particles as classes within an object-oriented programming framework adds complexity to the inter-process
communication of particle objects within a distributed memory environment.

The work in this chapter presents solutions to the posed challenges, providing the foundation for a particle tracking 
module. This module can then be incorporated within a detailed LES to extract Lagrangian dispersion statistics from the 
flow field. First, an introduction to the open source Nalu and STK code bases are presented. Following this, an overview
of the particle tracking module and its associated search methods are given. Finally, particle evolution, parallel
communication and boundary conditions are explained in detail. It is the author's intent that this chapter
will serve as documentation for the current particle tracking implementation and present a thorough introduction
for future developers of the module. 

\subsection*{Brief STK Vocabulary Discussion}\label{sec:stk}
Part of what makes Nalu such a powerful tool is that it relies on several highly optimized
software packages and third-party libraries, allowing the code base to continually improve in
simulation fidelity, memory management, portability, etc. A particularly important package
for development of Nalu's particle module is the Sierra ToolKit, or STK \cite{Report2016}.
The STK was created by Sandia National Laboratories to aid in the construction of high-performance 
software for engineering analysis, and offers modules to support the use of unstructured meshes and
geometric searching, among other utilities, in a distributed memory environment.

While most of the particle module development may be explained in the absence of a working knowledge 
of Nalu's usage of STK, defining a few terms up front may provide significant clarity in the following 
sections:
\begin{itemize}
  \item\textbf{Mesh} - The mesh is defined as a set of entities, parts, fields and field data, which is divided
        between MetaData and BulkData.

  \item\textbf{Part} - A part defines a subset of the mesh and its associated entities. Various kinds of parts exist
        within a mesh; examples include the locally-owned part, the aura part and the sideset part (if the mesh is read 
        from an Exodus file).

  \item\textbf{Field} - A field represents a set of data associated with mesh entities. Fields may be created across
        the entire mesh or on subsets (i.e. parts) of the mesh. Examples include a nodal field which stores the volume
        of a control volume or an area vector field stored on element edges.

  \item\textbf{Entity} - An entity is a general term for an object in the discretization including the following 
        types: node, edge, face, element and constraint.

  \item\textbf{MetaData} - MetaData is the component of the mesh which holds definitions of parts, fields and relationships
        among and between the two. This data is duplicated on every process and must be created before the BulkData is 
        constructed.

  \item\textbf{BulkData} - BulkData is the component of the mesh which holds definitions of entities and their ownership 
        (including ghosting information), connectivity data and data associated with the fields defined in the MetaData.
        Unlike the MetaData, BulkData is distributed amongst processes rather than duplicated, and must be constructed
        after the MetaData has been finalized.

  \item\textbf{Parallel Consistency} - STK Mesh is used for several engineering disciplines (e.g. thermal/fluid mechanics and
        structural dynamics). Because the application using the mesh is not known a priori by the developers, the mesh must be 
        consistent, i.e. it must always observe specific rules, regardless of the governing application. Enforcing consistency
        becomes significantly harder in parallel, and doing this requires a strict set of rules regarding the distribution of
        mesh data. The primary rule the Nalu particle module is concerned with is that of parallel ownership, which ensures
        that each mesh entity is only owned by one process, and that every process with a copy that entity (whether from 
        sharing or ghosting) knows the entity's owning process. 

  \item\textbf{Aura} - The STK Mesh module offers an automatic single-element thick ghosting layer around each process, denoted
                       as the aura. Though more intricate ghosting may be created through custom ghosting, the aura is turned on
                       easily through the Nalu input file and is sufficient for the particle module. 

\end{itemize}

Note that these definitions and more formal explanations for the above terms are available in \cite{stkManual}, 
which provides documentation including conceptual overviews and code examples covering the STK modules. 

%% file: texFiles/mainMatter/chapter3/naluStructure.tex
\section{An Informal Tour of Nalu}\label{sec:naluStructure}
Before the particle module may be explained in detail, an elementary understanding of
the Nalu code base must be developed. In this section, important structures within
Nalu are discussed and the flow of data over the course of a simulation is presented
at a high-level. Developers and those interested in a more complete picture are referred
to \cite{Domino2015}, which provides a more formal treatment of the topic. 

Nalu is capable of modeling various flow phenomena, either in isolated or multiphysics
simulation. Thus, the code base must support several equation systems
(e.g. momentum transport, thermal heat conduction, filtered mixture fraction), which
may need to be handled on different meshes with disparate material properties. To manage
the organizational challenges associated with this task, Nalu implements \textit{realms} 
and \textit{transfers} to separate each physics' data and allow for coupling. 

Nalu creates a realm for each type of physics to be modeled. Realms then develop their 
own understanding of the discretized domain and associated fields, equation systems and 
other information relevant to their simulation environment. This level of encapsulation
not only provides ease of use for the user/developer, but allows for distinct boundary
conditions, initial conditions and output frequency of data to be specified.

\begin{figure}
  \centering
    \includegraphics[width=0.15\textwidth]{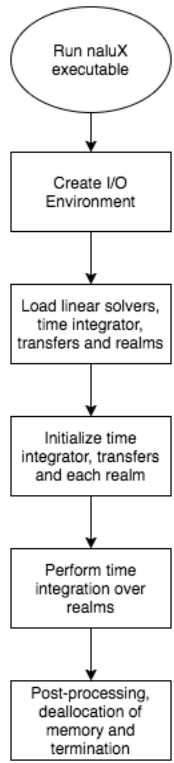}
    \caption{Abbreviated overview of a Nalu simulation.}
    \label{fig:naluSimulation} 
\end{figure}

In a multiphysics simulation, the realms will have some degree of coupling. Nalu takes
advantage of STK's transfer module to support sharing of data between realms. When 
coupling is required, the fields to be communicated and the sending/receiving realm pair
are specified by the user. Transfers will then take place before, during and after time 
integration. This ensures consistency of state in a realm's field data throughout system 
evolution. A simplified view of a Nalu simulation is given in Figure \ref{fig:naluSimulation}.

%% file: texFiles/mainMatter/chapter3/overview.tex
\section{Overview of the Particle Classes}
Within Nalu, the particles module exists as a collection of classes handling
particle physics, evolution and I/O. A note on the role of each class is given in 
Table \ref{tab:particleClasses}. This section is meant to be a brief overview
of the module's structure, addressing where various functions are carried out and how 
data is handled over the course of a simulation at a high level.

As mentioned in Section \ref{sec:naluStructure}, Nalu separates physical models by realm. 
The particle module implements a specific particle-based realm, ParticleRealm, derived from the standard 
Nalu realm, to handle the numerical environment for particles. While the particle realm follows a load 
procedure similar to the other realms, it is unique in that it doesn't have to manage assembly of an equation 
system. This allows the particle realm to have its own set of rules for initialization, directed 
towards constructing the mesh objects and loading particle data into the domain.

Managers are those objects which are only created once within the particle realm 
and are responsible for overseeing the most complicated tasks during the simulation.
Of the managers in the particle module, ParticleManager has primary oversight. 
Owning the other particle managers (e.g. ParticleCommManager and ParticlePeriodicManager) and the 
methods responsible for both particle integration and host element determination, the particle 
manager is the first place to explore when studying the module in-depth.

Inter-process particle communication is governed by ParticleCommManager. At each timestep,
this manager handles the packing, communication and unpacking of particles using tools from
the STK Utils module. In the presence of periodic boundary conditions, the ParticleCommManager
relies on the ParticlePeriodicManager to create a map for boundary-to-boundary communication.
This mapping feature is still under development, so the
user is recommended to approach the application of periodic boundary conditions with caution.

Finally, VtuManager handles the output of particle data. Currently, this manager
is only able to output coordinate and timestep data in a format suitable for visualization,
although extension to output additional data (e.g. particle temperature, unique identifier) is under
development. Write operations and the structure of the output are further explained in Section 
\ref{sec:initAndIo}. 
%
\begin{table}[h!]
  \centering
  \caption{Class overview of Nalu particle module. (TBI - To be implemented)}
  \label{tab:particleClasses}
  \begin{tabular}{ll}  
    \toprule
    Class                    & Purpose\\

    \midrule
    Particle                 & Defines a particle object.\\
    ParticleRealm            & Governs simulation environment and multiphysics interactions.\\
    ParticleManager          & Governs particle evolution and host-cell determination.\\
    ParticleCommManager      & Handles particle multi-process particle communication.\\
    ParticlePeriodicManager  & Handles application of periodic boundary conditions.\\
    ParticleEquation         & Defines particle equations (TBI).\\
    ParticlePhysics          & Defines coupling and physics governing particle (TBI).\\
    ParticleVariables        & Handles variables defining particle state.\\
    ParticleType             & Defines material properties and physics (TBI).\\
    ParticleIntegrator       & Governs particle integration (TBI).\\
    TracerParticlePhysics    & Defines physics for tracer particles (TBI).\\
    ParticleInsertion        & Stores and loads particle data from input file.\\
    VtuManager               & Outputs particle data.\\

    \bottomrule
  \end{tabular}
\end{table}

%% file: texFiles/mainMatter/chapter3/particleSearch.tex
\section{Host Element Determination}
At all times during a simulation, a particle must be aware of its host element to ensure
accurate tracing of its immediate environment. Thus, host-cell determination, as discussed
in Section \ref{sec:particleIntro}, is a critical component of initialization and time 
integration. The following subsections briefly discuss the methods used for particle search within
Nalu, as well as discussing another application which may find such methods useful.

\subsection*{Face-crossing Search}\label{sec:faceSearch}
While several methods exist to query an entity's location within a domain, 
the challenge of search is greatly simplified when the initial host element is known. 
Assuming a small time step such that a particle does not traverse several elements 
in a single update but has left its previous host-element, the particle may be tracked 
via a face-crossing search. This type of search method is the primary tool used within 
the particles module for host-element determination, due to its compact search domain.

In the face-crossing search, a particle's coordinates are known at both the previous and 
updated time levels. With this information, 
the face of the particle's initial host through which the particle exited may be determined. 
If this face only corresponds to one element, its clear that the particle has left the domain, and
will be sent to the coarse search tool for deletion. If the exit face is shared by two elements,
however, then the particle is moved to the element opposite the original host. If this element is 
the proper host, the particle will continue to integrate. In the instance where the particle has
traveled through multiple elements, then the particle will continue through this process, moving
from element to element without integrating, until it finds its proper host.

\subsection*{Nalu Coarse Search}
During particle initialization, the host element of the particle is not known. In this case, or in the 
instance the face-crossing search fails to find a suitable host, a more powerful search tool is necessary. 
STK provides a suite of search tools for Nalu which is already optimized for use on decomposed, 
unstructured meshes. From this suite, the particle module implements an extension of the Boost R-Tree 
spatial index to execute a range search, herein deemed a domain 
search.

An R-tree is a tree-based data structure designed to organize a set of n-dimensional geometric entities
into a set of n-dimensional minimum bounding rectangles (MBRs) \cite{Theodoridis2006}.
In general, the root node of the tree bounds the entire domain, where child nodes bound subsequently smaller
subsets of the domain and the entities of interest are containing within the MBRs corresponding to the leaves of
the tree. This structure allows for efficient ($\log{(n)}$, on average)
proximity-based searching within a computational domain. 

\begin{figure}
  \centering
    \includegraphics[width=0.85\textwidth]{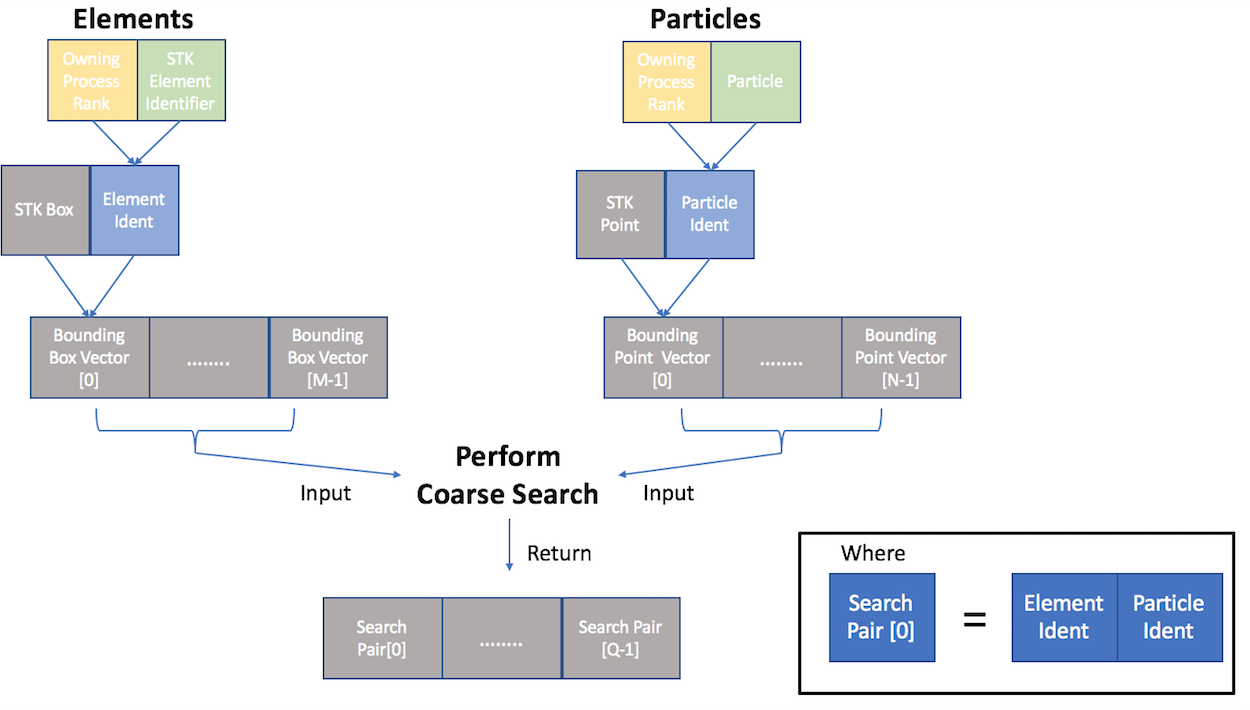}
    \caption{Schematic of the pair and vector data structures involved in Nalu's coarse search tool.}
    \label{fig:coarseSearch}
\end{figure}
 
Figure \ref{fig:coarseSearch} graphically details the use of domain search within Nalu for finding particles'
host elements. A unique identifier for the object (each particle or element) is first paired with the object's
owning process rank, creating an identifier, deemed an ident. Each ident is then paired with an STK geometric object, Boxes for 
elements and Points for particles in this case, which are the spatial constructs on which the domain search can 
operate. These pairs are then loaded into vectors, distinguished by the type of geometric object, and can be fed into the 
coarse search tool. Upon completion, a list of ident pairs is returned, providing a candidate host-element
(or multiple candidates) for each particle, along with each object's owning process. With a more rigorous check, 
the proper host-element may then be found from the candidates put forth by the search.

\subsubsection*{Application to Fluid-Structure Interaction}
Integration of Boost's RTree search method into STK and subsequently Nalu's utilities provides
a powerful tool for several applications, including host-cell determination for particles and 
entity mapping for periodic boundary conditions. One application of the RTree search worth 
mentioning briefly here, which has a scope that goes beyond this thesis, is in the development of 
fluid-structure algorithms within Nalu.

During a fluid-structure coupling algorithm, force will need to be transferred from an immersed solid onto
the fluid mesh. This loading may be distributed amongst several nodes neighboring the fluid-solid contact point. 
One way to locate the nodes 
in a desired neighborhood is to define a spherical or cubic volume about the contact point, and
use the RTree search to find any nodes within this bounding volume. Due to its flexibility in operating 
on either a local or global domain by simply specifying a different parallel communicator, the 
coarse search is an excellent tool for simplifying the implementation of the coupling algorithm on
unstructured meshes. 

%% file: texFiles/mainMatter/chapter3/evolutionAlg.tex
\section{The Particle Evolution Algorithm}\label{sec:evolutionAlg}
The primary utility gained from implementing a particle tracking framework in Nalu
is to capture a Lagrangian perspective of the flow field. Thus, an update algorithm 
is needed to advance the particles in time, given the conditions of their local 
environment. While the algorithm detailed in this section may be adapted to handle 
multiple-way coupling and particles experiencing varied physics, the current implementation
is designed for particles which are strictly one-way coupled to the fluid field, obtaining only the 
nodal velocity vectors from its environment.

\subsection*{Initialization and I/O}\label{sec:initAndIo}
In order to integrate particles through time, the particles must first be created
within the domain. Particle initialization currently occurs through the standard Nalu
input file. Data corresponding to each particle is read in and stored off into memory during the 
ParticleManager's load step, then used to create the particles during initialization. Each process begins with 
the same list of particles to create. After a global domain search and some accounting, each process creates
only those particles which reside within its piece of the distributed mesh. If a particle is to be created on
the boundary of multiple processes, it will be created only on the highest rank process.

Output of particle data is handled by the VtuManager class, which enables data to be visualized according
to the Visualization Toolkit (VTK) standard \cite{Schroeder2006}.
VTK is a widely supported, xml-based format for data visualization. The pvtu format, the only format currently supported
by the particles module, is meant for unstructured data 
created in a distributed memory computation environment, and thus provides an excellent solution for handling output 
from the particle module.

\begin{figure}
  \centering
    \includegraphics[width=0.85\textwidth]{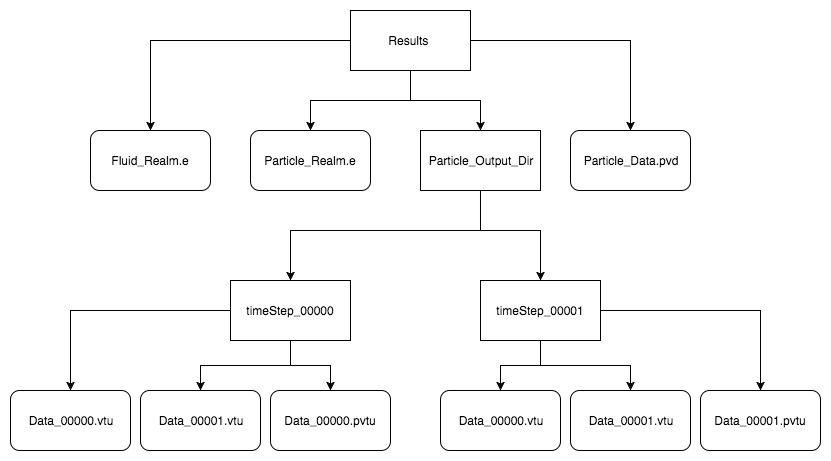}
    \caption{Example filesystem for a simulation with two timesteps on two processes. Rounded rectangles
             denote files and sharp-cornered rectangles denote directories.}
    \label{fig:outputFilesystem}
\end{figure}

The process for particle output is straight forward. At each timestep, including initialization,
a directory is created in a user-specified path. Within each directory, every process writes a .vtu file 
containing information regarding its locally-owned particles. Rank zero then creates a .pvtu file which links
all of the .vtu files together. Finally, the path of this .pvtu file and the corresponding timestep are written
to a .pvd (ParaView data) file, which may be read by ParaView alone to load in all particle data and corresponding
timesteps for the simulation. An example directory structure for a two-timestep two-process simulation 
are shown in Figure \ref{fig:outputFilesystem} for reference.

\subsection*{Local Time Integration}
Particle evolution can be thought of as a two-tier process. In the bottom tier, particle integration is a procedure
local to each process, ending when all particles on the process are either finished integrating and have found 
their respective host cells or waiting to find a new host cell via global communication. The top tier looks at 
particle integration as a global procedure, governing local advancement, particle communication and searching, 
and completion of the integration step. In this section, focus will be given towards explaining the lower tier,
i.e. the process of advancing particles locally. Following this discussion, explanations will
detail how local evolution fits in to the bigger picture when running a multi-process simulation.

\begin{figure}
  \centering
    \includegraphics[width=0.75\textwidth]{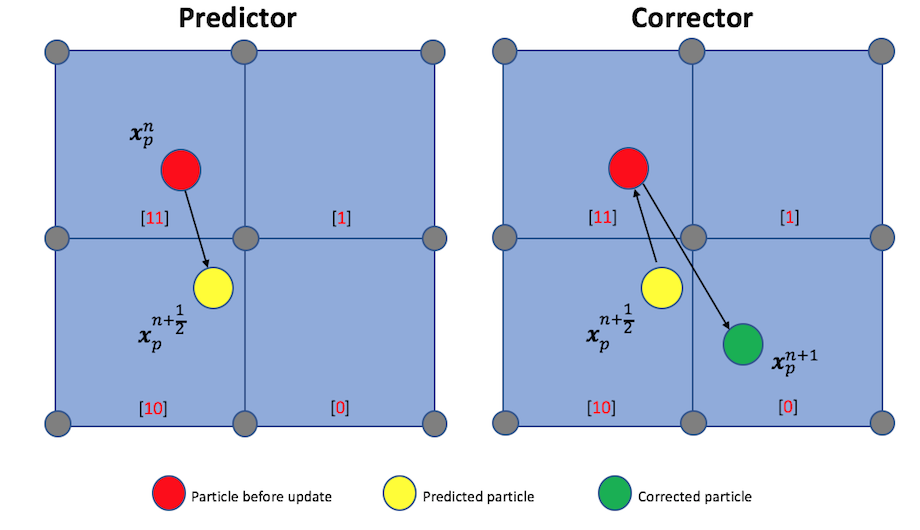}
    \caption{Depiction of the second order Runge-Kutta time integration scheme.}
    \label{fig:RK2}
\end{figure}

Time integration is carried out using a second order explicit Runge-Kutta scheme, defined as:
\begin{equation}
  \textbf{x}_{p}^{n + \frac{1}{2}} = 
    \textbf{x}_{p}^n + 
    \frac{\Delta{t}}{2}\textbf{u}_{f}^n(\textbf{x}_{p}^n)
  \label{eq:predictor}
\end{equation}
%
\begin{equation}
  \textbf{x}_{p}^{n + 1} = 
    \textbf{x}_{p}^{n + \frac{1}{2}} + 
    \Delta{t}[\textbf{u}_{f}^{n + \frac{1}{2}}(\textbf{x}_{p}^{n + \frac{1}{2}}) - 
    \frac{1}{2}\textbf{u}_{f}^n(\textbf{x}_{p}^n)]
  \label{eq:corrector}
\end{equation}
where $\Delta{t}$ and the superscript $n$ denotes the timestep and time level, respectively, subscripts $p$ and $f$ 
denote attributes of the particle and fluid, respectively, and $\textbf{x}$ and $\textbf{u}$ denote position and velocity
vectors, respectively. In the predictor step, a particle is first integrated over $\frac{1}{2}\Delta{t}$ using the fluid
velocity at its start position. By obtaining the fluid velocity at  $\textbf{x}_{p}^{n + \frac{1}{2}}$, the particle
can then return to its initial position and be integrated over a full time step using this intermediate velocity in the 
corrector step. A diagram detailing the integration of a single particle is provided in Figure \ref{fig:RK2}. For the
remainder of the thesis, the following nomenclature will be used in association with this integration scheme:
\begin{itemize}
 \item\textbf{Predicted} - A particle has been through the predictor step, but has yet to go through the corrector step. 
 \item\textbf{Corrected} - A particle has gone through both the predictor and corrector steps.
 \item\textbf{Active} - A particle is either predicted or yet to be predicted.
 \item\textbf{Done} - A particle is corrected.
 \item\textbf{Found} - A particle knows its current host element's identifier.
 \item\textbf{Lost} - A particle does not know its current host element's identifier.
\end{itemize}

\begin{figure}
  \centering
    \includegraphics[width=0.3\textwidth]{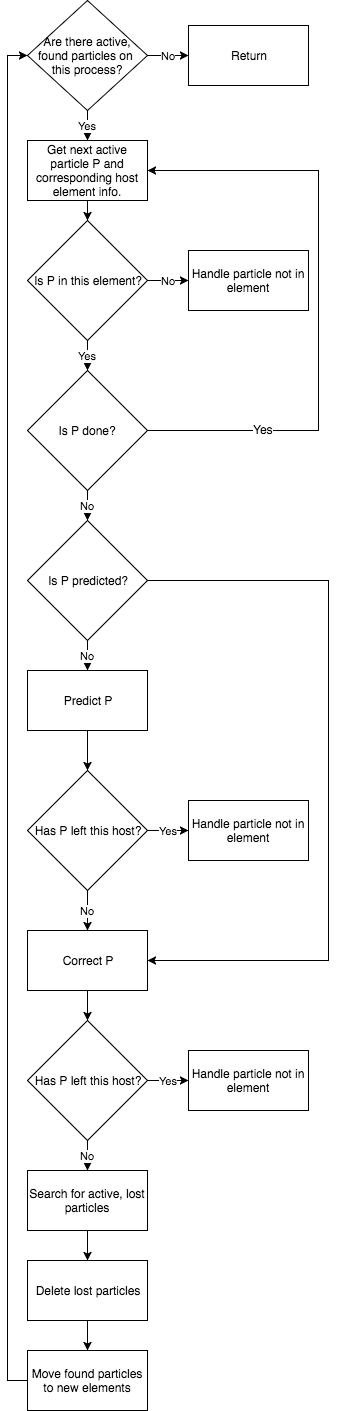}
    \caption{Simplified schematic of the local evolution algorithm.}
    \label{fig:localUpdate}
\end{figure}

An abbreviated diagram of the local particle time integration scheme is shown in Figure \ref{fig:localUpdate}.  
A process begins local integration by checking to see if it owns any active, found particles. If it does
not, then the integration procedure is complete and the process returns to the calling method. In the 
more interesting case where active, found particles exist, particles are iterated over and integrated.
For a particle going through this process, its containing element's information (e.g. nodal information
related to interpolation) is gathered. The particle is then checked to see if it is contained by 
the element it believes it is owned by. If not, a new host element must be determined. This is done 
through a method called handle{\_}particle{\_}exits{\_}element (HPEE), and is described below.

If a particle is indeed contained by the element it believes to be contained by, a check is done to see if
the particle is finished integrating. Assuming this is not the case (the next particle is drawn on the contrary),
the particle must be either active and predicted or active and yet to be predicted. If the former, the particle is
updated according to Equation \ref{eq:predictor}. If the latter, the particle is updated according to Equation 
\ref{eq:corrector}. In either case, a check is done to see if the particle has left its host, and if so it
enters HPEE.

Once all of the particles on each process have been iterated over, a local domain search is done on the set of
active, lost particles to find their new host elements. Particles that have left the domain are then deleted, 
and any particles found locally by the HPEE method are moved into their new hosts. This procedure continues
until the process has no remaining active, found particles, at which point it will return to the calling 
method and get involved with the top tier, global aspects of particle time integration.  

\subsection*{Handling A Particle That Exits an Element}\label{sec:HPEE}
Whenever a particle exits its host element, it must enter the handle{\_}particle{\_}exits{\_}element method,
which will attempt to find the particle's proper host. HPEE does this by executing a face-crossing search
with the element's current and past coordinates, as described in Section \ref{sec:faceSearch}. Based on
the results of this search, the particle may take one of several paths.

If the face-crossing search cannot find any exit side or new containing element, the particle is added to a 
set of particles that will undergo either a local or global domain search in a final attempt to find the proper 
host. This setup allows the domain search to act only as a support to the face-crossing search, which minimizes
the computational resources spent locating particles. Note that if a particle has no proper host (i.e. it has left the 
domain), it will be marked for deletion during the global domain search and deleted immediately after.

When the face-crossing search is successful in finding a new element, the lost particle is assigned the target element.
Once all active, found particles have been iterated over, the particle is then moved into its new host and integration
may resume. Note that it is possible that the new element is not the proper host, but an intermediate element between 
the previous containing element and the new host. In this case, the particle will continue in the integration loop, 
continually getting sent into HPEE until the proper host is determined.

According to this logic, every time a particle moves beyond its owning process' boundaries, it would be subjected to a
global domain search in order to find its new host. Although this is possible with a small number of particles, this 
frequency of global communication is not feasible with particle numbers of engineering interest. To mitigate this 
expense, we can take advantage of the aura feature provided by STK. Since the aura provides a single-element layer of 
ghosting around process boundaries, a particle will always enter a ghosted element before leaving the owning process' 
domain, assuming a sufficiently small time step and that the boundary is shared. As mentioned in Section \ref{sec:stk}, 
STK ensures unique ownership of each entity (e.g. an element) within the discretization. Thus, every time the 
face-crossing search returns an element, the element's ownership is assessed and determined to be either locally owned 
or ghosted, providing a third path for particles passed to HPEE.

\begin{figure}
  \centering
    \includegraphics[width=0.75\textwidth]{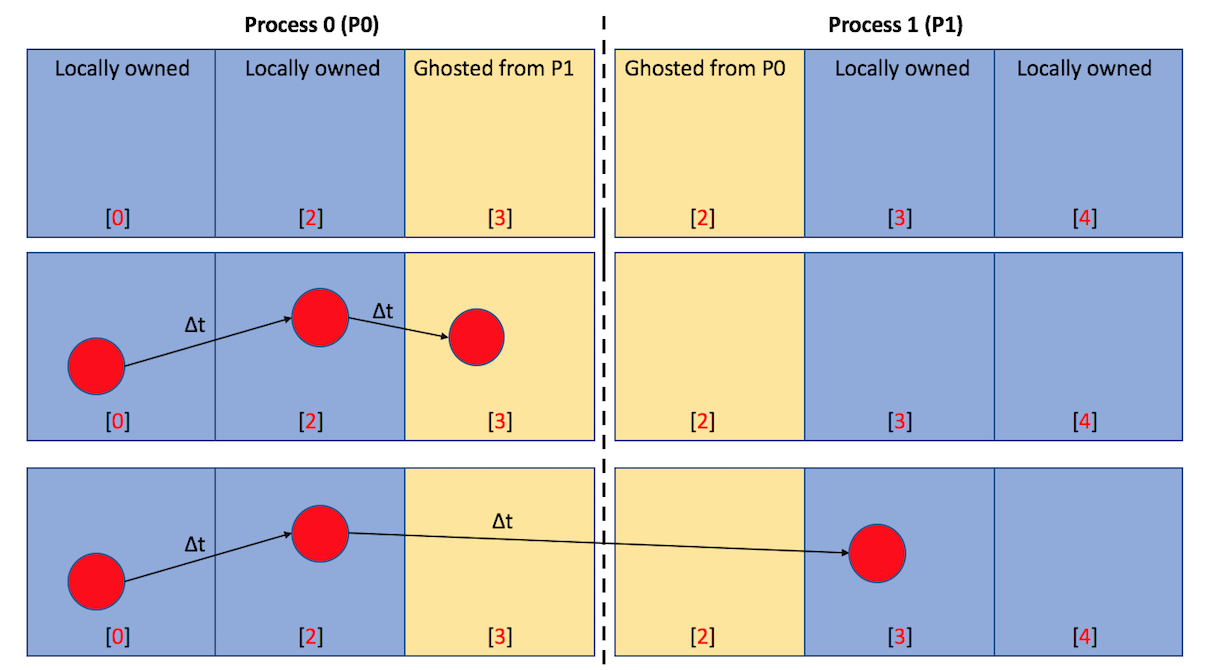}
    \caption{Particle uses aura to determine communication path.}
    \label{fig:auraGhosting}
\end{figure}

In the instance a particle enters a ghosted element, the element's owning process is queried from the STK entity database
and both this process and the ghosted host's identifier are stored on the particle. By storing this data, the particle 
may be communicated to the correct process and assume integration in the locally owned copy of the ghost element 
entered previously. This particle is then added to a set of particles to be communicated once the set of active, found 
particles is empty, and integration at the current timestep continues. Host identification using the face-crossing search
and aura layer is shown in Figure \ref{fig:auraGhosting}, and the communication procedure for a traveling particle is 
described in greater detail below.  

\subsection*{Global Time Integration}\label{sec:globalIntegration}
Evolution at the top-tier involves coordinating processes through time integration in the presence of inter-process
particle communication and global domain searching. Since the global domain search requires all processes to be 
present for communication, a method is needed to ensure that every process remains in the integration procedure 
until all particles within the entire domain have finished evolving for the current timestep. To meet this
requirement, a flag is set on each process at the onset of integration which denotes the ``advancing status'' of
the process. 

Beginning with a status of 1 (i.e. done advancing), each process integrates its particles locally 
(bottom tier integration). A global domain search follows local integration, finding done, lost particles
and marking those particles which have left the domain for deletion. Then, particles marked for communication
are communicated via the ParticleCommManager. The manager does this by packaging up the critical particle 
information, including target host element and coordinates, and communicating the data with 
send/receive methods provided by the STK Utils module. Communicated particle contents are then unpacked
on the receiving process, where integration of the traveling particle is resumed if necessary. The original copy
of the particle is marked for deletion, and any particles marked as such are deleted following the communication 
step.

If at any time a particle is found locally via the face-crossing search or is communicated to another process, 
the advancing status of the process owning the new host is set to 0 (i.e. not ready to advance). Parallel reduction
is used following the communication step to determine if every process in the parallel group is finished integrating.
If so, specified data is written to a file and control is returned to the fluid realm for the next timestep. If the group 
is not ready to advance, all processes continue through the integration loop, regardless of the number of particles owned, 
until all particles reach the proper time level.

\subsection*{Boundary Conditions}
With the implementation detailed for particles existing in the interior of the domain,
a quick discussion on boundary conditions is necessary. Currently, particles interact with
domain boundaries in two ways. As a general rule, particles experience boundaries as one-way outflow
conditions. Due to the requirement that a particle must always know its host element at the
end of every timestep, a search is always performed at the conclusion of each time integration step.
When this happens, any particle that has moved beyond the limits of the domain is left hostless,
and thus gets deleted. This approach is suitable for simple models where flow is either 
one-dimensional in nature or in simulations where particles approach impenetrable walls. However,
in more intricate geometries, integration error may allow for particles to move beyond
walls in an unphysical manner, requiring a more involved treatment of the boundaries.

In addition to the general outflow condition, support for periodic boundary conditions within the particle
module is currently under development. Periodic boundary conditions are necessary to track particle trajectories 
over multiple flow-through times, and thus critical in computing bed residence times or gathering accurate 
dispersion statistics. While a proof of concept implementation has been demonstrated (detailed below), 
further work must be done before it is production-ready. 

\begin{figure}
  \centering
    \includegraphics[width=0.75\textwidth]{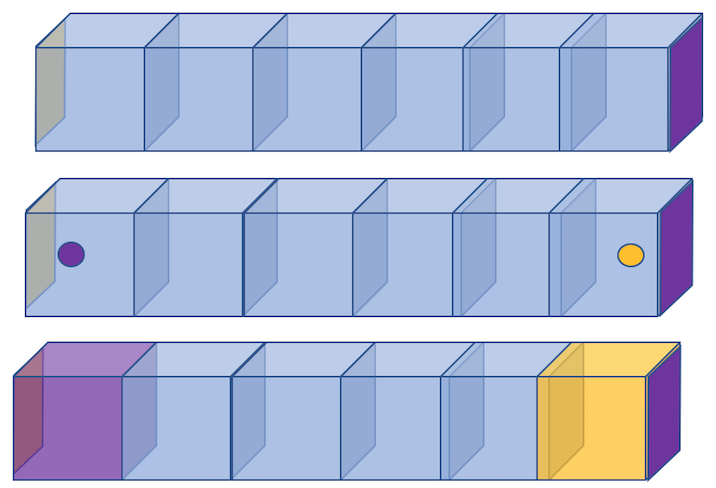}
    \caption{Process to create a face-element map for particle communication at a periodic boundary.}
    \label{fig:faceEltMap}
\end{figure}

Nalu enforces periodic boundary conditions through a master-slave mapping of boundary nodes, which is applied within 
the equation systems. Largely due to the particles being objects distinct from the STK Mesh, the particles cannot 
implicitly respect the periodic boundary condition applied to the fluid realm. Thus, these conditions must be 
enforced explicitly within the particle realm. Currently, this is achieved by constructing a face-element map,
depicted in Figure \ref{fig:faceEltMap}. 

To construct this map, boundary element faces are first found for each periodic boundary via the sideset-based
part from STK. Then, a centroid for each face is computed and offset by the length of the periodic dimension, thus
placing it in a boundary element on the opposite side of the domain. Using a global domain search, each point 
(an offset boundary-face centroid), which knows the face identifier and corresponding process rank it belongs to,
is mapped to an element which contains a face on the opposing periodic boundary. This map may then be queried
any time a particle exits a face on the boundary of the domain to identify if periodic boundary conditions 
have been turned on, and, if so, both which element and process rank it should be sent to. 

\begin{figure}[!tbp]
    \centering  
    \subfloat[]
        {\includegraphics[width=0.51\textwidth]{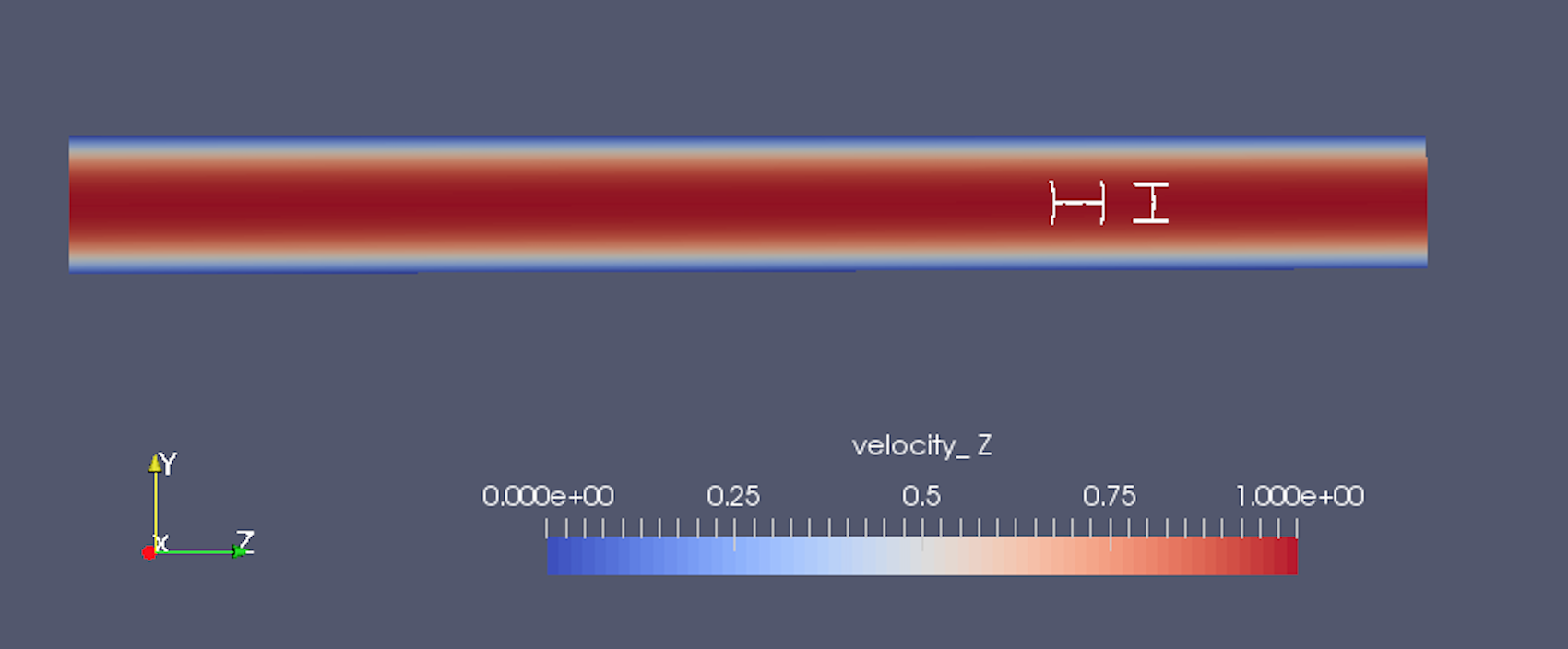}
         \label{fig:periodicBc1}}
    \subfloat[]
        {\includegraphics[width=0.5\textwidth]{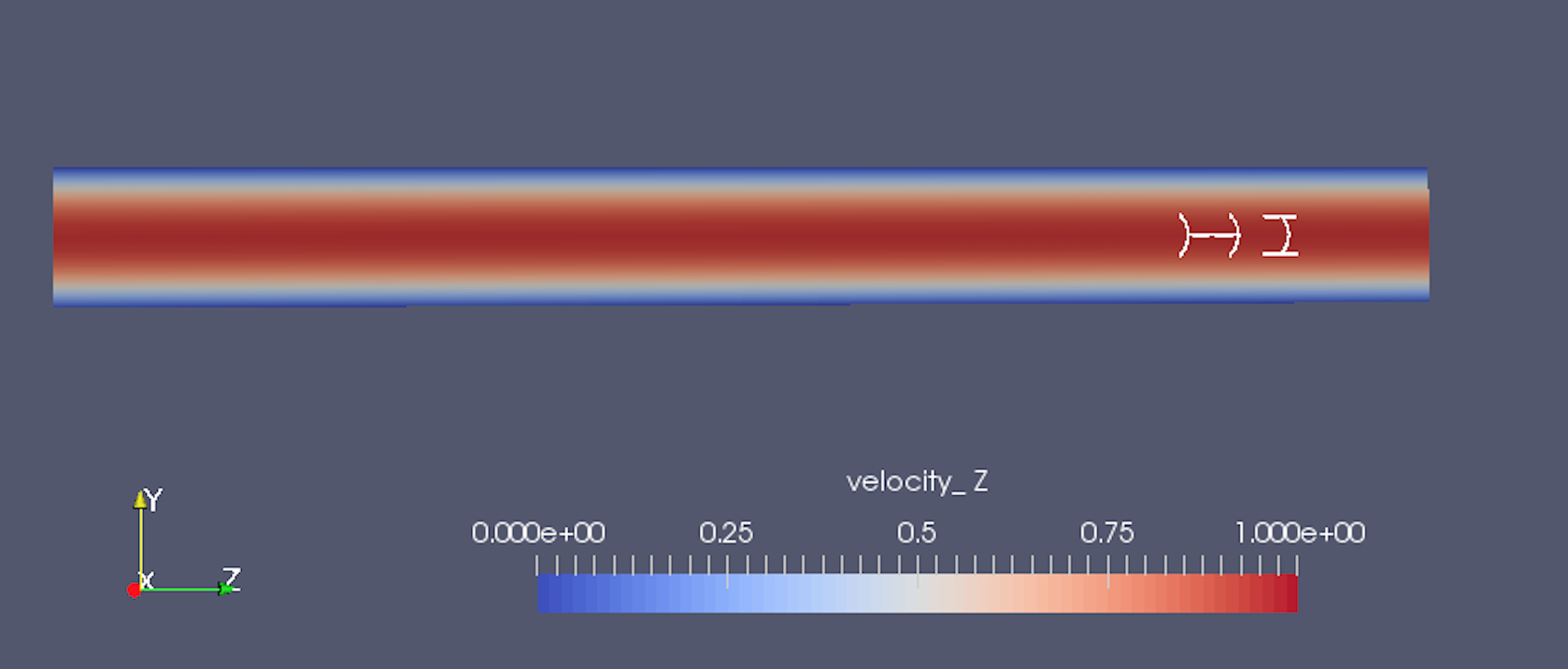}
         \label{fig:periodicBc2}}
    \hfill
    \subfloat[]
        {\includegraphics[width=0.5\textwidth]{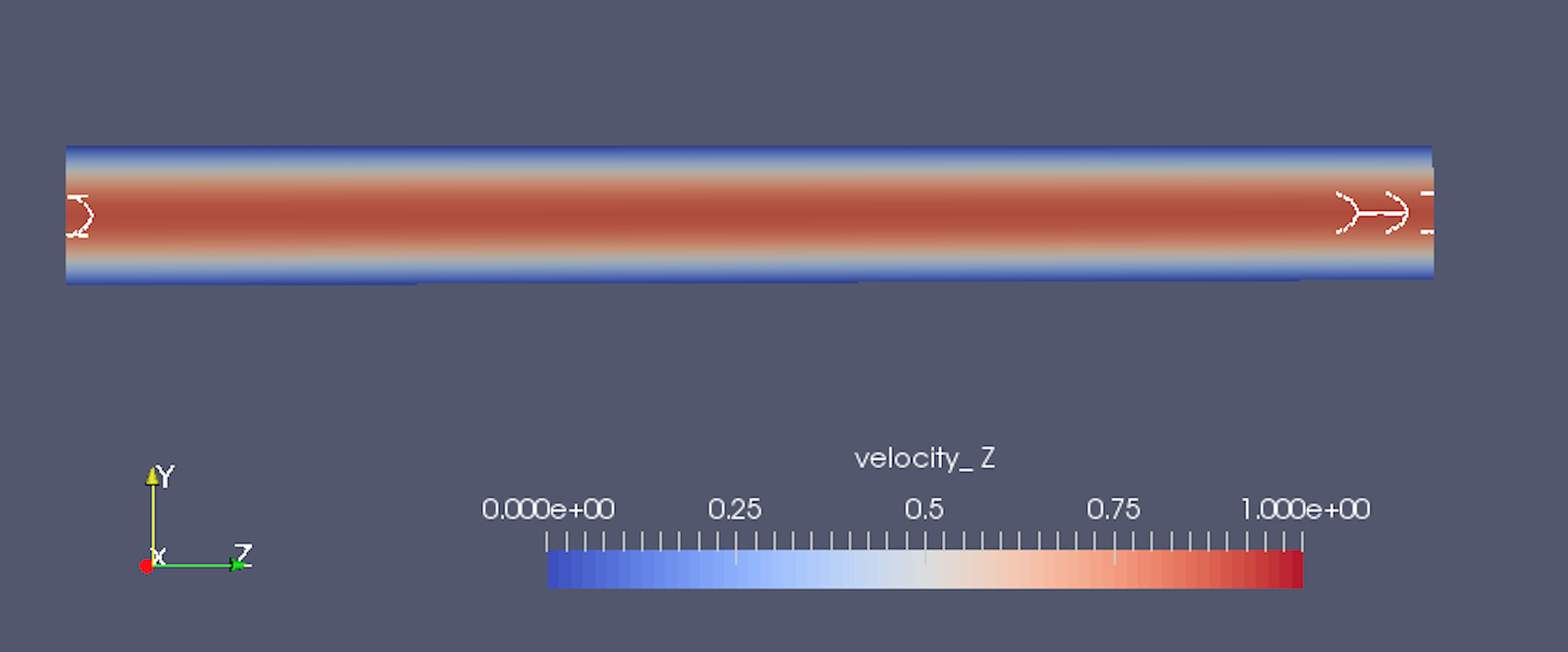}
         \label{fig:periodicBc3}}
    \subfloat[]
        {\includegraphics[width=0.51\textwidth]{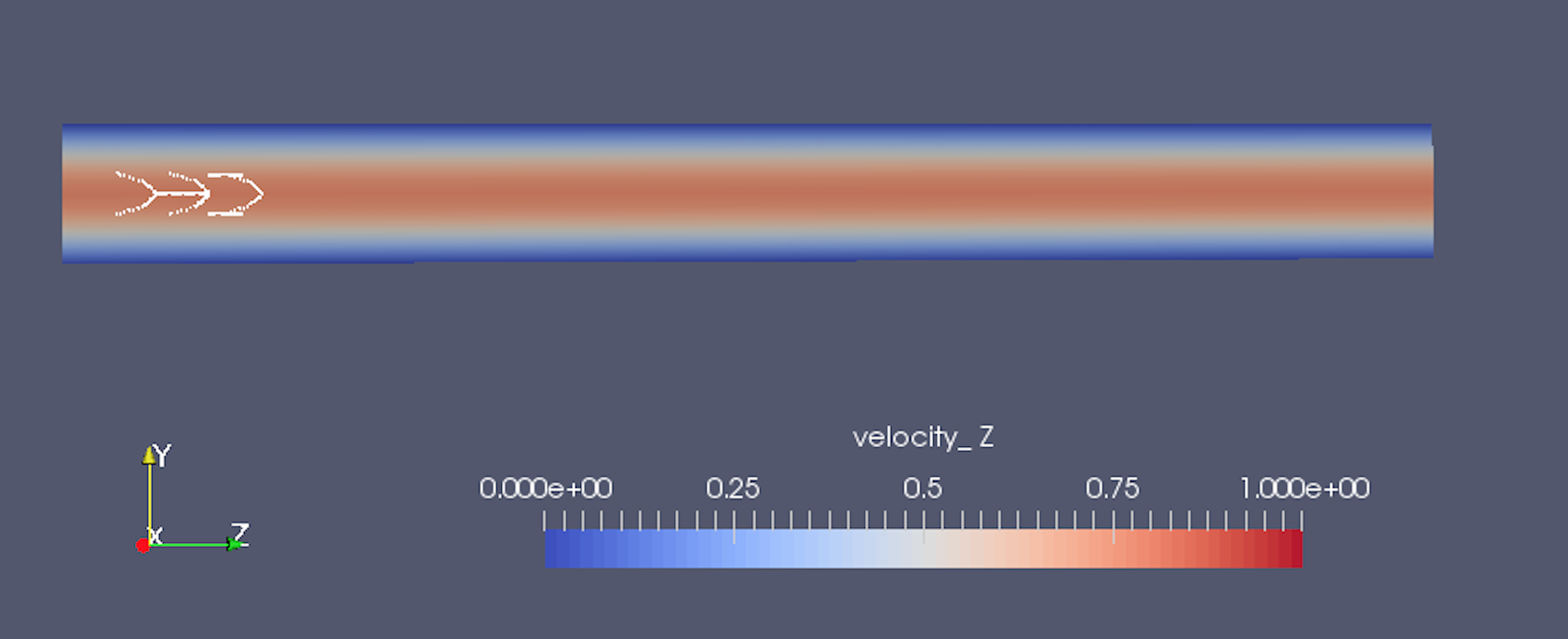}
         \label{fig:periodicBc3}}
    \caption{A demonstration of particles moving through a cylindrical channel flow with periodic boundary conditions
             in the longitudinal direction. Several snapshots are shown, detailing: 
      \protect\subref{fig:periodicBc1} particle creation,
      \protect\subref{fig:periodicBc2} initial movement,
      \protect\subref{fig:periodicBc3} motion through the periodic boundary and
      \protect\subref{fig:periodicBc3} continued evolution from the left side of the channel.}
    \label{fig:periodicBc}
\end{figure}

A demonstration of the periodic boundary conditions may be seen in Figure \ref{fig:periodicBc}. A set of particles
move downstream in a cylindrical channel flow, slowly spreading out according to the parabolic velocity profile
carried by the fluid. While these conditions appear to be working well in Figure \ref{fig:periodicBc}, 
the method proposed for constructing the face-element map allows for duplicate entries in the map, as a 
face may be mapped to multiple elements. This problem arises from performing the domain search
on elements and face centroids without executing a secondary search to distinguish perfect element-face matches 
from candidate pairs. A more robust solution may be to directly
create a face-to-face mapping structure utilizing a search procedure, similar to the one implemented by Nalu for 
establishing periodic boundary conditions, although this algorithm is beyond the scope of the present work.

%% file: texFiles/mainMatter/chapter3/verification.tex
\section{Verification of the Particle Evolution Algorithm}
In order to verify the evolution algorithm detailed in Section \ref{sec:evolutionAlg}, 
particles have been simulated in flow through a laminar cylindrical channel such that their 
computed paths may be compared against exact theoretical trajectories. Although this test 
lacks an examination of numerical convergence and removes many of the physical and numerical
considerations associated with problems of industrial or academic interest (e.g. boundary layers,
turbulence), it provides an adequate arena to assess the essential components of the implementation
discussed previously. Eventually, more rigorous testing (e.g. tracing of a rotating flow) will be 
useful in verifying extensions to the modules and its ability to perform in more dynamic 
environments.

\begin{figure}[!tbp]
    \centering  
    \subfloat[]
        {\includegraphics[width=0.4\textwidth]{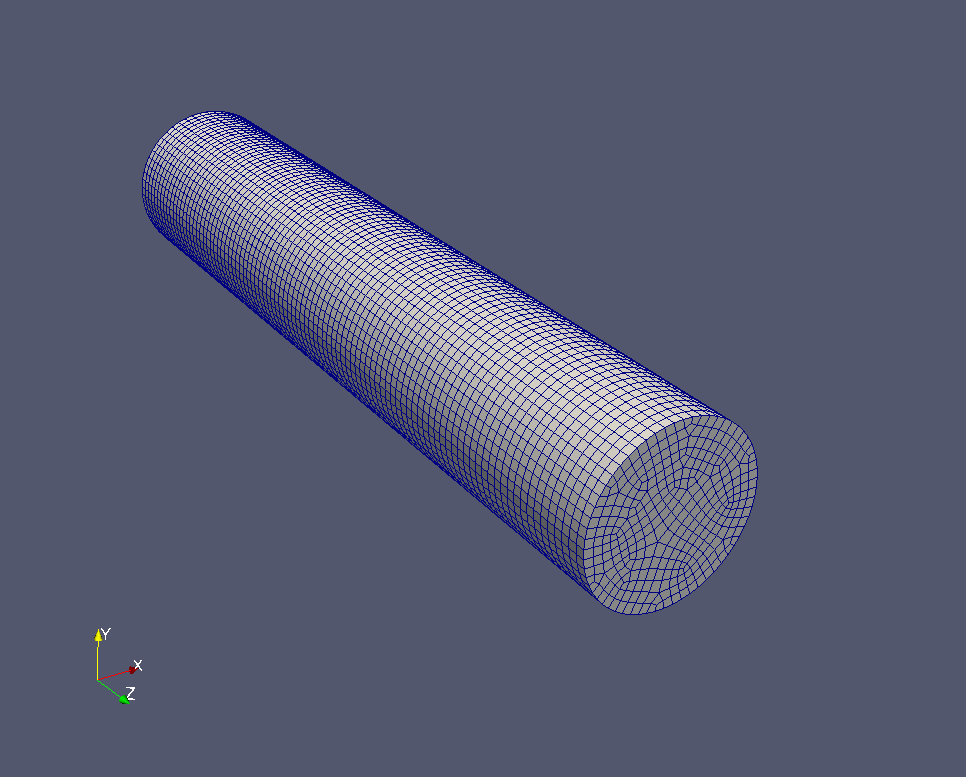}
         \label{fig:testingMesh}}
    \subfloat[]
        {\includegraphics[width=0.45\textwidth]{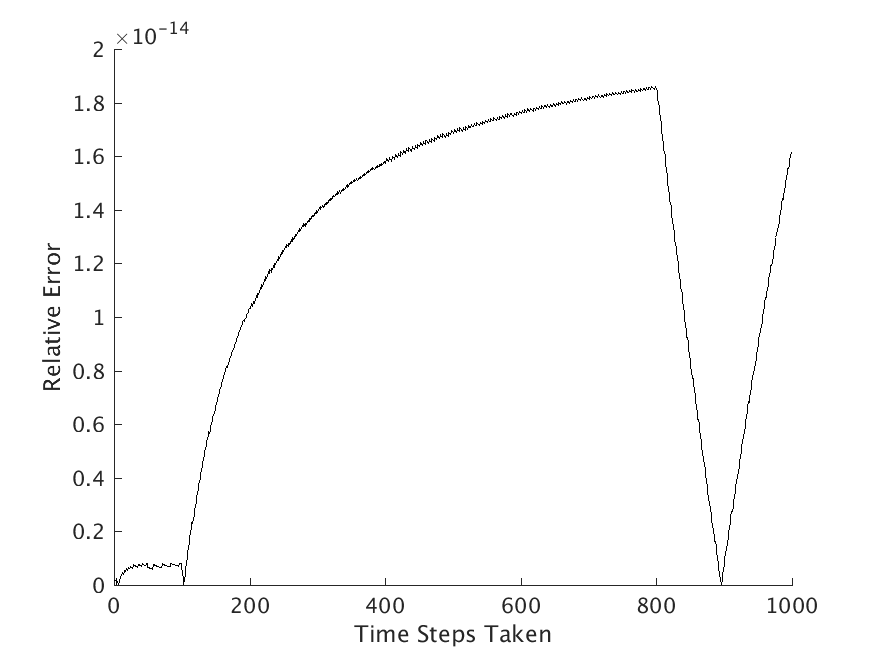}
         \label{fig:particleErrors}}
    \caption{Details regarding the initial verification test, including
      \protect\subref{fig:testingMesh} the unstructured hex-mesh used and 
      \protect\subref{fig:particleErrors} relative error in the computed particle position.}
    \label{fig:particleTest}
\end{figure}

The channel mesh is unstructured and consists of $\approx 30,000$ hexahedral elements, seen 
in Figure \ref{fig:testingMesh}. The fluid is incompressible and Newtonian with a density
$\rho = 1.0$ $kg/m^3$ and viscosity $\mu = 1.0\cdot{10^{-3}}$ $kg/(m\cdot{s})$. In an effort to 
remove boundary layer effects from this test and maintain a constant, unidirectional velocity 
throughout the domain, the inflow condition, wall condition and initial condition for the 
rest of the domain have been set to a velocity $\textbf{u} = (0, 0, 0.01)$ $m/s$ along 
the longitudinal direction of the 
channel. Additionally, an outflow condition is set to the surface opposite the inflow. 
By maintaining a unidirectional constant velocity within the domain, the 
distance any particle should travel in a given time step should be equal to a fixed fraction 
of the size of the time step. This provides a simple scenario to study the evolution of 
particles through an unstructured mesh in a parallel-processing environment.

Results of this test are shown in Figure \ref{fig:particleErrors} for simulation taking
one thousand time steps with a domain decomposed among ten processors. The relative error
between the theoretical and computed positions is on the order of machine error, as expected
from the constant velocity field. This simple test demonstrates the successful function of the 
particle module, showing that a particle is properly communicated between processors and interpolating
its velocity appropriately from an unstructured, distributed mesh.

\begin{figure}[!tbp]
    \centering  
    \subfloat[]
        {\includegraphics[width=0.305\textwidth]{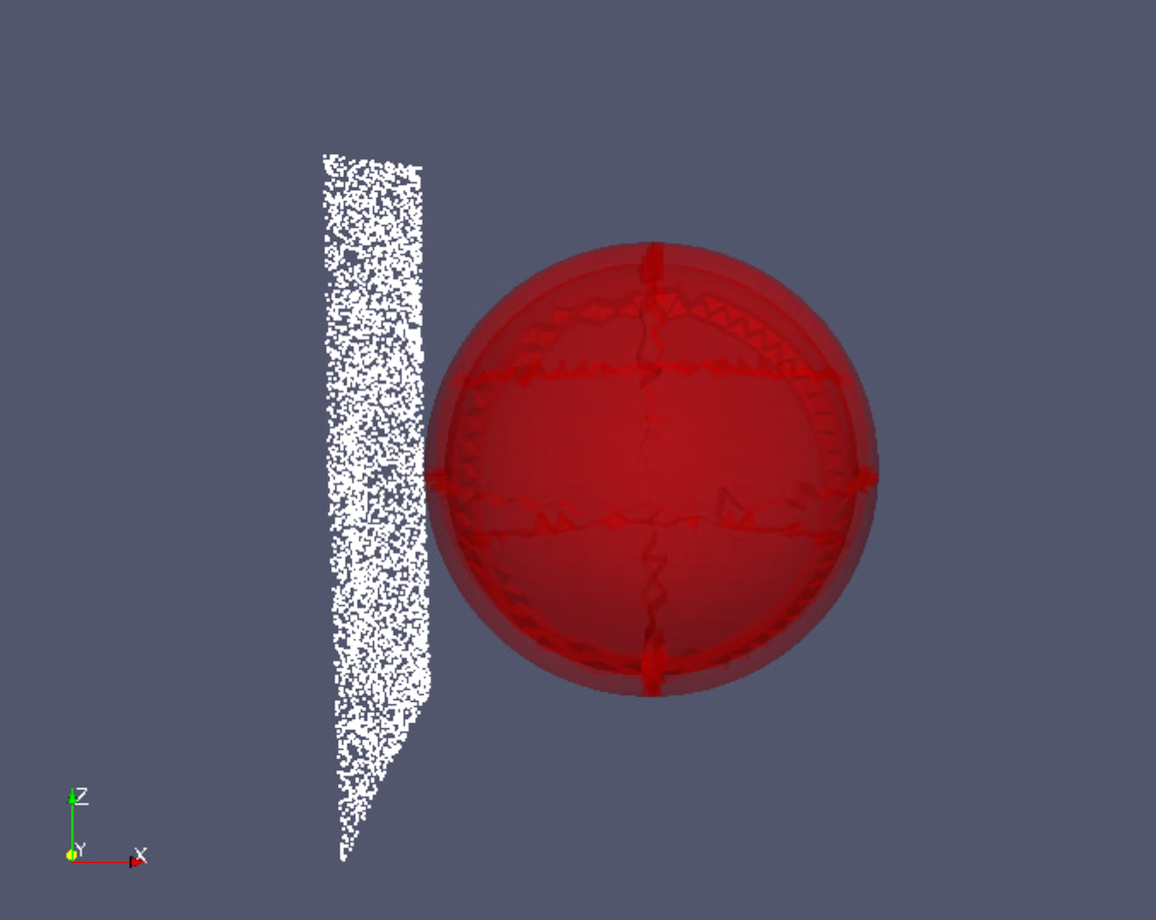}
         \label{fig:flowPastSphere1}}
    \subfloat[]
        {\includegraphics[width=0.3\textwidth]{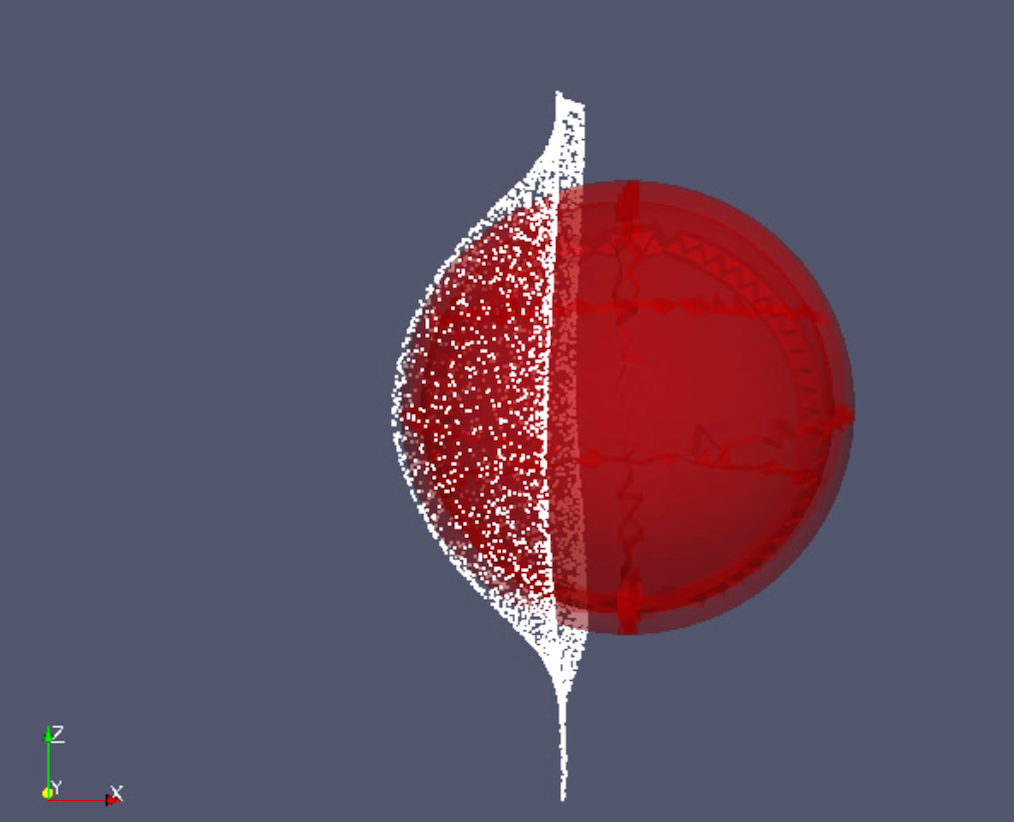}
         \label{fig:flowPastSphere2}}
    \subfloat[]
        {\includegraphics[width=0.3\textwidth]{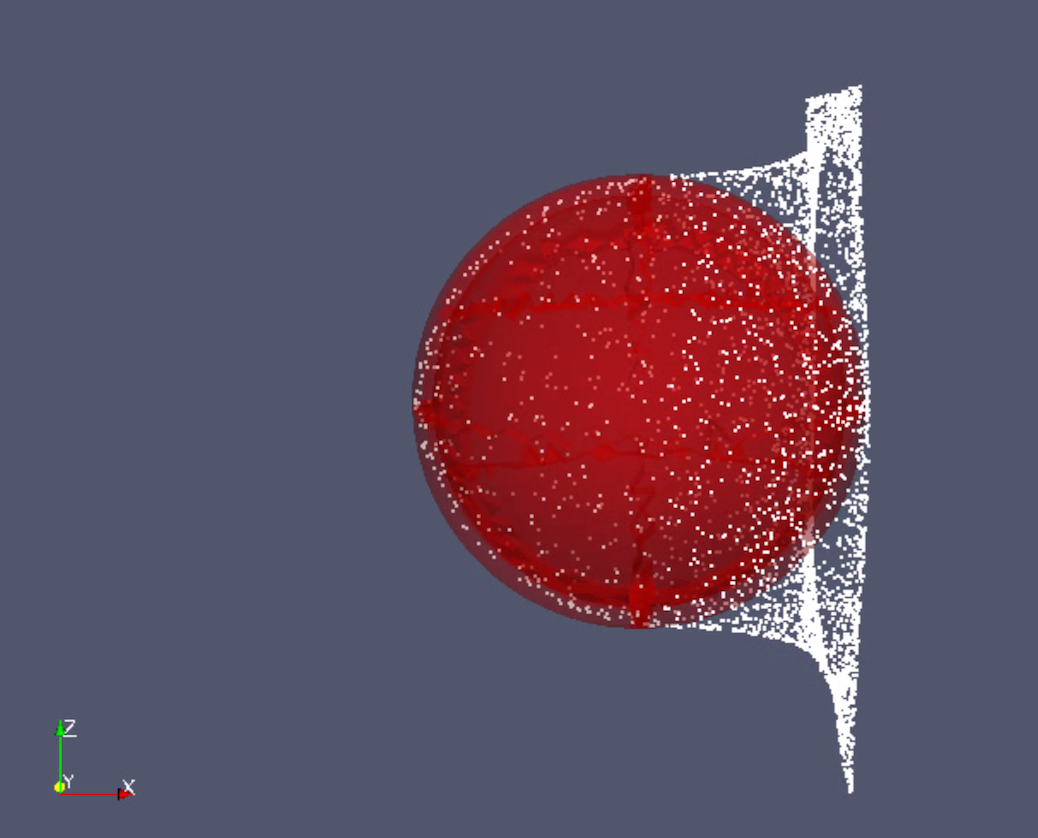}
         \label{fig:flowPastSphere3}}
    \caption{Snapshots of particles tracing flow past a sphere when the bulk of the particles are
      \protect\subref{fig:flowPastSphere1} upstream of the sphere,
      \protect\subref{fig:flowPastSphere2} moving past the upstream end of the sphere,
      \protect\subref{fig:flowPastSphere3} past the sphere.}
    \label{fig:flowPastSphere}
\end{figure}

In addition to the test above, images are shown in Figure \ref{fig:flowPastSphere} of particles 
tracking laminar flow past a sphere. Particles can be seen tracking the flow around the sphere, as well as
getting stuck in the low-pressure zone at the upstream stagnation point. 
These results are shown to indicate qualitatively that
the first order behavior of the fluid is captured by the particles, 
and that this module is on its way to being used in more applicable simulation
environments. 

%% file: texFiles/mainMatter/conclusion/conclusions.tex
\section{Introduction}
In this final chapter, the primary conclusions and their implications will be reaffirmed,
recommendations for future work on both the detailed LES and the particle tracking module are discussed,
and plans for their unity and a suggested path forward towards studying scalar transport are detailed.

\section{Regarding the Detailed Large Eddy Simulation}
Throughout Chapters 2 and 3 of this thesis, aspects of a detailed LES model for turbulent 
flow over a permeable bed have been discussed, examined and tested. Analysis of a mesh refinement study
clearly demonstrates the feasibility of using a graded mesh to reduce computational cost, provided the 
mesh in the transition region is sufficiently resolved to avoid numerical instability. By comparing
averaged velocity profiles and autocorrelations predicted by simulations with varying domain/bed size, periodic
boundary conditions have been shown to artificially limit the development of large turbulent structures. The 
double-averaging methodology has been applied to study interfacial momentum transport in the presence of significant 
spatial heterogeneity, notably showing that form drag always dominates viscous drag within the bed and that the 
Reynolds shear stress rapidly decays within and below the transition layer. Lastly, two different turbulence closures
have been studied, highlighting the role of turbulent viscosity in creating smaller, and possibly weaker, turbulent
structures in the Smagorinsky model compared to the WALE model.

While these findings are largely focused on details regarding the development of a high fidelity numerical
model, several aspects of the physical problem are still ripe for exploration. First and foremost,
a tool with which one can more closely investigate the presence of turbulence is needed. Whether this tool
be turbulent kinetic energy budget calculations, sophisticated conditional sampling of velocity fluctuations 
or measures of vorticity and swirling strength, elucidating the size, strength and motion of turbulent structures
within the transition layer will prove useful in quantifying interfacial momentum transport.

One way to more thoroughly probe the pore space is with quadrant analysis at locations that vary in both
depth and streamwise position. Several studies have put effort
towards studying the structure of turbulence within pore throats across different layers of the bed
(e.g. \cite{Blois2012,Pokrajac2009}). By leveraging the hydrodynamic detail present in LES,
one might be able to more clearly explain the origin, evolution and influence of bed turbulence at various 
depths. 

Finally, parametric studies of several model parameters may prove useful in understanding interfacial
transport. Such parameters include the body force, the sphere diameter, the interparticle gap spacing and 
the surface flow height. By examining the changes in the flow field as a function of these parameters, one
may gain insight into which Reynolds number (e.g. $Re_{K}$, $Re_{D}$) best characterizes the flow, or
how the depth of the surface flow height and the particle diameter affects inner and outer scaling of the
turbulent flow.

Although expensive, LES has been shown to be a suitable alternative to DNS, enabling one to resolve the 
dominant physics at play while reducing computational requirements. While many questions remain open in the study 
of this type of system, the work in this thesis further demonstrates that LES can be a useful tool in
helping to find answers.

\section{Regarding Nalu's Particle Tracking Module}
Within Chapter 4, the implementation of a particle tracking module within the Nalu
code base has been proposed and detailed. Critical aspects of I/O, time integration and 
distributed memory computation have been documented. The module has been shown to 
successfully integrate a particle across a distributed, unstructured mesh with machine
precision and capture the first order behavior of flow past a spherical obstruction with
5,000 particles. Additionally, a proof-of-concept implementation of periodic boundary 
conditions has been demonstrated, laying the groundwork for simulations of
academic and industrial interest.

Several components of the particle tracking module must be improved before particles are ready to
be used within an LES. Primarily, a robust algorithm for establishing periodic boundary conditions must be 
implemented (currently underway), which will enable particle tracking over time scales much larger
than a single flow-through time. Once this is accomplished, a method to enable a particle to keep track
of its previous positions will be needed. Giving the particles this awareness will allow one to statistically
evaluate particle trajectories in post-processing, providing for the extraction of dispersion information for
upscaling to lower-order models. An example of this upscaling procedure is seen in \cite{Sund2015}, where 
detailed information is gathered from two-dimensional simulations. To the best of the author's knowledge, 
this procedure has not been done with detailed, three-dimensional turbulence modeling at the pore scale.  

Eventually, the particle module should be expanded to allow for particles which may experience
different kinds of physics (e.g. electrodynamics, thermodynamics). Further degrees of coupling may
be explored, enabling particle-particle contact or computing heat transfer between the particles
and fluid. Advanced extensions may be developed, such as linking against molecular dynamics libraries
to study molecular gas dynamics within Nalu. Many possibilities for expansion exist, and it seems that
virtually all of them lead to exciting methods of inquiry.

\section{Regarding the Combination of the LES and Particles}

One of the primary goals of developing the detailed LES and the particle tracking
module is to incorporate them together. By doing this, an extensive data set, rich in
both Eulerian and Lagrangian information, may be gathered. This data set can then be 
used both to interrogate the hydrodynamics throughout the domain and
more easily upscale dispersion information to reduced-order models.

While this plan cannot be executed until the particle module has been given the necessary feature
set, one alternative to this plan has yet to be explored. Knowing that the particles are purely passive,
and their trajectories are entirely governed by the surrounding fluid motion, it might be possible, without the 
particles, to extract the same dispersion information that is intended to be captured by the particles. The path
to do this is not immediately clear. A reasonable place to start may be looking at the Reynolds shear stresses.
As lower-order models often look only to treat the wall-normal and streamwise flow directions, perhaps 
the turbulent momentum transport described by $\tau^{R}_{13}$ can provide some sort of parameterization 
for the probability and magnitude of vertical displacement in a stochastic particle tracking model.

With either this method or the use of particle tracking, this suggested path forward offers both an opportunity to gain direct
insight into the mechanisms driving interfacial transport in flows over permeable beds, and inform 
reduced-order, less expensive models which can be used to predict transport across a host of environmentally applicable
length and time scales.

%% file: ms.bbl
\begin{thebibliography}{10}

\bibitem{Stonedahl2010}
S.~H. Stonedahl, J.~W. Harvey, A.~W{\"{o}}rman, M.~Salehin, and A.~I. Packman.
\newblock {A multiscale model for integrating hyporheic exchange from ripples
  to meanders}.
\newblock {\em Water Resources Research}, 46(12), 2010.

\bibitem{Berkowitz2006}
B.~Berkowitz, A.~Cortis, M.~Dentz, and H.~Scher.
\newblock {Modeling non-Fickian transport in geological formations as a
  continuous time random walk}.
\newblock {\em Reviews of Geophysics}, 44(2), 2006.

\bibitem{Chakraborty2009}
P.~Chakraborty, M.~M. Meerschaert, and C.~Y. Lim.
\newblock {Parameter estimation for fractional transport: A particle-tracking
  approach}.
\newblock {\em Water Resources Research}, 45(10), 2009.

\bibitem{Boano2014}
F.~Boano, J.~W. Harvey, A.~Marion, A.~I. Packman, R.~Revelli, L.~Ridolfi, and
  A.~W{\"{o}}rman.
\newblock {Hyporheic flow and transport processes: Mechanisms, models, and
  biogeochemical implications}.
\newblock {\em Reviews of Geophysics}, 52(4):603--679, 2014.

\bibitem{Manes2012}
C.~Manes, L.~Ridolfi, and G.~Katul.
\newblock {A phenomenological model to describe turbulent friction in
  permeable-wall flows}.
\newblock {\em Geophysical Research Letters}, 39(14), 2012.

\bibitem{WHITE2007}
B.~L. White and H.~M. Nepf.
\newblock {Shear instability and coherent structures in shallow flow adjacent
  to a porous layer}.
\newblock {\em Journal of Fluid Mechanics}, 593:1--32, 2007.

\bibitem{Giometto2016}
M.~G. Giometto, A.~Christen, C.~Meneveau, J.~Fang, M.~Krafczyk, and M.~B.
  Parlange.
\newblock {Spatial characteristics of roughness sublayer mean flow and
  turbulence over a realistic urban surface}.
\newblock {\em Boundary-Layer Meteorology}, 160(3):425--452, 2016.

\bibitem{Shiozawa2016}
S.~Shiozawa and M.~McClure.
\newblock {Simulation of proppant transport with gravitational settling and
  fracture closure in a three-dimensional hydraulic fracturing simulator}.
\newblock {\em Journal of Petroleum Science and Engineering}, 138:298--314,
  2016.

\bibitem{Schumer2009}
R.~Schumer, M.~M. Meerschaert, and B.~Baeumer.
\newblock {Fractional advection-dispersion equations for modeling transport at
  the Earth surface}.
\newblock {\em Journal of Geophysical Research}, 114(F4), 2009.

\bibitem{Sund2015}
N.~Sund, D.~Bolster, S.~Mattis, and C.~Dawson.
\newblock {Pre-asymptotic transport upscaling in inertial and unsteady flows
  through porous media}.
\newblock {\em Transport in Porous Media}, 109(2):411--432, 2015.

\bibitem{Cardenas2008a}
M.~B. Cardenas.
\newblock {Three-dimensional vortices in single pores and their effects on
  transport}.
\newblock {\em Geophysical Research Letters}, 35(18), 2008.

\bibitem{Nikora2007a}
V.~Nikora, I.~McEwan, S.~McLean, S.~Coleman, D.~Pokrajac, and R.~Walters.
\newblock {Double-averaging concept for rough-bed open-channel and overland
  flows: Theoretical background}.
\newblock {\em Journal of Hydraulic Engineering}, 133(8):873--883, 2007.

\bibitem{Goharzadeh2005}
A.~Goharzadeh, A.~Khalili, and B.~B. Jo{\o}rgensen.
\newblock {Transition layer thickness at a fluid-porous interface}.
\newblock {\em Physics of Fluids}, 17(5), 2005.

\bibitem{Pokrajac2007}
D.~Pokrajac, C.~Manes, and I.~McEwan.
\newblock {Peculiar mean velocity profiles within a porous bed of an open
  channel}.
\newblock {\em Physics of Fluids}, 19(9), 2007.

\bibitem{Manes2009}
C.~Manes, D.~Pokrajac, I.~McEwan, and V.~Nikora.
\newblock {Turbulence structure of open channel flows over permeable and
  impermeable beds: A comparative study}.
\newblock {\em Physics of Fluids}, 21(12):1--12, 2009.

\bibitem{Manes2011a}
C.~Manes, D.~Pokrajac, V.~I. Nikora, L.~Ridolfi, and D.~Poggi.
\newblock {Turbulent friction in flows over permeable walls}.
\newblock {\em Geophysical Research Letters}, 38(3), 2011.

\bibitem{Manes2011}
C.~Manes, D.~Poggi, and L.~Ridolfi.
\newblock {Turbulent boundary layers over permeable walls: Scaling and
  near-wall structure}.
\newblock {\em Journal of Fluid Mechanics}, 687:141--170, 2011.

\bibitem{BenMeftah2016}
M.~B. Meftah and M.~Mossa.
\newblock {A modified log-law of flow velocity distribution in partly
  obstructed open channels}.
\newblock {\em Environmental Fluid Mechanics}, 16(2):453--479, 2016.

\bibitem{Breugem2006}
W.~P. Breugem, B.~J. Boersma, and R.~E. Uittenbogaard.
\newblock {The influence of wall permeability on turbulent channel flow}.
\newblock {\em Journal of Fluid Mechanics}, 562:35--72, 2006.

\bibitem{Breugem2005}
W.~P. Breugem and B.~J. Boersma.
\newblock {Direct numerical simulations of turbulent flow over a permeable wall
  using a direct and a continuum approach}.
\newblock {\em Physics of Fluids}, 17(2):1--15, 2005.

\bibitem{Kuwata2015}
Y.~Kuwata and K.~Suga.
\newblock {Lattice Boltzmann direct numerical simulation of interface
  turbulence over porous and rough walls}.
\newblock {\em International Journal of Heat and Fluid Flow}, 61:145--157,
  2016.

\bibitem{Kuwata2016}
Y.~Kuwata and K.~Suga.
\newblock {Transport mechanism of interface turbulence over porous and rough
  walls}.
\newblock {\em Flow, Turbulence and Combustion}, 97(4):1071--1093, 2016.

\bibitem{Chandesris2016}
M.~Chandesris, A.~D'Hueppe, B.~Mathieu, D.~Jamet, and B.~Goyeau.
\newblock {Direct numerical simulation of turbulent heat transfer in a
  fluid-porous domain}.
\newblock {\em Physics of Fluids}, 25(12):1--21, 2013.

\bibitem{SMAGORINSKY1963}
J.~Smagorinsky.
\newblock {General circulation experiments with the primitive equations}.
\newblock {\em Monthly Weather Review}, 91(3):99--164, 1963.

\bibitem{Deardorff1970}
J.~W. Deardorff.
\newblock {A numerical study of three-dimensional turbulent channel flow at
  large Reynolds numbers}.
\newblock {\em Journal of Fluid Mechanics}, 41(2):453--480, 1970.

\bibitem{Pope2004}
S.~B. Pope.
\newblock {Ten questions concerning the large-eddy simulation of turbulent
  flows}.
\newblock {\em New Journal of Physics}, 6(1):35--58, 2004.

\bibitem{Stoesser2007}
T.~Stoesser, J.~Fr{\"{o}}hlich, and W.~Rodi.
\newblock {Turbulent open-channel flow over a permeable bed}.
\newblock {\em 32nd IAHR Congress}, 2007.

\bibitem{Temmerman2003}
L.~Temmerman, M.~A. Leschziner, C.~P. Mellen, and J.~Fr{\"{o}}hlich.
\newblock {Investigation of wall-function approximations and subgrid-scale
  models in large eddy simulation of separated flow in a channel with
  streamwise periodic constrictions}.
\newblock {\em International Journal of Heat and Fluid Flow}, 24(2):157--180,
  2003.

\bibitem{Mendez2008}
S.~Mendez and F.~Nicoud.
\newblock {Large-eddy simulation of a bi-periodic turbulent flow with
  effusion}.
\newblock {\em Journal of Fluid Mechanics}, 598:27--65, 2008.

\bibitem{Frohlich2005}
J.~Fr{\"{o}}hlich, C.~P. Mellen, W.~Rodi, L.~Temmerman, and M.~A. Leschziner.
\newblock {Highly resolved large-eddy simulation of separated flow in a channel
  with streamwise periodic constrictions}.
\newblock {\em Journal of Fluid Mechanics}, 526:19--66, 2005.

\bibitem{Nikora2007b}
V.~Nikora, S.~Mclean, S.~Coleman, D.~Pokrajac, I.~McEwan, L.~Campbell,
  J.~Aberle, D.~Clunie, and K.~Koll.
\newblock {Double-averaging concept for rough-bed open-channel and overland
  flows: Applications}.
\newblock {\em Journal of Hydraulic Engineering}, 133(8):884--895, 2007.

\bibitem{Blois2012}
G.~Blois, G.~H. {Sambrook Smith}, J.~L. Best, R.~J. Hardy, and J.~R. Lead.
\newblock {Quantifying the dynamics of flow within a permeable bed using
  time-resolved endoscopic particle imaging velocimetry (EPIV)}.
\newblock {\em Experiments in Fluids}, 53(1):51--76, 2012.

\bibitem{Pokrajac2009}
D.~Pokrajac and C.~Manes.
\newblock {Velocity measurements of a free-surface turbulent flow penetrating a
  porous medium composed of uniform-size spheres}.
\newblock {\em Transport in Porous Media}, 78:367--383, 2009.

\bibitem{Schneider1987}
G.~E. Schneider and M.~J. Raw.
\newblock {Control volume finite-element method for heat transfer and fluid
  flow using colocated variables— 1. Computational procedure}.
\newblock {\em Numerical Heat Transfer}, 11(4):363--390, 1987.

\bibitem{Domino2015}
S.~Domino.
\newblock {Sierra Low Mach Module: Nalu Theory Manual 1.0}.
\newblock SAND2015-3107W, Sandia National Laboratories Unclassified Unlimited
  Release (UUR). 2015. \url{https://github.com/NaluCFD/NaluDoc}.

\bibitem{Ducros1998}
F.~Ducros, F.~Nicoud, and T.~Poinsot.
\newblock {Wall-adapting local eddy-viscosity models for simulations in complex
  geometries}.
\newblock {\em Conference on Numerical Methods in Fluid Dynamics}, 1998.

\bibitem{schlichting2000}
H.~Schlichting and K.~Getsten.
\newblock {\em {Boundary Layer Theory}}.
\newblock Springer-Verlag, Berlin, 8th edition, 2000.

\bibitem{Galmarini1999}
S.~Galmarini and P.~Thunis.
\newblock {On the validity of Reynolds assumptions for running-mean filters in
  the absence of a spectral gap}.
\newblock {\em Journal of the Atmospheric Sciences}, 56(12):1785--1796, 1999.

\bibitem{Whitaker1999}
S.~Whitaker.
\newblock {\em {The Method of Volume Averaging}}.
\newblock Springer, Dordrecht, 1999.

\bibitem{Kundu2016}
P.~K. Kundu, I.~M. Cohen, and D.~R. Dowling.
\newblock {\em {Fluid Mechanics}}.
\newblock Academic Press, Boston, 6th edition, 2016.

\bibitem{Meyers2016}
W.~Munters, C.~Meneveau, and J.~Meyers.
\newblock {Shifted periodic boundary conditions for simulations of wall-bounded
  turbulent flows}.
\newblock {\em Physics of Fluids}, 28(2), 2016.

\bibitem{Martin2009}
G.~D. Martin, E.~Loth, and D.~Lankford.
\newblock {Particle host cell determination in unstructured grids}.
\newblock {\em Computers and Fluids}, 38(1):101--110, 2009.

\bibitem{Cheng1996}
H.~Cheng, J.~Cheng, and G.~Yeh.
\newblock {A particle tracking technique for the Lagrangian-Eulerian finite
  element method in multi-dimensions}.
\newblock {\em International Journal for Numerical Methods in Engineering},
  39:1115--1136, 1996.

\bibitem{Lohner1990}
R.~L{\"{o}}hner and J.~Ambrosiano.
\newblock {A vectorized particle tracer for unstructured grids}.
\newblock {\em Journal of Computational Physics}, 91(1):22--31, 1990.

\bibitem{Pokrajac2002}
D.~Pokrajac and R.~Lazic.
\newblock {An efficient algorithm for high accuracy particle tracking in finite
  elements}.
\newblock {\em Advances in Water Resources}, 25(4):353--369, 2002.

\bibitem{Report2016}
H.~Edwards, A.~Williams, G.~Sjaardema, D.~Baur, and W.~Cochran.
\newblock {Sierra toolkit computational mesh conceptual model}.
\newblock {\em Technical Report SAND2010-1192}, Sandia National Laboratories
  Unlimited Release, 2010.

\bibitem{stkManual}
Sierra Toolkit~Development Team.
\newblock {Sierra toolkit manual version 4.42}.
\newblock {\em Technical Report SAND2016-9964}, Sandia National Laboratories
  Unlimited Release, 2016.

\bibitem{Theodoridis2006}
Y.~Manolopoulos, A.~Nanopoulos, A.~N. Papadopoulos, and Y.~Theodoridis.
\newblock {\em R-Trees: Theory and Applications}.
\newblock Springer-Verlag, London, 2006.

\bibitem{Schroeder2006}
W.~Schroeder, K.~Martin, and B.~Lorensen.
\newblock {\em The Visualization Toolkit}.
\newblock Kitware, 4th edition, 2006.

\end{thebibliography}
